\documentclass[twocolumn,epjc3]{svjour3}
\usepackage[T1]{fontenc}
\usepackage{microtype}
\usepackage{mathptmx}
\usepackage{flushend}
\usepackage[numbers,sort&compress]{natbib}
\journalname{Eur. Phys. J. C}

\usepackage{graphicx}
\graphicspath{{data/}}
\usepackage[caption=false]{subfig}
\usepackage{tikz}

\usepackage{amsfonts,amssymb,amsmath,bbm,dsfont,mathrsfs,mathtools}

\usepackage[colorlinks,citecolor=blue,urlcolor=blue,linkcolor=blue]{hyperref}
\usepackage{physics}
\usepackage{mleftright}\mleftright
\usepackage{enumitem}
\setlist[enumerate,1]{nosep,align=left}
\setlist[enumerate,2]{nosep,align=left,leftmargin=*,labelsep=0pt}
\setlist[itemize]{nosep}
\usepackage{booktabs}

\newcommand{\Rho}{\text{P}}
\newcommand{\csubref}[1]{\protect\subref{#1}}

\begin{document}
\title{Geodesic motion around a supersymmetric AdS$_5$ black hole}
\author{Jens-Christian Drawer\thanksref{e1} \and Saskia Grunau\thanksref{e2}}
\institute{Institut f\"ur Physik, Universit\"at Oldenburg, D--26111 Oldenburg,
Germany}
\thankstext{e1}{e-mail: jens-christian.drawer@uni-oldenburg.de}
\thankstext{e2}{e-mail: saskia.grunau@uni-oldenburg.de}

\maketitle

\begin{abstract}
	In this article the geodesic motion of test particles in the spacetime of a
	supersymmetric AdS$_5$ black hole is studied. The equations of motion are
	derived and solved in terms of the Weierstrass $\wp$, $\sigma$, and $\zeta$
	functions. Effective potentials and parametric diagrams are used to analyze
	and characterize timelike, lightlike, and spacelike particle motion and a list
	of possible orbit types is given. Furthermore, various plots of orbits are
	presented.
\end{abstract}

\section{Introduction}
The famous anti-de Sitter/conformal field theory (AdS/CFT) correspondence
provides a relation between gravity and quantum field theory, in particular,
Maldacena \cite{Maldacena:1997re} connected compactifications of string theory
on anti-de Sitter to a conformal field theory. Therefore, black holes that are
asymptotically anti-de Sitter are very interesting to study.

A few years after Kerr \cite{Kerr:1963ud} presented an asymptotically flat
rotating black hole, Carter \cite{Carter:1968ks} came up with the first rotating
asymptotically anti-de Sitter black hole. In five dimensions Hawking et al.
\cite{Hawking:1998kw} found an AdS black hole with two rotation parameters. Five
dimensional AdS black holes are especially interesting since the AdS$_5$/CFT$_4$
correspondence is very well understood and CFT can be described as
$\mathcal{N}=4$ SU(\textit{N}) super Yang-Mills theory. One of the first
supersymmetric AdS$_5$ black hole solutions were found by Gutowski and Reall
\cite{Gutowski:2004ez}. They considered the minimal $D=5$ gauged supergravity
theory described in \cite{Gauntlett:2003wb} and found an asymptotically AdS$_5$
black hole parameterized by its mass, charge, and two equal angular momenta.
Many more black hole solutions in supergravity theories were found, see, e.g.,
\cite{Cvetic:2005zi}.

The motion of test particles is a useful tool to study black holes in various
theories of gravity. The solutions of the equations of motion can be applied to
calculate observable quantities like the shadow of a black hole or the
periastron shift of a bound orbit. Geodesics also provide information on the
structure of a spacetime. In the framework of AdS/CFT, geodesics correspond to
two-point correlators \cite{Balasubramanian:1999zv}. In particular, spacelike
geodesics with both endpoints on the boundary (i.e., escape orbits of particles
with imaginary rest mass) are related to the eikonal approximation of
holographic two-point functions. CFT correlators describe observables on the AdS
boundary.

The Hamilton-Jacobi formalism represents an efficient method to derive the
equations of motion for test particles. In the four-dimensional Kerr spacetime,
Carter \cite{Carter:1968ks} showed that the Hamilton-Jacobi equation for test
particles separates. The resulting equations of motion can be solved
analytically in terms of elliptic functions. In higher dimensions, or in
spacetimes with a cosmological constant, the analytical solutions of the
geodesic equations often require hyperelliptic functions
\cite{Kraniotis:2005zm,Kraniotis:2006ux,Fujita:2009bp,Hackmann:2009nh,
Hackmann:2010zz,Grunau:2017uzf}. The geodesics in a rotating supersymmetric
black hole spacetime were analyzed in \cite{Gibbons:1999uv,Diemer:2013fza},
where the complete analytical solution of the geodesics equations in the
supersymmetric Breckenridge-Myers-Peet-Vafa (BMPV) \cite{Breckenridge:1996is}
spacetime was presented. Here we will study the geodesic motion of test
particles around the supersymmetric, asymptotically AdS$_5$ black hole of
Gutowski and Reall \cite{Gutowski:2004ez}.

The article is structured as follows. We derive the equations of motion in
Section~\ref{sec:derivation} and give a complete classification of the geodesics
in Section~\ref{sec:classification}. In Section~\ref{sec:solution} we solve the
equations of motion analytically in terms of the Weierstrass $\wp$, $\sigma$,
and $\zeta$ functions. Finally we present some example plots of the orbits in
Section~\ref{sec:orbits} and conclude in Section~\ref{sec:conclusion}.

\section{\label{sec:derivation}The supersymmetric AdS\texorpdfstring{$_5$}{5}
black hole}
Gutowski and Reall \cite{Gutowski:2004ez} found a one-parameter family of
supersymmetric AdS$_5$ black holes. The metric is given by
\begin{equation}
	\begin{split}
		\dd{s^2}=&-{f}^{2}{\dd{t}}^{2}-2{f}^{2}\Psi{\dd{t}}\sigma_L^3
			+{U(R)}^{-1}{\dd{R}}^{2}\\
		&+\frac{{R}^{2}}{4}\left[(\sigma_L^1)^2+(\sigma_L^2)^2
			+\Lambda(R)(\sigma_L^3)^2\right]\,, \label{eq:metricequation}
	\end{split}
\end{equation}
where the $\sigma_L^i$ can be expressed in terms of the Euler angles
$(\theta,\phi,\psi)$ as
\begin{subequations}
	\begin{align}
		\sigma_L^1&=\sin\phi\dd{\theta}-\cos\phi\sin\theta\dd{\psi}\,,\\
		\sigma_L^2&=\cos\phi\dd{\theta}+\sin\phi\sin\theta\dd{\psi}\,,\\
		\sigma_L^3&=\dd{\phi}+\cos\theta\dd{\psi}\,,
	\end{align}
\end{subequations}
and the metric functions are
\begin{align}
	f&=1-\frac{R_{0}^{2}}{{R}^{2}}\,,\\
	\Psi&=-\frac{\epsilon{R}^{2}}{2l}\left(1+{\frac{2R_{0}^{2}}{{R}^{2}}}
		+{\frac{3R_{0}^{4}}{2{R}^{2}\left({R}^{2}-R_{0}^{2}\right)}}\right)\,,\\
	U&=\left(1-\frac{R_0^{2}}{{R}^{2}}\right)^{\!2}\left(1+\frac{2R_0^{2}}{{l}^{2}}
		+\frac{{R}^{2}}{{l}^{2}}\right)\,,\\
	\Lambda&=1+\frac{R_0^{6}}{{l}^{2}{R}^{4}}-\frac{R_0^{8}}{4{l}^{2}{R}^{6}}\,.
\end{align}
The Maxwell potential is
\begin{equation}
	A=\frac{\sqrt{3}}{2}\left[\left(1-\frac{R^2}{R_0^2}\right)\dd{t}
		+\frac{\epsilon R_0^4}{4lR^2}\sigma_L^3\right]\,.
\end{equation}
Here $R_0$ is the radial coordinate of the black hole's degenerate horizon,
$\epsilon=\pm1$ is the sign of its angular momentum, and $l$ is the AdS radius.
Note that $\epsilon$ can be absorbed into $l$, consequently, in the following
analysis of the geodesics we set $\epsilon=1$ but examine the geodesic motion
for arbitrary sign of $l$.

It can be shown that the solution is asymptotically AdS$_5$, by using the
coordinate transformation $\phi'=\phi+\frac{2\epsilon}{l}t$, see
\cite{Gutowski:2004ez}. It has the $R\times S^3$ Einstein universe as its
conformal boundary and the $S^3$ has the radius $l$. Boundary as well as bulk
time translations are generated by $\pdv{t}$. However, as for all rotating AdS
black holes, there is another timelike Killing vector field in the bulk
\begin{equation}
	V=\pdv{t}+\frac{2\epsilon}{l}\pdv{\phi'}\,.
\end{equation}
If $V$ is used to generate time translations, we are working in a co-rotating
frame and there is no ergoregion. If, on the other hand, $\pdv{t}$ generates
time translations, then an ergoregion exists.

The metric (\ref{eq:metricequation}) is characterized by its conserved
quantities associated with symmetries of the conformal boundary, which was shown
for asymptotically AdS spacetimes of dimension $D\geq4$ by Ashtekar and Das
\cite{Ashtekar:1999jx}. In this case, the black hole's conserved quantities can
be defined by an Ashtekar and Das mass
\begin{equation}
	M=\frac{3\pi R_0^2}{4G}\left(1+\frac{3R_0^2}{2l^2}
		+\frac{2R_0^4}{3l^4}\right)\,,
\end{equation}
an angular momentum with respect to $\phi'=\phi+\frac{2\epsilon}{l}t$ of
\begin{equation}
	J'=\frac{3\epsilon\pi R_0^4}{8Gl}\left(1+\frac{2R_0^2}{3l^2}\right)\,,
\end{equation}
a vanishing angular momentum with respect to $\psi$, an energy of
\begin{equation}
	E=M+\frac{3\pi l^2}{32G}\,,
\end{equation}
and a charge of
\begin{equation}
	Q=\frac{\sqrt{3}\pi R_0^2}{2G}\left(1+\frac{R_0^2}{2l^2}\right)\,.
\end{equation}
We then obtain
\begin{equation}
	M-\frac{2}{l}|J'|=\frac{\sqrt{3}}{2}|Q|
\end{equation}
and therefore the solution saturates the BPS bound, see also
\cite{Gutowski:2004ez}. Note that the conserved charges by Ashtekar and Das are
only correct for a special class of solutions, where the non-normalizable modes
of all matter fields vanish and the Ricci curvature of the boundary metric also
vanishes, unless $D\leq 4$. In \cite{Papadimitriou:2005ii} the authors present
well defined conserved charges for general asymptotically AdS black holes in the
presence of matter.

The solution (\ref{eq:metricequation}) has a smooth event horizon at $R=R_0$ and
the spatial geometry of the horizon is a squashed $S^3$
\begin{equation}
	\dd{s_3^2}=\frac{R_0^2}{4}\left[\left({\sigma_L^1}''\right)^{\!2}
		+\left({\sigma_L^2}''\right)^{\!2}
		+\left(1+\frac{3R_0^2}{4l^2}\right)\left({\sigma_L^3}''\right)^{\!2}\right]\,,
\end{equation}
where the ${\sigma_L^i}''$ are defined as the $\sigma_L^i$ with
$\phi''=\phi+\frac{4f^2\Psi}{R^2U}r$ instead of $\phi$. Behind the event horizon
there is a curvature singularity at $R=0$ surrounded by a region of closed
timelike curves \cite{Gutowski:2004ez}.

\subsection{The equations of motion}
We use the Hamilton-Jacobi formalism to obtain the equations of motion for test
particles in the spacetime of a supersymmetric black hole. To solve the
Hamilton-Jacobi equation of an uncharged particle
\begin{equation}
	-2\pdv{S}{\tau}=g^{\mu\nu}\pdv{S}{x^\mu}\pdv{S}{x^\nu}\,,
	\label{eq:hamiltonjacobi}
\end{equation}
we make the ansatz for the action $S$
\begin{equation}
	S=\frac{1}{2}\delta\tau-Et+L\phi+J\psi+S_R(R)+S_\theta(\theta)\,.
	\label{eq:ansatz}
\end{equation}
Here $E$ is the particle's conserved energy, $L$ and $J$ are its conserved
angular momenta along $\phi$ and $\psi$, respectively, $\tau$ is an affine
parameter along the geodesic, and $\delta$ is equal to 0 for light, equal to 1
for particles of positive mass, and equal to $-1$ for particles of imaginary
mass. The case $\delta=-1$ corresponds to spacelike geodesics and is of
relevance for AdS/CFT if the geodesics' endpoints are on the boundary
$R\rightarrow\infty$. This is discussed in more detail in
Section~\ref{sec:spacelike}.

Using this ansatz and the metric (\ref{eq:metricequation}), the Hamilton-Jacobi
equation (\ref{eq:hamiltonjacobi}) becomes
\begin{equation}
	\begin{split}
		-\delta&=\frac{-R^2\Lambda E^2}{\left(4\Phi^2f^2+R^2\Lambda\right)f^2}
			+\frac{8\Phi EL}{4\Phi^2f^2+R^2\Lambda}\\
		&+\frac{\left(16\Phi^2\cos^2\theta f^2
			+4\Lambda\cos^2\theta R^2+4R^2\sin^2\theta\right)L^2}{
			R^2\sin^2\theta\left(4\Phi^2f^2+R^2\Lambda\right)}\\
		&-\frac{8\cos\theta JL}{R^2\sin^2\theta}+\frac{4J^2}{R^2\sin^2\theta}
			+\frac{4}{R^2}\left(\pdv{S_\theta}{\theta}\right)^{\!2}
			+U\left(\pdv{S_R}{r}\right)^{\!2}\,.
	\end{split}
\end{equation}
One can separate the Hamilton-Jacobi equation (\ref{eq:hamiltonjacobi}) by terms
in $R$ and $\theta$ in
\begin{equation}
	\begin{split}
		\delta R^2-\frac{R^4\Lambda E^2}{\left(4\Phi^2f^2+R^2\Lambda\right)f^2}
			+\frac{8R^2\Phi EL}{4\Phi^2f^2+R^2\Lambda}&\\
		+\frac{4R^2L^2}{4\Phi^2f^2+R^2\Lambda}+R^2U\left(\pdv{S_R}{r}\right)^{\!2}&=K
		\label{eq:separated-R}
	\end{split}
\end{equation}
and
\begin{equation}
	-\frac{4\cos^2\theta L^2}{\sin^2\theta}+\frac{8\cos\theta JL}{\sin^2\theta}
		-\frac{4J^2}{\sin^2\theta}-4\left(\pdv{S_\theta}{\theta}\right)^{\!2}=K\,.
 	\label{eq:separated-theta}
\end{equation}
Here we introduced $K$ as a separation constant known as the Carter
\cite{Carter:1968rr} constant. Now we can solve Eq.~(\ref{eq:separated-R}) for
$\pdv{S_R}{r}$ and Eq.~(\ref{eq:separated-theta}) for $\pdv{S_\theta}{\theta}$,
which can then be used to substitute the functions $S_R$ and $S_\theta$ in the
ansatz (\ref{eq:ansatz}) (no need to actually compute the integrals here).
Finally, the equations of motion can be deduced with a variational method; the
derivatives of the action $S$ with respect to the constants of motion can be set
to zero.

With the help of the Mino \cite{Mino:2003yg} time $\gamma$ given by
$R^2\dd{\gamma}=\dd{\tau}$ to remove the factor $R^2$ from all equations and the
substitution $r=R^2$, this yields five differential equations of motion
\begin{equation}
	\left(\dv{r}{\gamma}\right)^{\!2}=\Rho(r)\,, \label{eq:motionr}
\end{equation}
\begin{equation}
	\left(\dv{\theta}{\gamma}\right)^{\!2} = \Theta(\theta)\,,
	\label{eq:motiontheta}
\end{equation}
\begin{equation}
	\begin{split}
		\dv{\phi}{\gamma}=&-l{\frac{2E{r}^{2}+(2E-Ll)r+Ll-E}{({l}^{2}+r+2)(r-1)}}\\
		&-\frac{\cos\theta}{\sin^2\theta}\left(J-L\cos\theta\right)\,,
		\label{eq:motionphi}
	\end{split}
\end{equation}
\begin{equation}
	\dv{\psi}{\gamma}=\frac{1}{\sin^2\theta}\left(J-L\cos\theta\right)\,,
	\label{eq:motionpsi}
\end{equation}
\begin{equation}
	\dv{t}{\gamma}=
		\frac{(4E{l}^{2}+2Ll){r}^{3}+(4E-3Ll)r+Ll-E}{4({l}^{2}+r+2)(r-1)^{2}}\,.
	\label{eq:motiont}
\end{equation}
The polynomial $\Rho$ and the function $\Theta$ are
\begin{equation}
	\begin{split}
		\Rho=&-4{\frac{\delta{r}^{4}}{{l}^{2}}}+4\left({E}^{2}+{\frac{EL}{l}}
			-\delta-\frac{1}{4}{\frac{K}{{l}^{2}}}\right){r}^{3}\\
		&+\left(-{L}^{2}-K+8\delta+12{\frac{\delta}{{l}^{2}}}\right){r}^{2}\\
		&+2\left({L}^{2}-3{\frac{EL}{l}}+K-2\delta+2{\frac{{E}^{2}}{{l}^{2}}}
			+\frac{3}{2}{\frac{K}{{l}^{2}}}-4{\frac{\delta}{{l}^{2}}}\right)r\\
		&-{L}^{2}+2{\frac{EL}{l}}-K-{\frac{{E}^{2}}{{l}^{2}}}-2{\frac{K}{{l}^{2}}}\,,
		\label{eq:Rho}
	\end{split}
\end{equation}
\begin{equation}
	\Theta=K-\frac{1}{\sin^2\theta}\left(J-L\cos\theta\right)^2\,. \label{eq:Theta}
\end{equation}
To simplify the equations of motion, dimensionless quantities were introduced by
scaling with $R_0$
\begin{align}
	R&\rightarrow R_0R\,, & t&\rightarrow R_0t\,, & \tau&\rightarrow R_0\tau\,, &
		l&\rightarrow R_0l\,, \nonumber\\
	L&\rightarrow \frac{1}{4}R_0L\,, & J&\rightarrow \frac{1}{4}R_0J\,, &
		K&\rightarrow \frac{1}{4}R_0^2K\,.
\end{align}
This was achieved by setting $R_0=1$ and canceling factors of 4 in front of $L$,
$J$, and $K$ for convenience.

\section{\label{sec:classification}Classification of the geodesics}
The properties of the geodesics are determined by the polynomial $\Rho$ in
Eq.~(\ref{eq:Rho}) and the function $\Theta$ in Eq.~(\ref{eq:Theta}). The
characteristics of $\Theta$ and $\Rho$ are given by the particle's constants of
motion (energy, angular momenta, Carter constant, $\delta$ parameter) and the
metric's (positive or negative) AdS radius. In this section, features of the
function $\Theta$ and the polynomial $\Rho$ -- and therefore the types of orbits
-- for various sets of constants of motion are studied. This is done analogously
to \cite{Grunau:2010gd}, where geodesic motion of electrically and magnetically
charged test particles in the Reissner-Nordstr\"om spacetime has been examined.

\subsection{\label{sec:classificationtheta}The \texorpdfstring{$\theta$}{theta}
motion}
To obtain real values of $\theta$, the requirement $\Theta\geq0$ has to be met.
From this it follows that $K\geq0$. The substitution $\xi=\cos\theta$ turns
Eq.~(\ref{eq:motiontheta}) into
\begin{equation}
	\left(\dv{\xi}{\gamma}\right)^{\!2} = \Theta_\xi \quad \text{with} \quad
	\Theta_\xi\coloneqq a\xi^2+b\xi+c\,, \label{eq:derivationtheta}
\end{equation}
where $a=-L^2-K$, $b=2LJ$, and $c=K-J^2$. Since $K\geq0$ holds, it follows
$a\leq0$. The zeros of the second degree polynomial $\Theta_\xi$ correspond to
angles that confine the particle's $\theta$ motion. Note that for vanishing $L$
in Eqs.~(\ref{eq:motiontheta}) and (\ref{eq:motionpsi}) the test particle's
motion is planar in the 3-dimensional subspace given by the spherical
coordinates $(R,\theta,\psi)$ as in the Schwarzschild case.

\begin{figure} 
	\subfloat[Case \ref{itm:case1}, $K=14$.]{%
		\includegraphics[width=.333\linewidth]{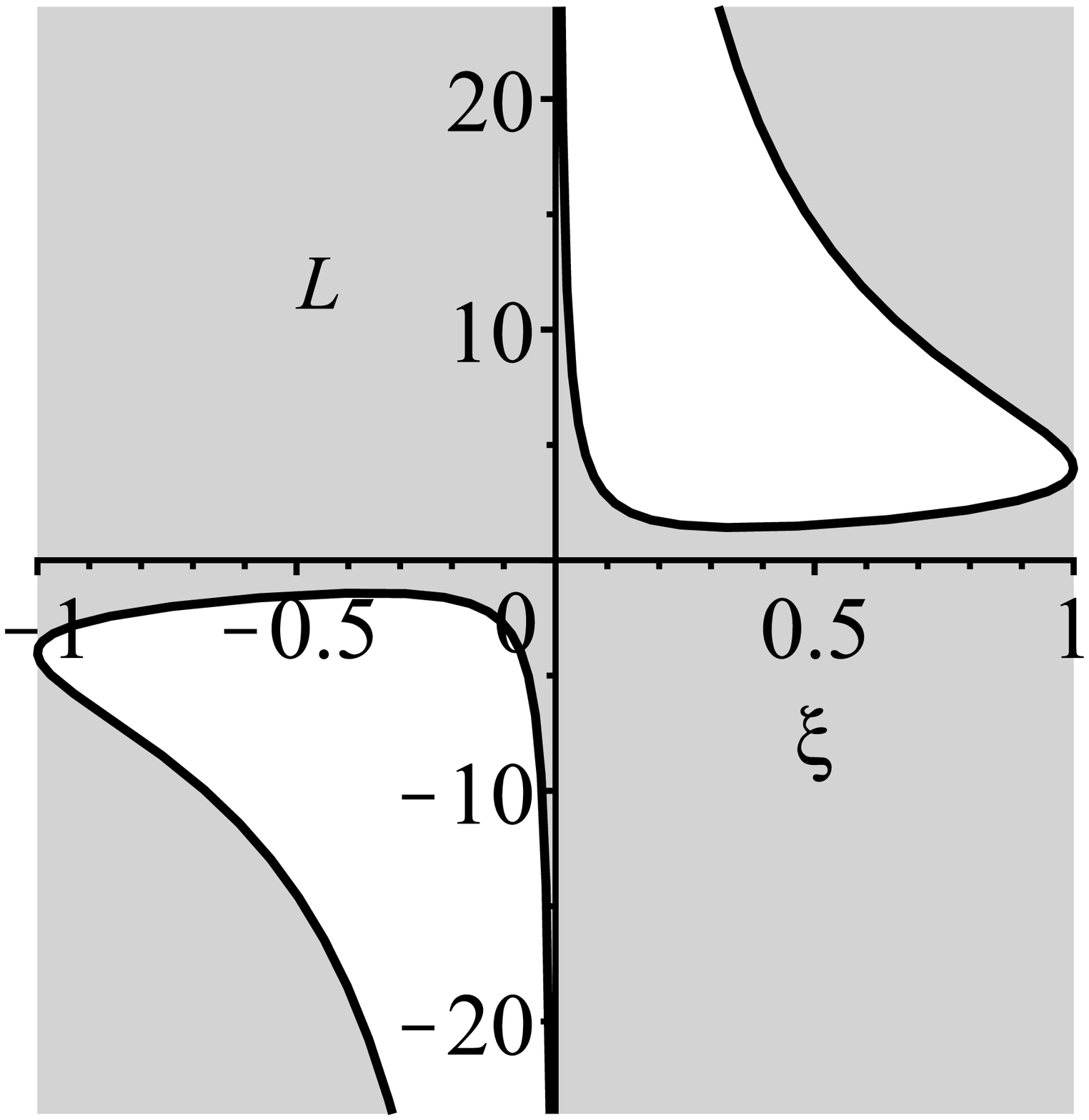}%
	}\hfill
	\subfloat[Case \ref{itm:case2}, $K=16$.]{%
		\includegraphics[width=.333\linewidth]{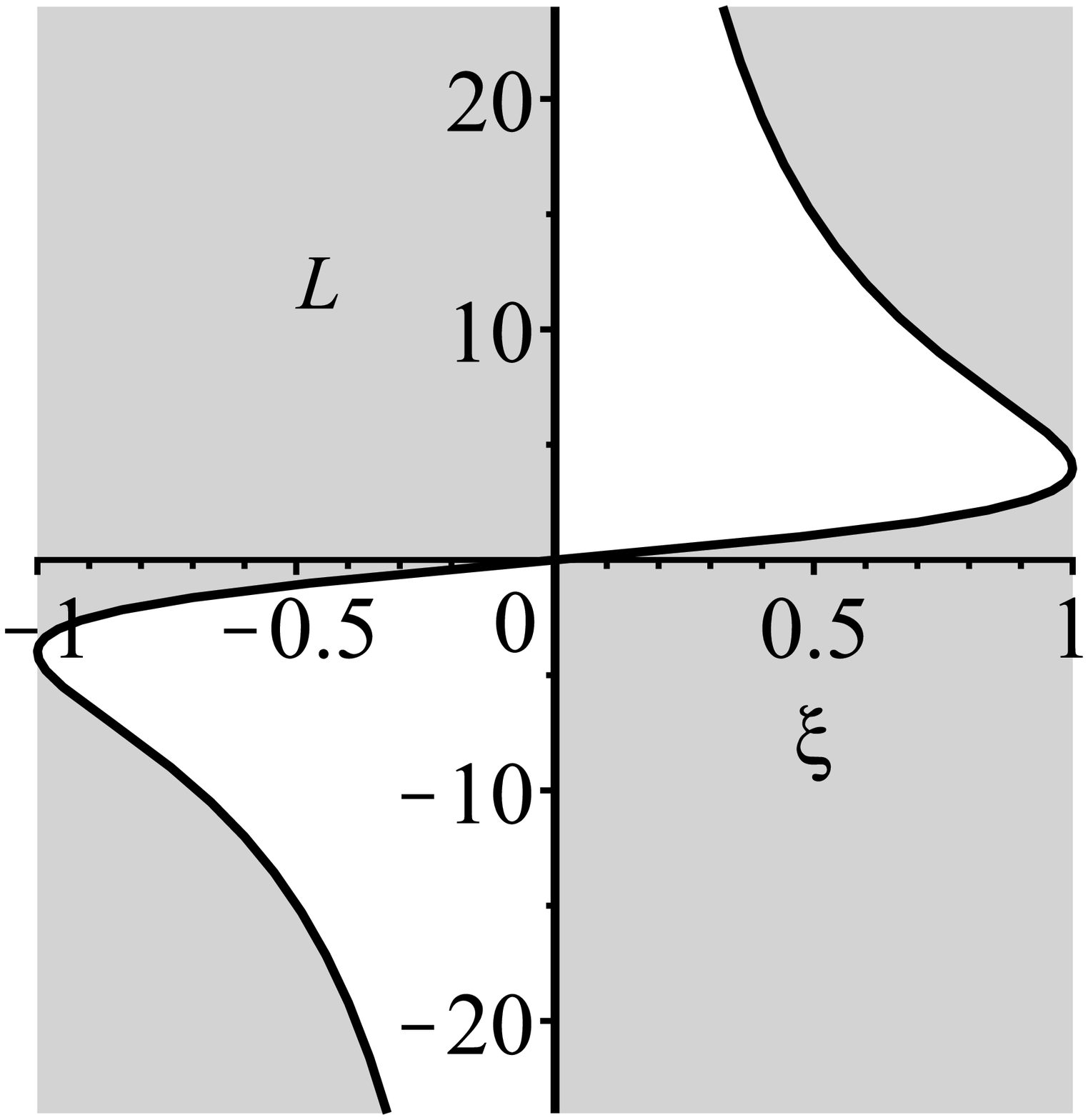}%
		\label{fig:effpotthetab}%
	}\hfill
	\subfloat[Case \ref{itm:case3}, $K=18$.]{%
		\includegraphics[width=.333\linewidth]{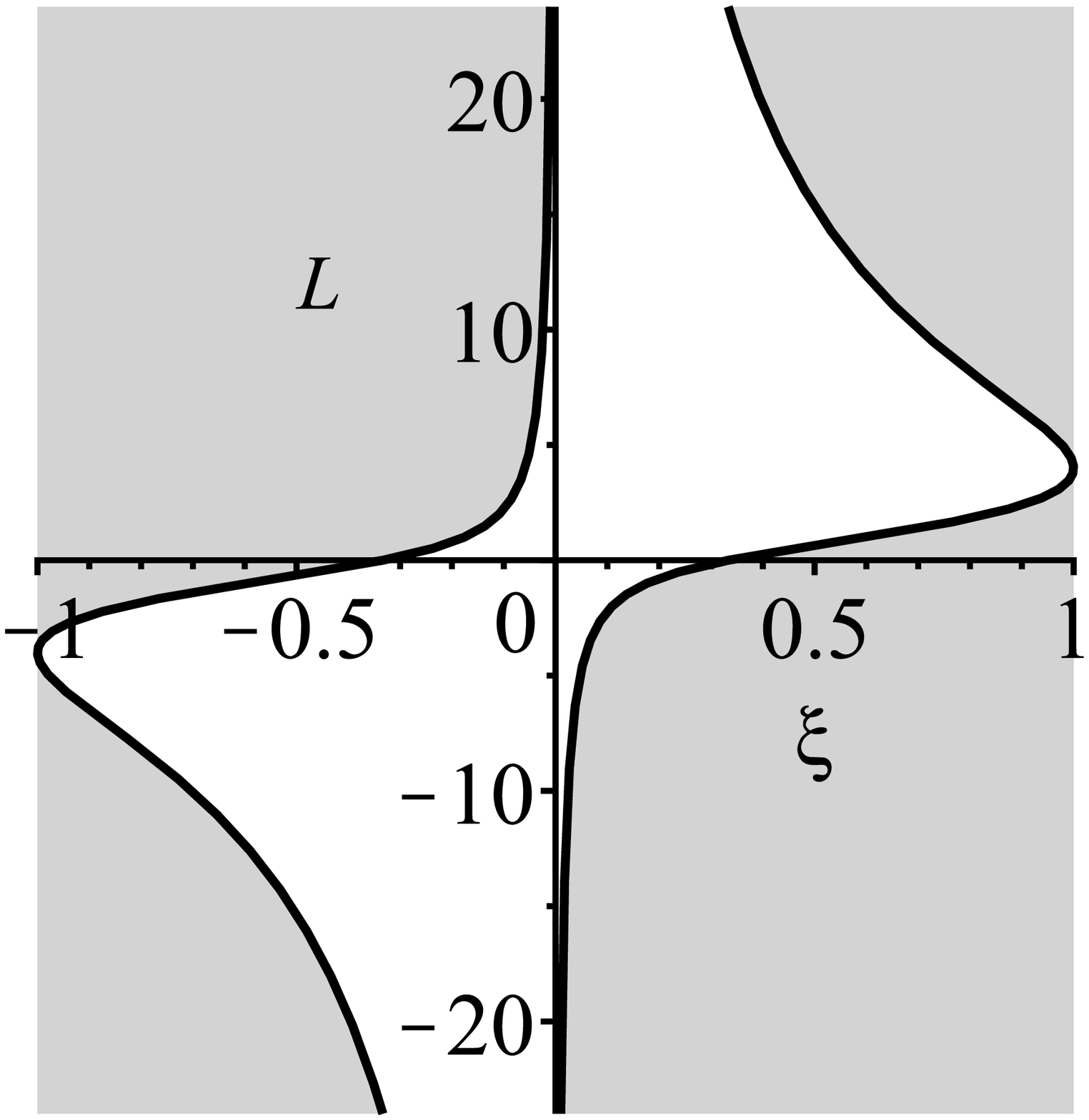}%
	}
	\caption{\label{fig:effpottheta}%
		Allowed $\xi$ motion in dependence of $L$ for $J=4$ and varying $K$.
		Physically forbidden areas are marked in gray.}
\end{figure}

The $\Theta_\xi$ polynomial's discriminant is given by $D=b^2-4ac$ and can be
expressed as $D=4K\kappa$ with $\kappa=K+L^2-J^2$. $\Theta_\xi$ describes a
downward opened parabola with zeros
\begin{equation}
	\xi_0=\frac{LJ\pm\sqrt{K\kappa}}{L^2+K} \in[-1,1] \label{eq:zerostheta}
\end{equation}
and maximum at $\big(\frac{LJ}{L^2+K},\frac{K\kappa}{L^2+K}\big)$. A real
solution $\theta$ implies real zeros of $\Theta_\xi$ and thus requires $D\geq0$.
While $|L|\geq|J|$ is sufficient, other cases require an upper limit of $|J|$
given by $J_{\text{max}}=\sqrt{K+L^2}$. For symmetric motion with respect to the
equatorial plane $L$ or $J$ have to vanish. Other cases are depending on the
sign of $K-J^2$:
\begin{enumerate}
	\item\label{itm:case1} $K<J^2$: The zeros of $\Theta_\xi$ are either both
	positive or both negative, which confines the particle's motion to
	$\theta\in[0,\pi/2)$ for $LJ>0$ and $\theta\in(\pi/2,\pi]$ for $LJ<0$.
	\item\label{itm:case2} $K=J^2$: The zeros of $\Theta_\xi$ are
	$\big\lbrace0,\frac{2LJ}{L^2+K}\big\rbrace$ with $\theta\in[0,\pi/2]$ for
	$LJ>0$ and $\theta\in[\pi/2,\pi]$ for $LJ<0$. With the additional condition
	$|L|=|J|$, the orbit fills an entire hemisphere [see
	Fig.~\ref{fig:effpotthetab} at $L=\pm4$].
	\item\label{itm:case3} $K>J^2$: One zero of $\Theta_\xi$ is positive and one
	is negative, allowing the particle to cross the equatorial plane and
	$\theta\in[0,\pi]$.
\end{enumerate}
Similarly to an effective potential, this behavior can be seen in
Fig.~\ref{fig:effpottheta}, where the allowed area of motion with respect to $L$
is shown for different choices of $K$.

In case of a double zero $\xi_0$ of $\Theta_\xi$, the particle's motion is
confined to a cone of opening angle $\arccos\xi_0$, which simplifies
Eqs.~(\ref{eq:motionphi}) and (\ref{eq:motionpsi}). This is possible for $K=0$
or $\kappa=0$ in three cases:
\begin{enumerate}
	\item $K=0,\,\kappa>0$: $\xi_0=J/L$ for $|L|>|J|$.
	\item $K>0,\,\kappa=0$: $\xi_0=L/J$ for $|L|<|J|$.
	\item $K=0,\,\kappa=0$: $\xi_0=\text{sign}LJ$ for $|L|=|J|$.
\end{enumerate}

Since the $\theta$ motion is not depending on the particle's mass parameter
$\delta$, all results hold for all particle types.

\subsection{The \texorpdfstring{$r$}{r} motion}
\subsubsection{Possible types of orbits}
For a degenerate horizon at $r=1$ the following types of orbits can be found for
this spacetime:
\begin{enumerate}
	\item \textit{Escape orbits} (EO) with range $[r_1,\infty)$ and $1<r_1$.
	\item \textit{Two-world escape orbits} (TEO) with range $[r_1,\infty)$ and
	$r_1<1$.
	\item \textit{Periodic bound orbits} (BO) with range $[r_1,r_2]$ and
	$r_1<r_2<1$ or $1<r_1<r_2$.
	\item \textit{Many-world periodic bound orbits} (MBO) with range $[r_1,r_2]$
	and $r_1<1<r_2$.
	\item \textit{Terminating orbits} (TO) with range $[0,r_2]$ and $r_2<1$ or
	with range $[0,\infty)$.
\end{enumerate}

\subsubsection{\label{sec:classificationr}Analysis of the radial motion}
\begin{figure*} 
	\subfloat[$\delta = 1$.]{%
		\includegraphics[width=.333\linewidth]{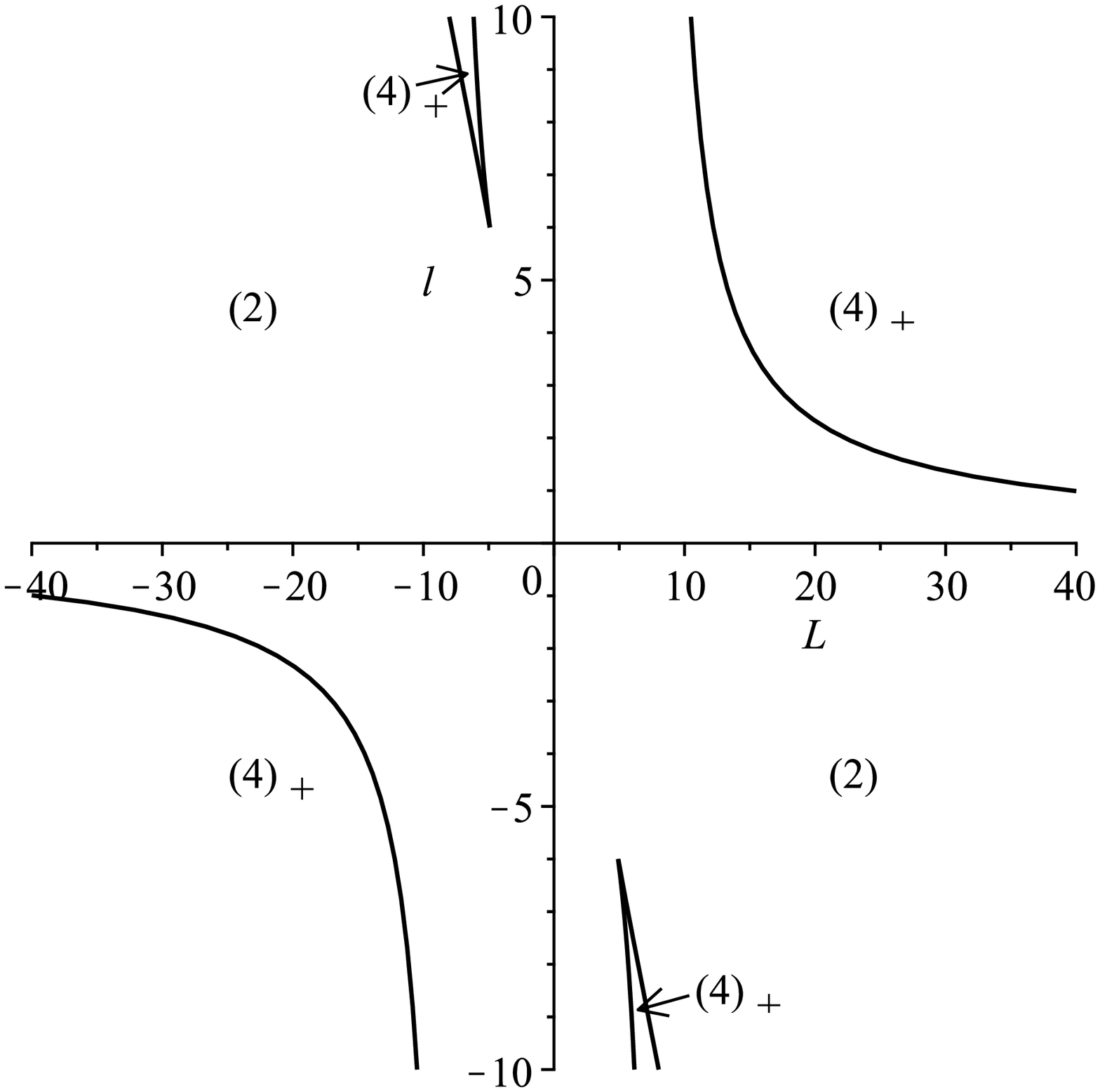}\label{fig:paraplot2a}%
	}\hfill
	\subfloat[$\delta = 0$.]{%
		\includegraphics[width=.333\linewidth]{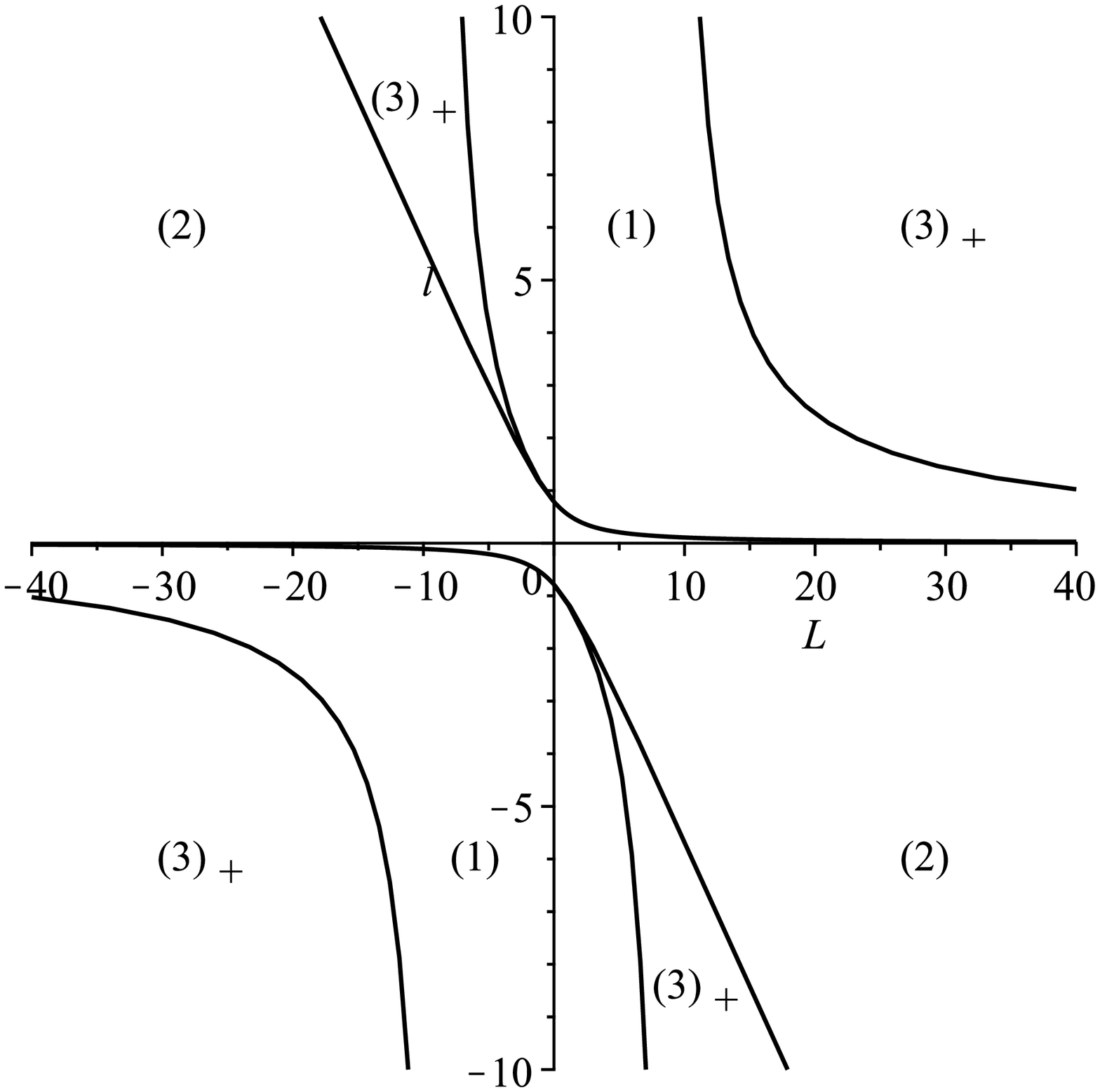}%
	}\hfill
	\subfloat[$\delta = -1$.]{%
		\includegraphics[width=.333\linewidth]{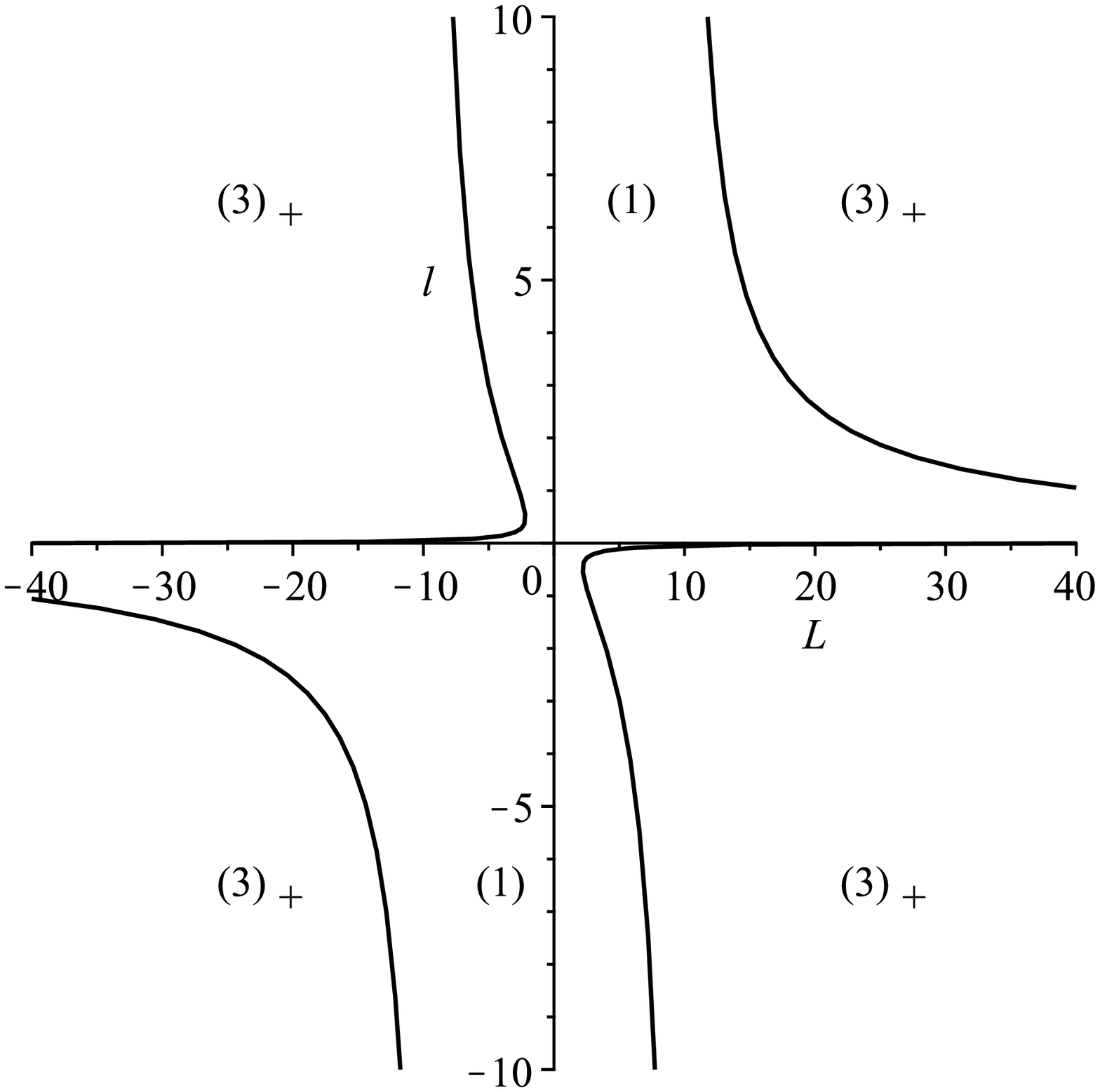}\label{fig:paraplot2c}%
	}\\
	\subfloat[$\delta = 1$.]{%
		\includegraphics[width=.333\linewidth]{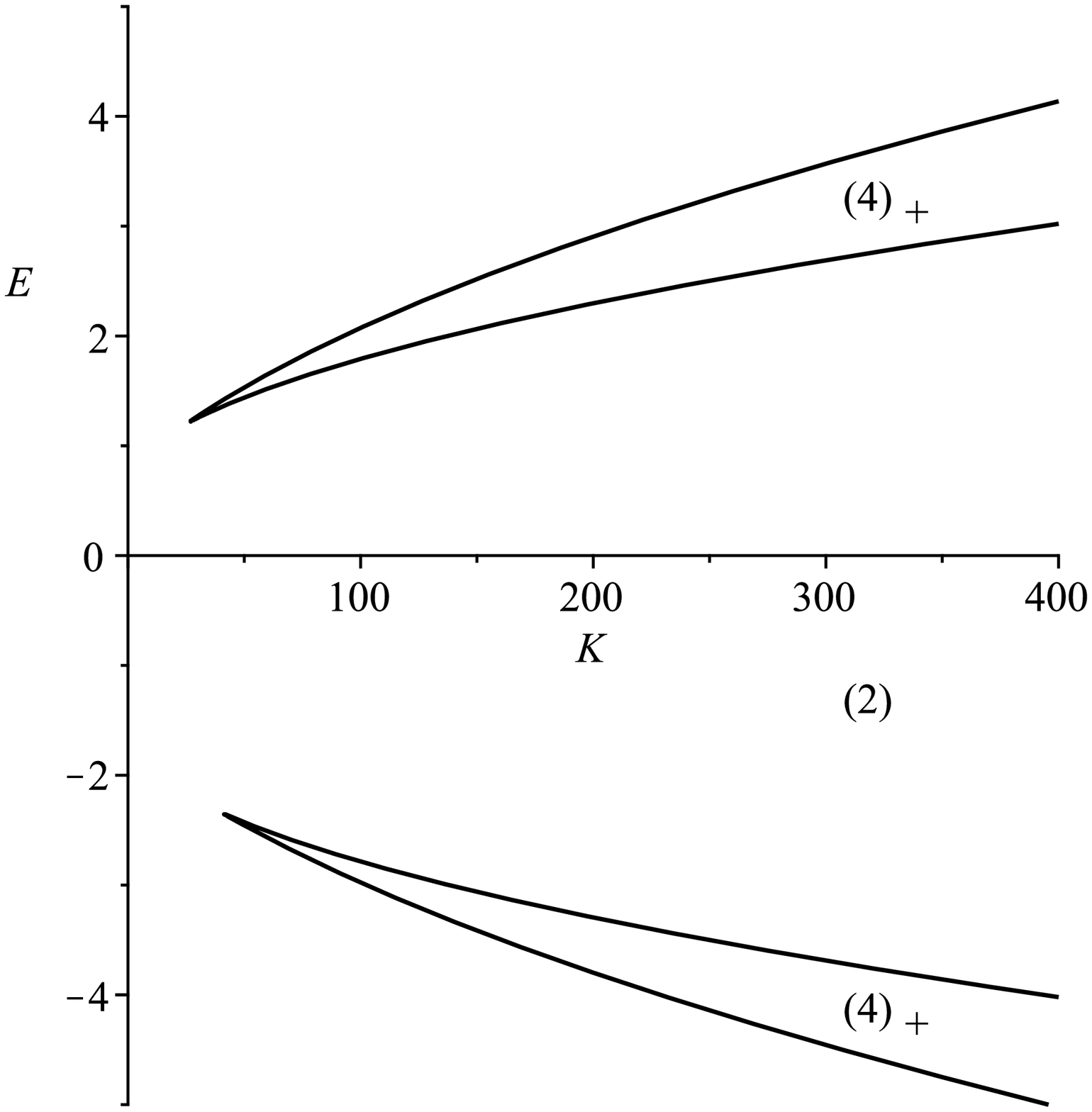}\label{fig:paraplot1a}%
	}\hfill
	\subfloat[$\delta = 0$.]{%
		\includegraphics[width=.333\linewidth]{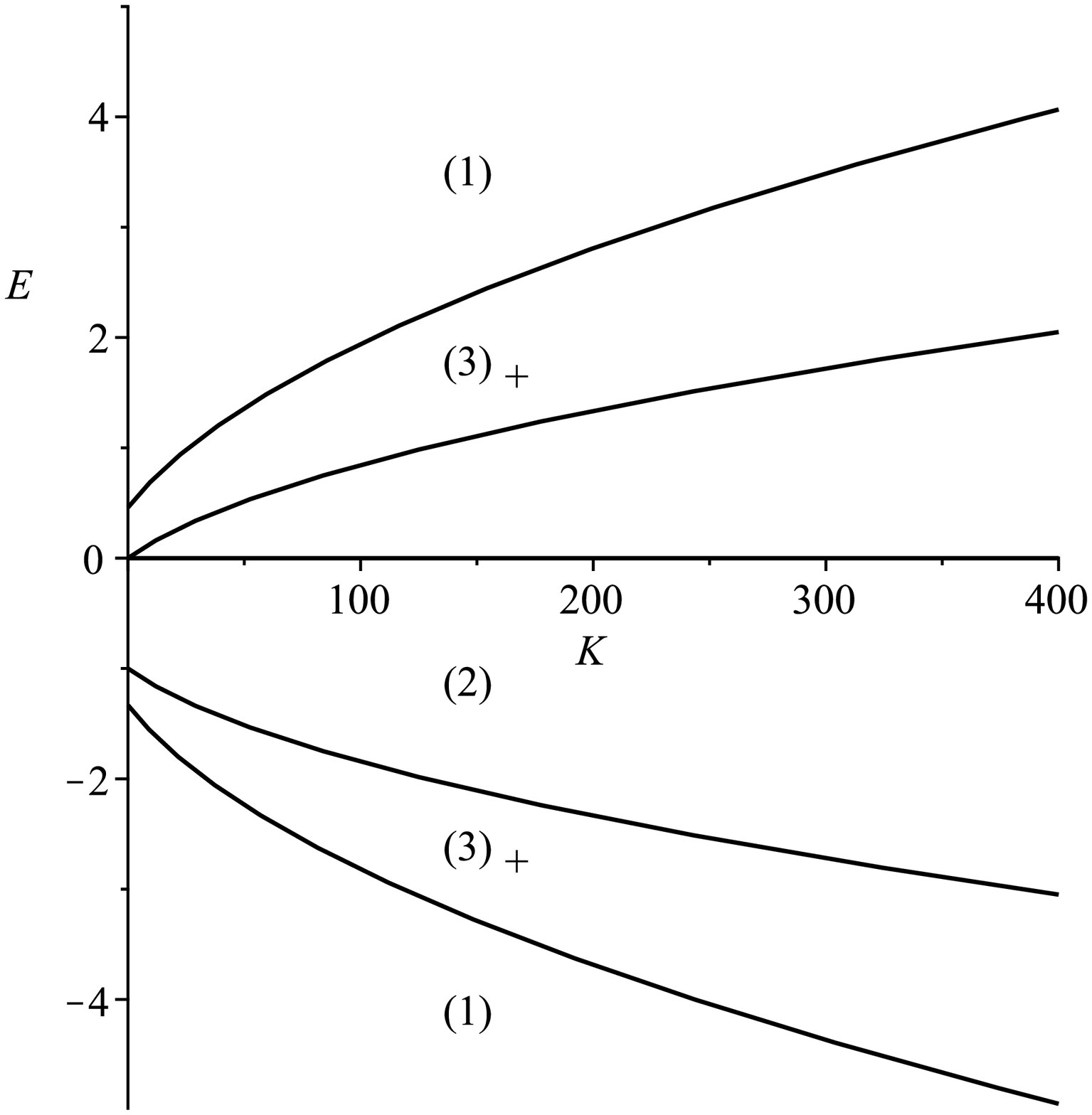}\label{fig:paraplot1b}%
	}\hfill
	\subfloat[$\delta = -1$.]{%
		\includegraphics[width=.333\linewidth]{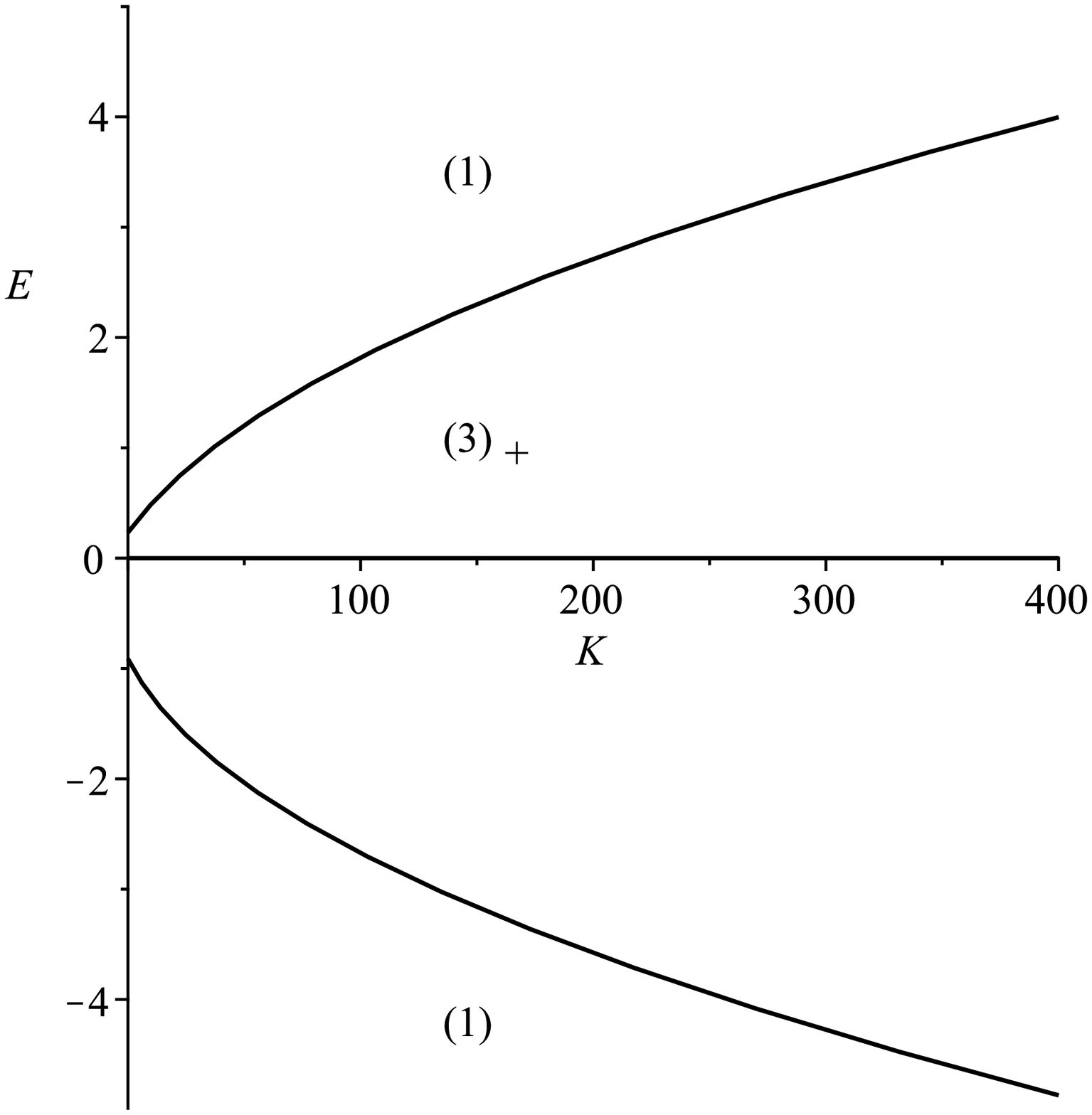}\label{fig:paraplot1c}%
	}\\
	\subfloat[$\delta = 1$.]{%
		\includegraphics[width=.333\linewidth]{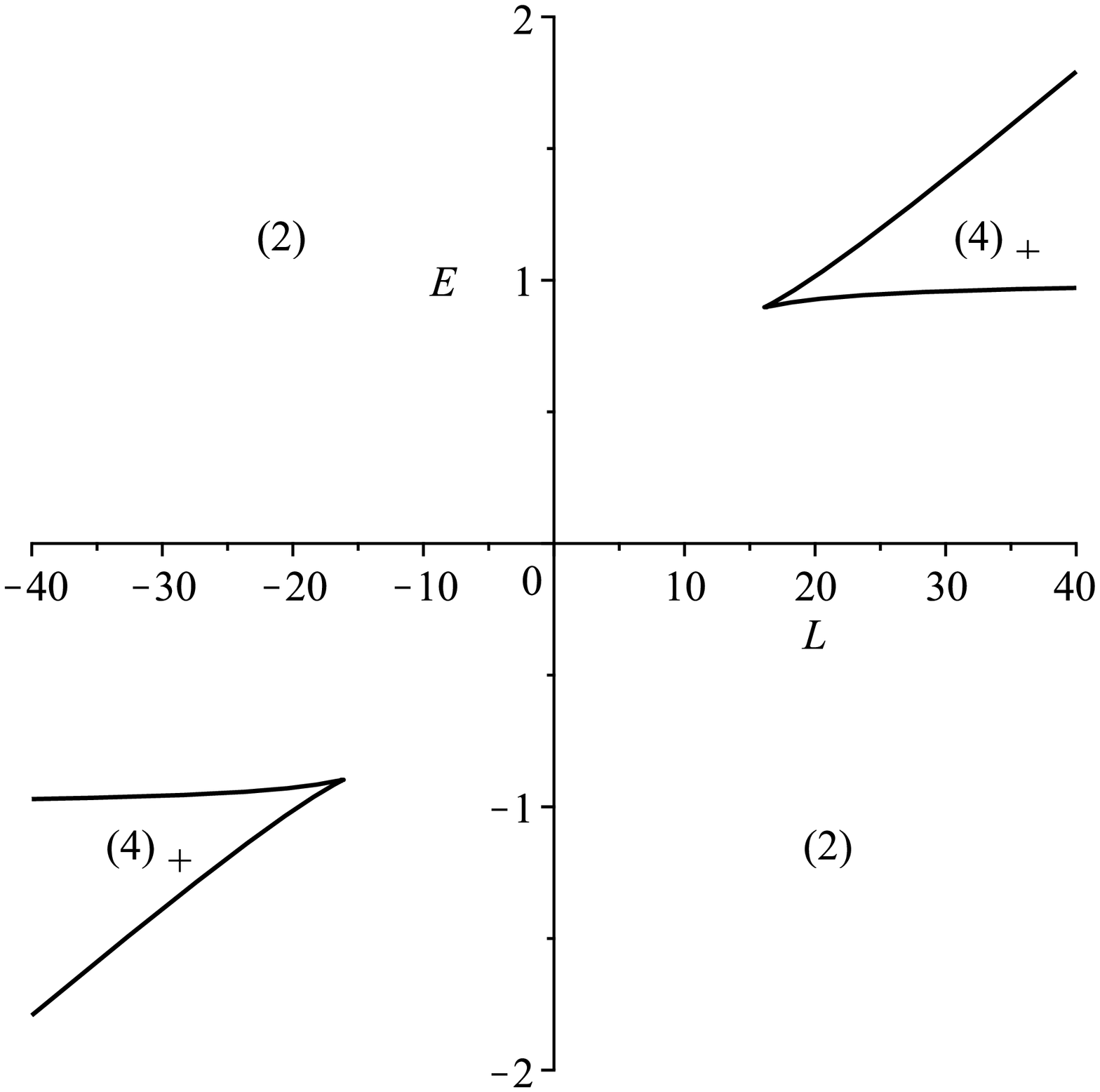}\label{fig:paraplot3a}%
	}\hfill
	\subfloat[$\delta = 0$.]{%
		\includegraphics[width=.333\linewidth]{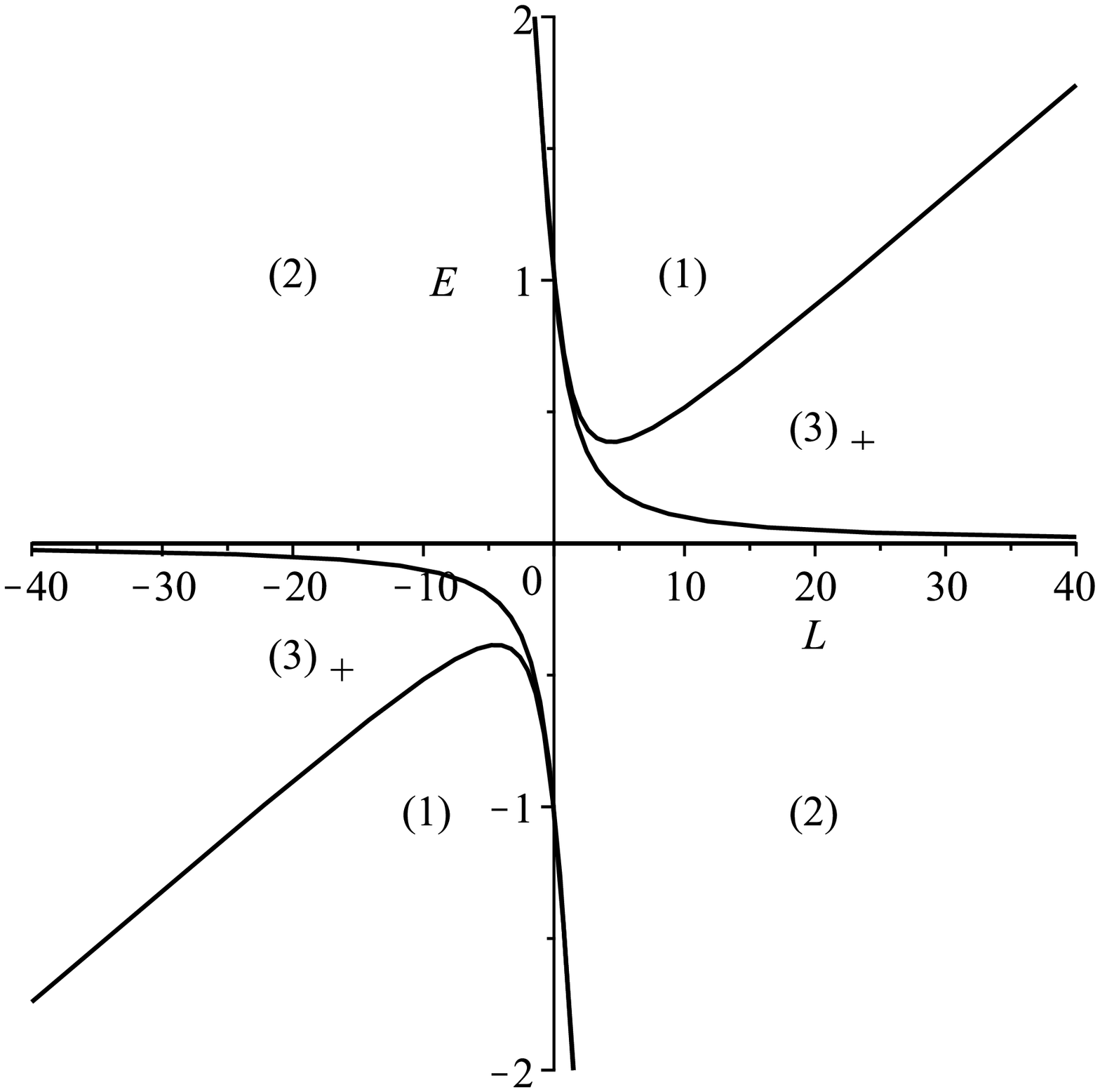}%
	}\hfill
	\subfloat[$\delta = -1$.]{%
		\includegraphics[width=.333\linewidth]{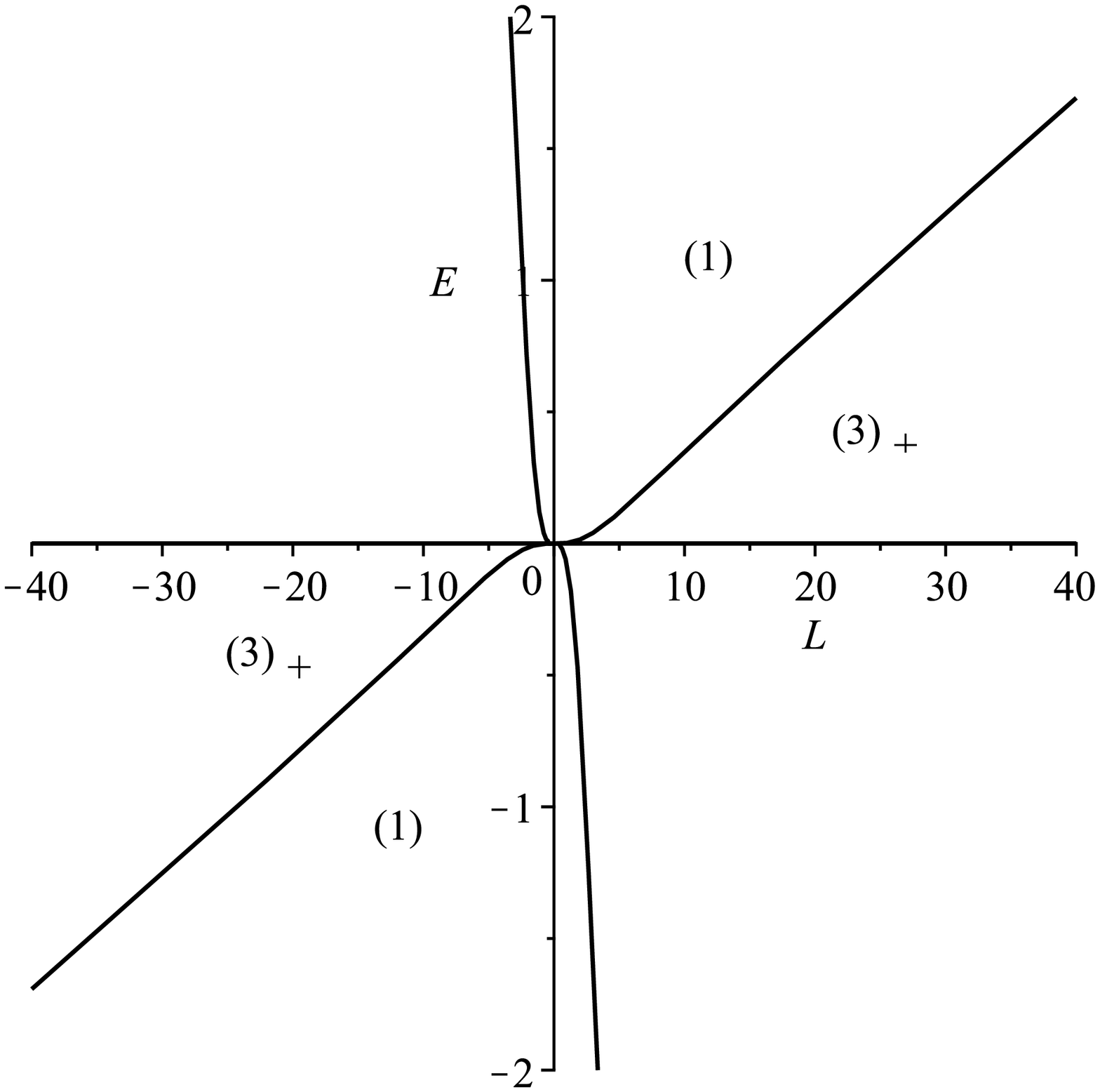}\label{fig:paraplot3c}%
	}
	\caption{\label{fig:paraplot}%
		Parametric $L$-$l$, $K$-$E$, and $L$-$E$ diagrams for varying mass parameter
		$\delta$ and parameters $E = 1.8,\, K = 8$
		[\csubref{fig:paraplot2a}--\csubref{fig:paraplot2c}], $L = 4,\, l = 4$
		[\csubref{fig:paraplot1a}--\csubref{fig:paraplot1c}], and $l = 1,\, K = 4$
		[\csubref{fig:paraplot3a}--\csubref{fig:paraplot3c}], showing regions
		(1)--(4). This marks the number of positive zeros of the polynomial $\Rho$ in
		Eq.~(\ref{eq:Rho}), i.e., lines corresponding to transitions of negative roots
		to complex ones are not shown. Associated orbit types can be found in
		Table~\ref{tab:orbits} and are described in Section~\ref{sec:classificationr}.
		In cases \csubref{fig:paraplot1a}--\csubref{fig:paraplot3c} one double zero of
		$\Rho$ for $E=0$ and all $K$ or $L$ can be found. Since no regions of
		different numbers of zeros are separated by these lines (see
		Fig.~\ref{fig:effpotrab} for $E=0$), they are omitted.}
\end{figure*}

To obtain real values of $r$ from Eq.~(\ref{eq:motionr}), the requirement
$\Rho\geq0$ has to be met. Radial regions of physically allowed motion are
separated from forbidden ones by the positive zeros of $\Rho$, which correspond
to the orbits' turning points. Whenever the polynomial has a non negative double
zero, that is,
\begin{equation}
	\Rho(r)=0 \quad \text{and} \quad \dv{\Rho}{r}{(r)}=0\,, \label{eq:paraplots}
\end{equation}
a variation of parameters is expected to change the number of positive zeros. By
plotting the zeros of the resultant of the two expressions in
Eq.~(\ref{eq:paraplots}), one obtains parameter plots showing the boundaries
between regions of 1, 2, 3 or 4 zeros of $\Rho$. This is shown for parametric
$L$-$l$, $K$-$E$, and $L$-$E$ diagrams in Fig.~\ref{fig:paraplot}.

As $\Rho$ is a polynomial of second degree in $E$, one can define the two-part
effective potential $V_\pm(r)$ as the values of energy that yield $\Rho=0$,
i.e., $\Rho$ can be rearranged in the form
\begin{equation}
	\Rho=f(r)(E-V_+)(E-V_-)\,.
\end{equation}

\begin{figure*} 
	\subfloat[$K = 4,\, L = 20,\, \delta = 1,\, l = 1$.]{%
		\includegraphics[width=.333\linewidth]{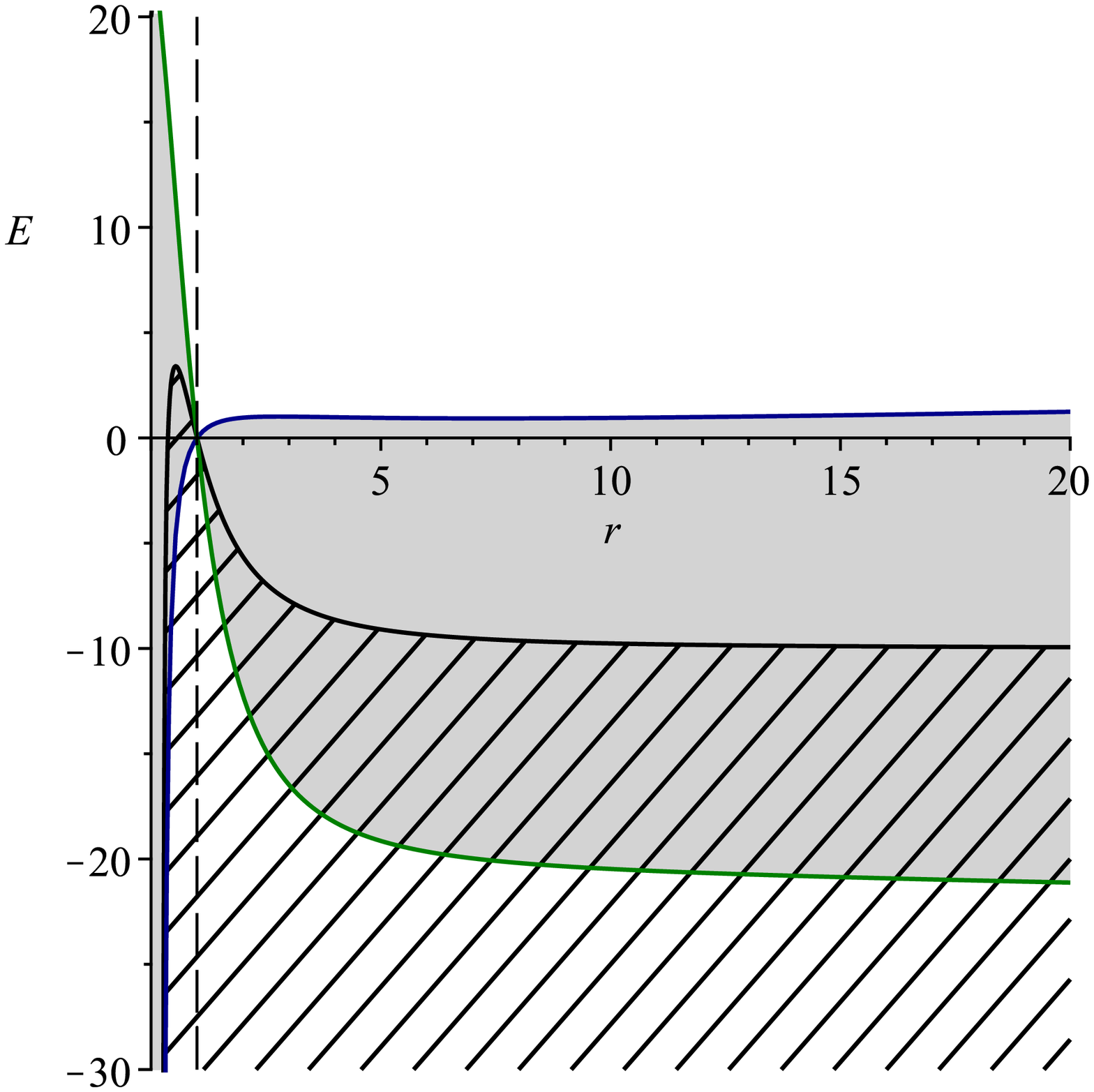}\label{fig:effpotr1a}%
	}\hfill
	\subfloat[$K = 4,\, L = 8,\, \delta = 0,\, l = -20$.]{%
		\includegraphics[width=.333\linewidth]{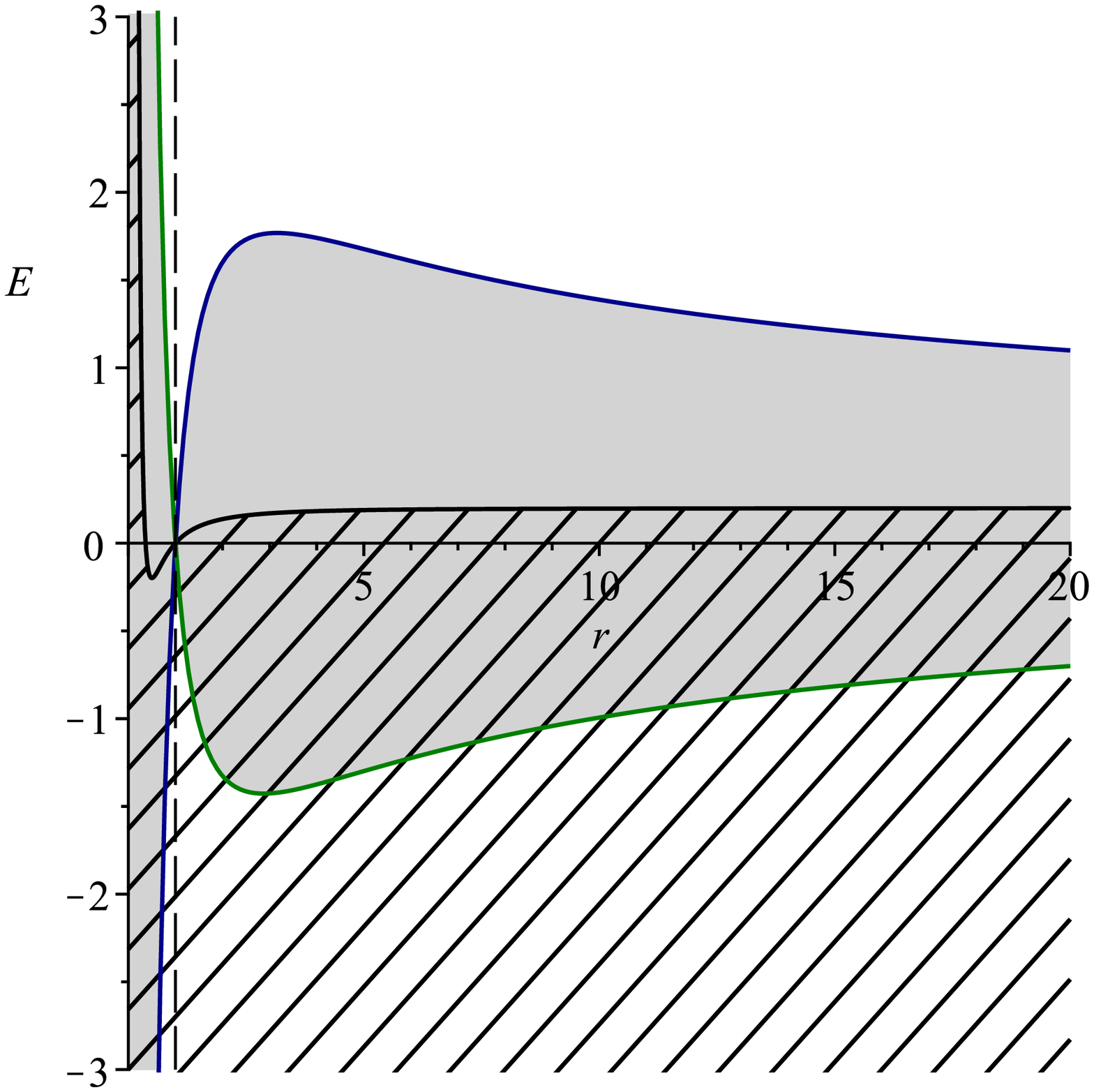}\label{fig:effpotr2a}%
	}\hfill
	\subfloat[$K = 4,\, L = 8,\, \delta = -1,\, l = -20$.]{%
		\includegraphics[width=.333\linewidth]{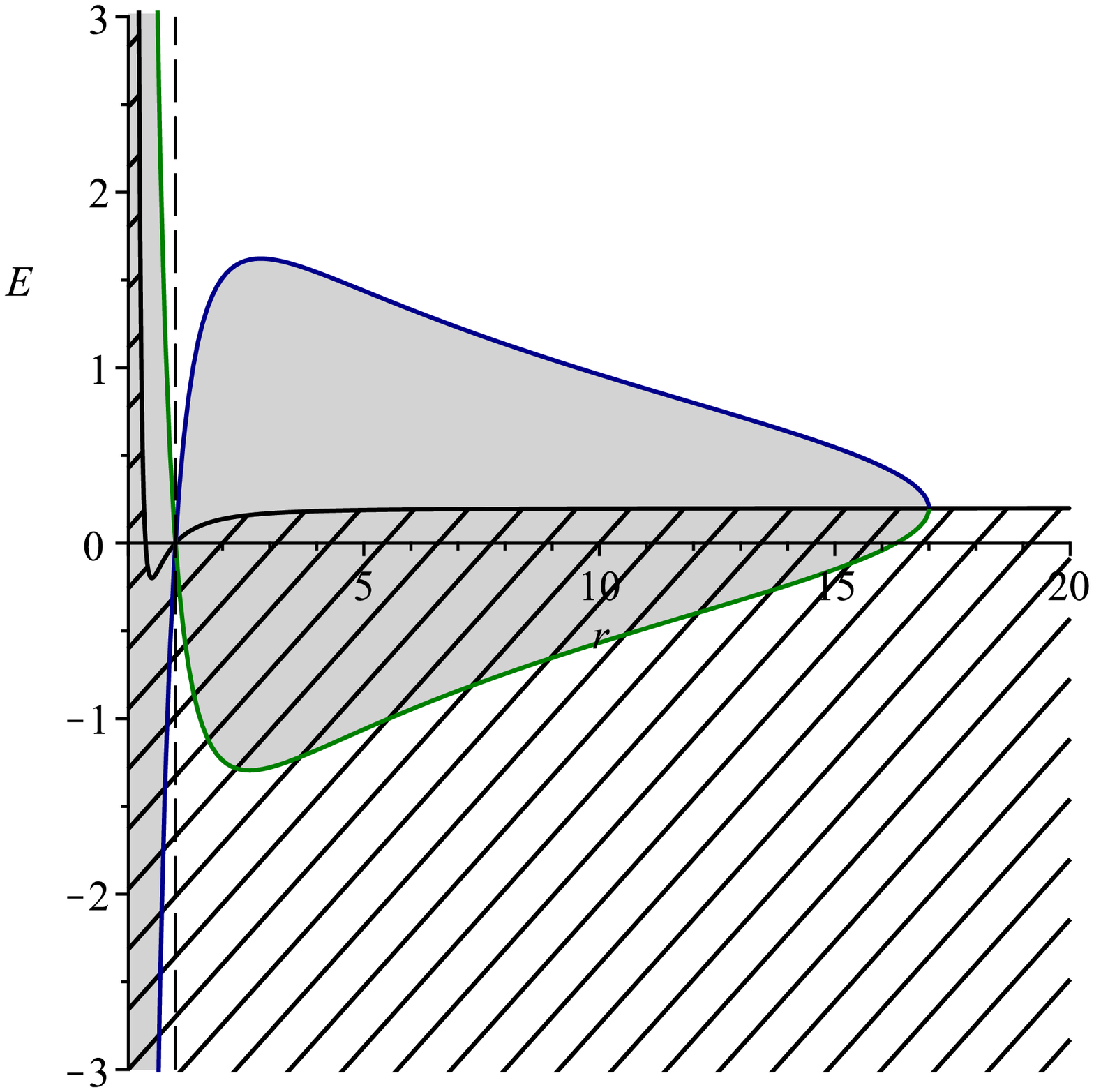}\label{fig:effpotr3a}%
	}\\
	\subfloat[Detail of \csubref{fig:effpotr1a}.]{%
		\includegraphics[width=.333\linewidth]{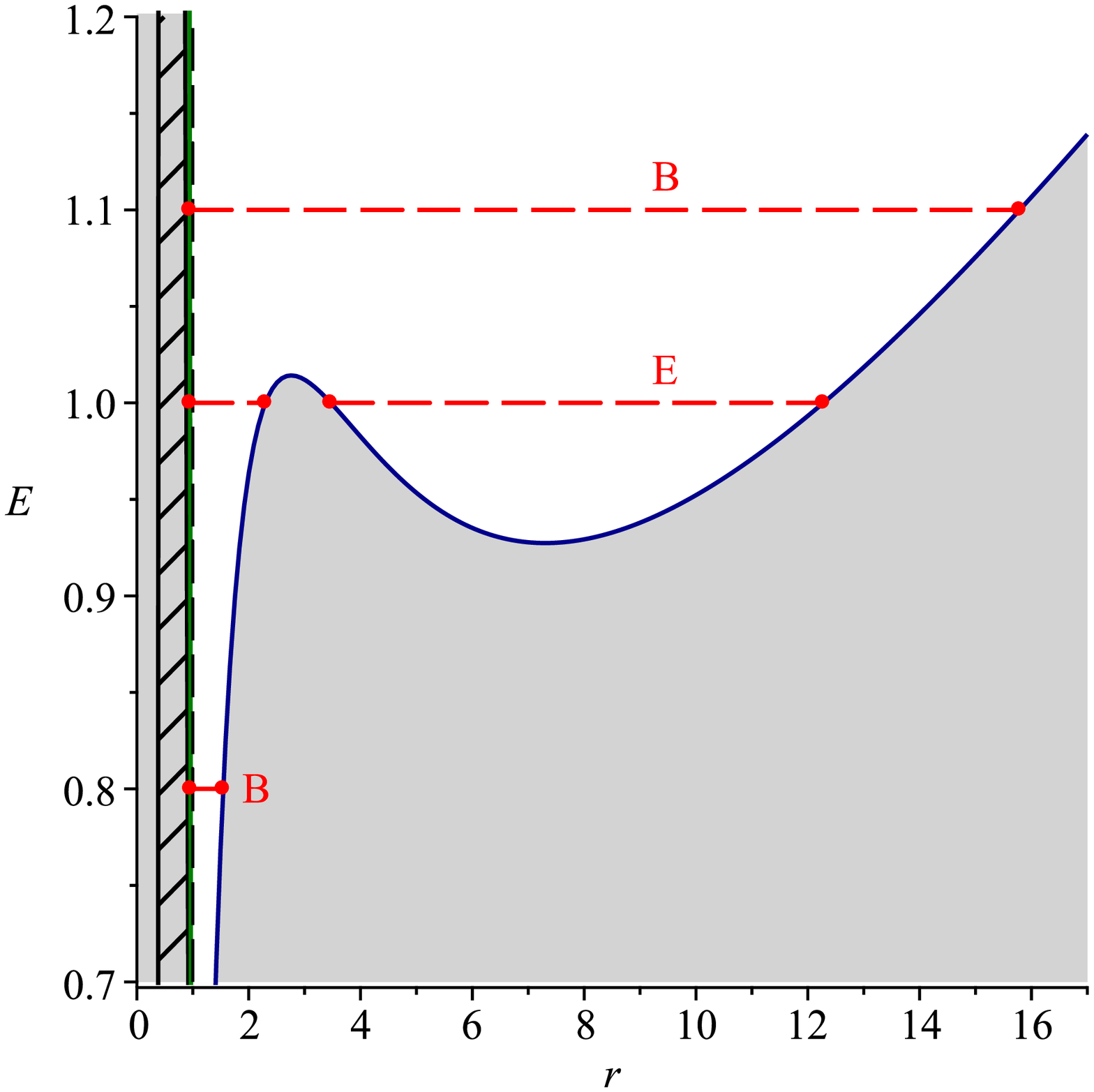}\label{fig:effpotr1b}%
	}\hfill
	\subfloat[Detail of \csubref{fig:effpotr2a}.]{%
		\includegraphics[width=.333\linewidth]{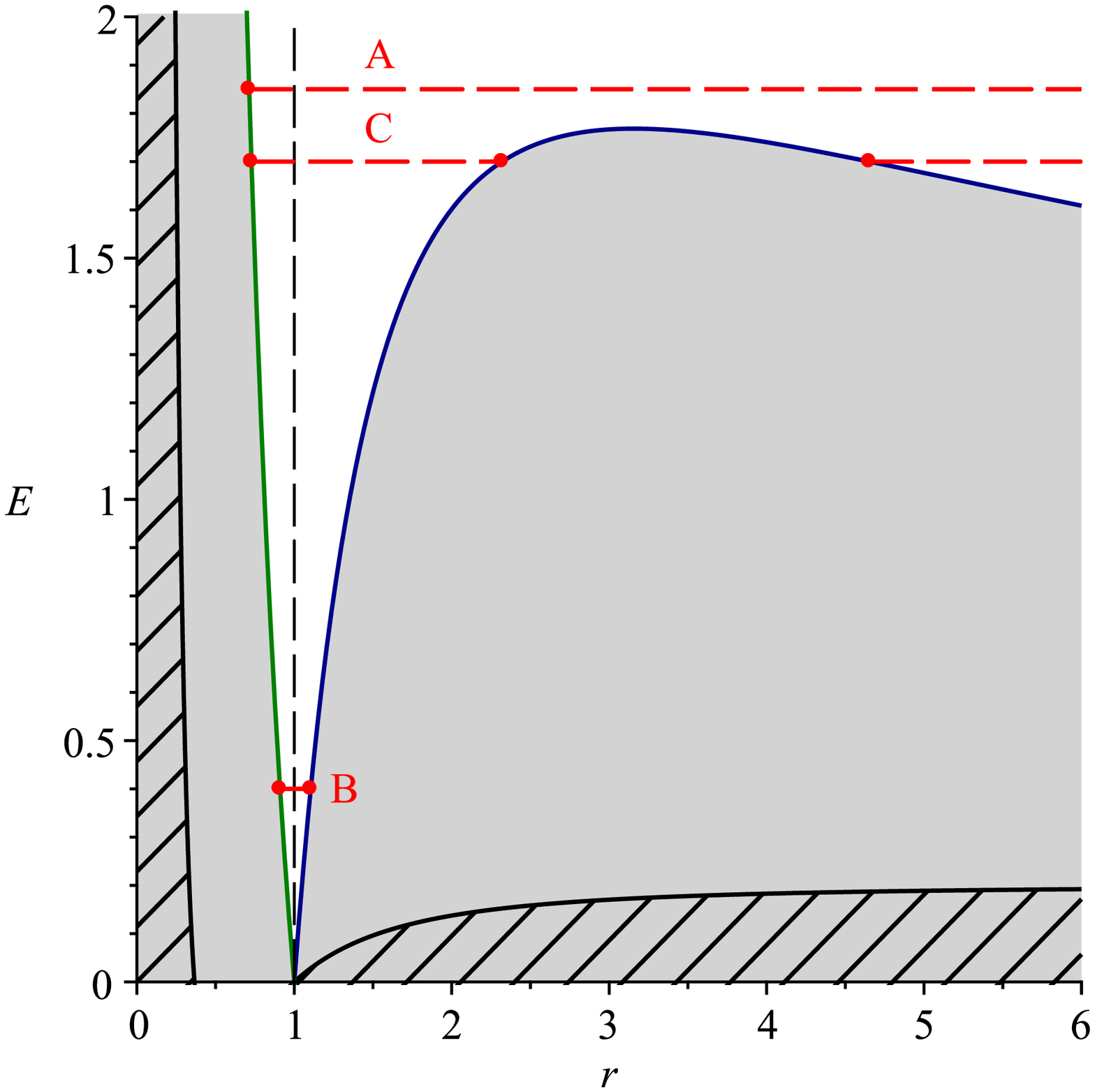}%
	}\hfill
	\subfloat[Detail of \csubref{fig:effpotr3a}.]{%
		\includegraphics[width=.333\linewidth]{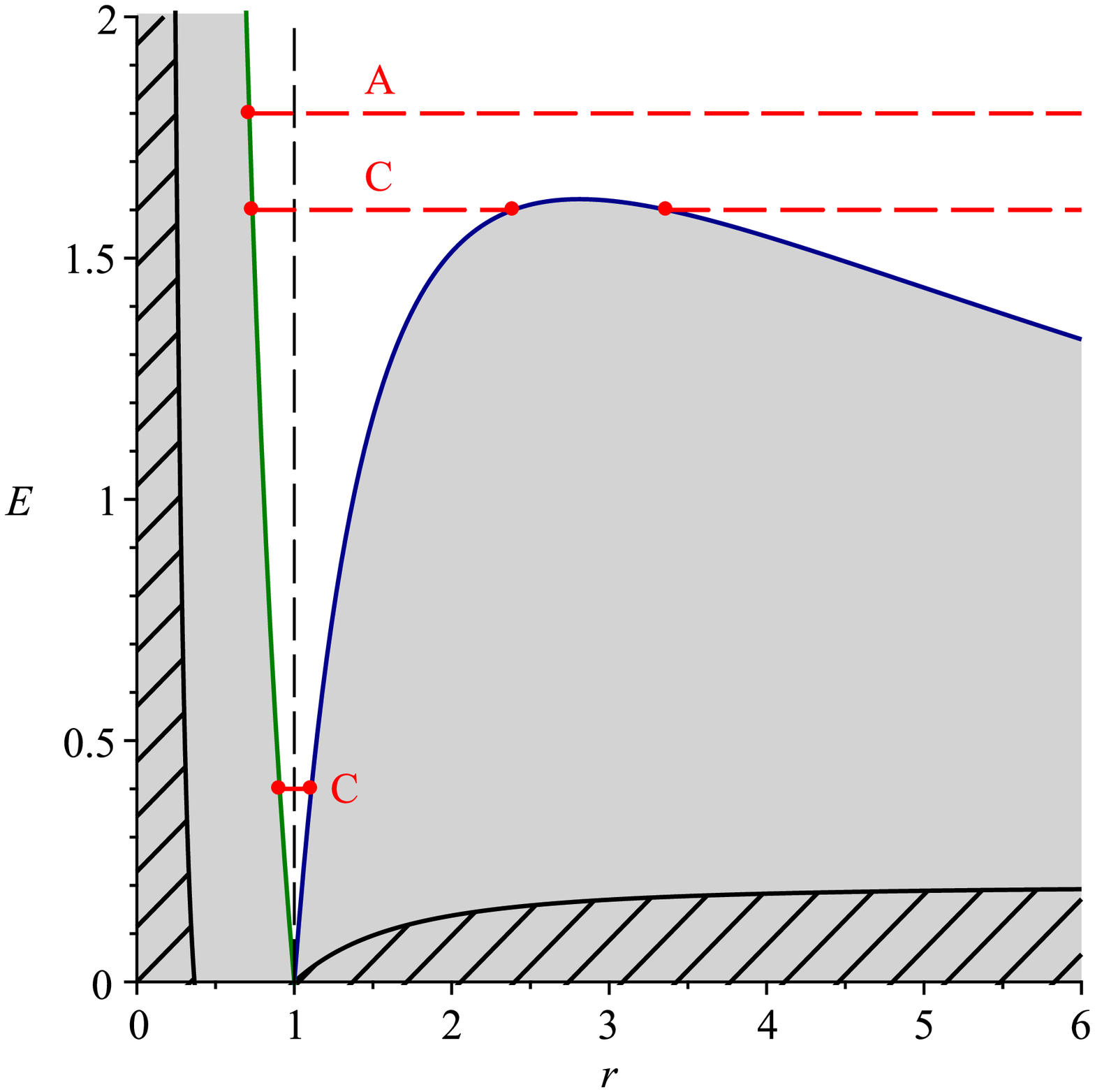}%
	}
	\caption{\label{fig:effpotrab}%
		Effective potential $V_\pm$ of the $r$ motion for various sets of parameters.
		The blue (dark gray) line indicates $V_+$, the green (light gray) one $V_-$.
		Physically forbidden areas are marked in gray and in the hatched areas it
		holds $\dd{t}/\dd{\gamma}<0$. The horizon at $r=1$ is shown by a vertical
		dashed line. Some characteristic orbits of type A, B, C, E from
		Table~\ref{tab:orbits} with 1 to 4 turning points are shown. Points denote the
		turning points and horizontal dashed lines correspond to the particle's energy
		$E$. Comparing \csubref{fig:effpotr1b} and the $L$-$E$ diagram in
		Fig.~\ref{fig:paraplot3a} at $L=20$, orbits of type B and E can be found in
		regions with 2 and 4 zeros, respectively.}
\end{figure*}

\begin{figure*} 
	\subfloat[$\delta = 1$.]{%
		\includegraphics[width=.333\linewidth]{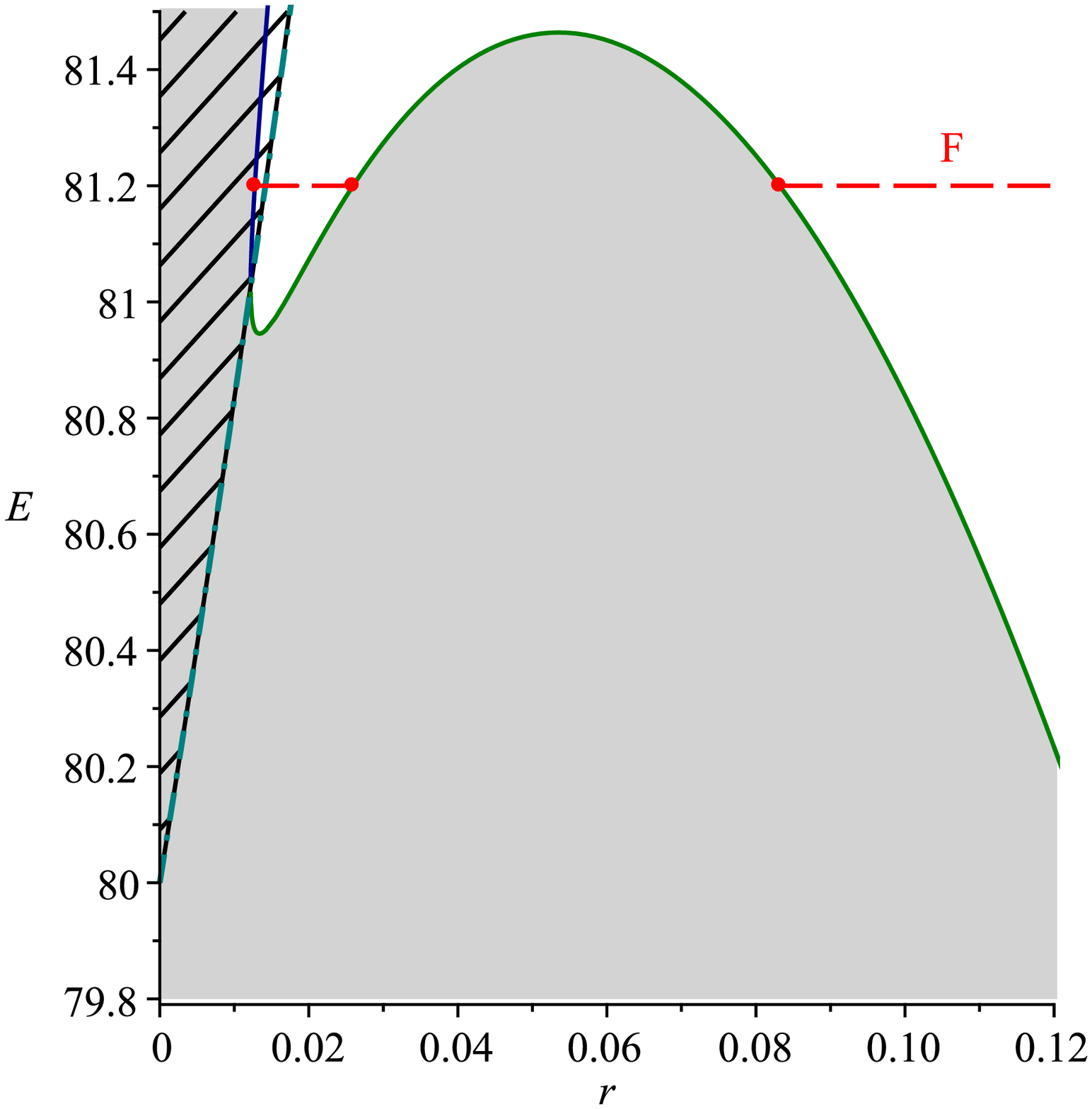}\label{fig:effpotr1d}%
	}\hfill
	\subfloat[$\delta = 0$.]{%
		\includegraphics[width=.333\linewidth]{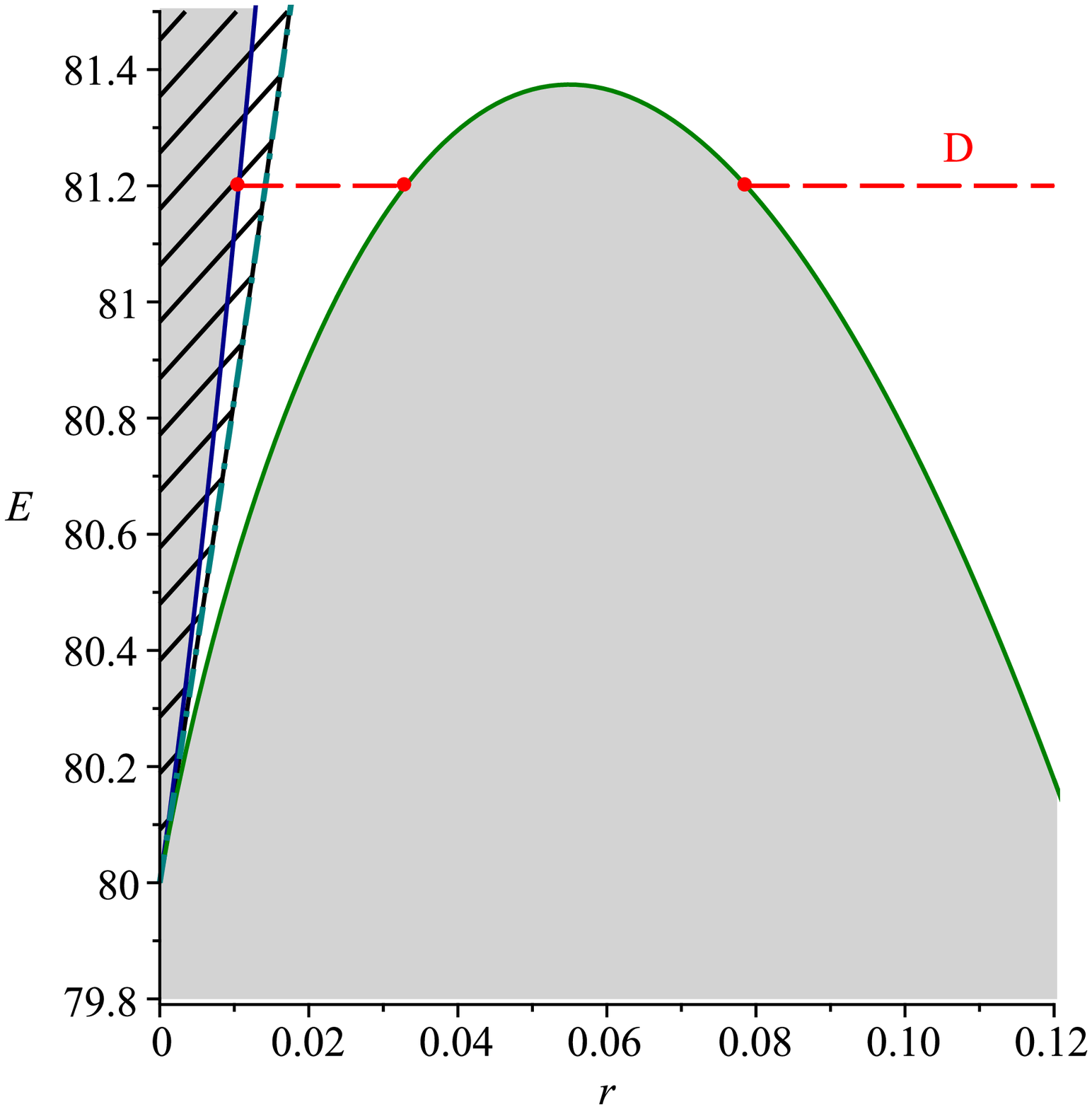}%
	}\hfill
	\subfloat[$\delta = -1$.]{%
		\includegraphics[width=.333\linewidth]{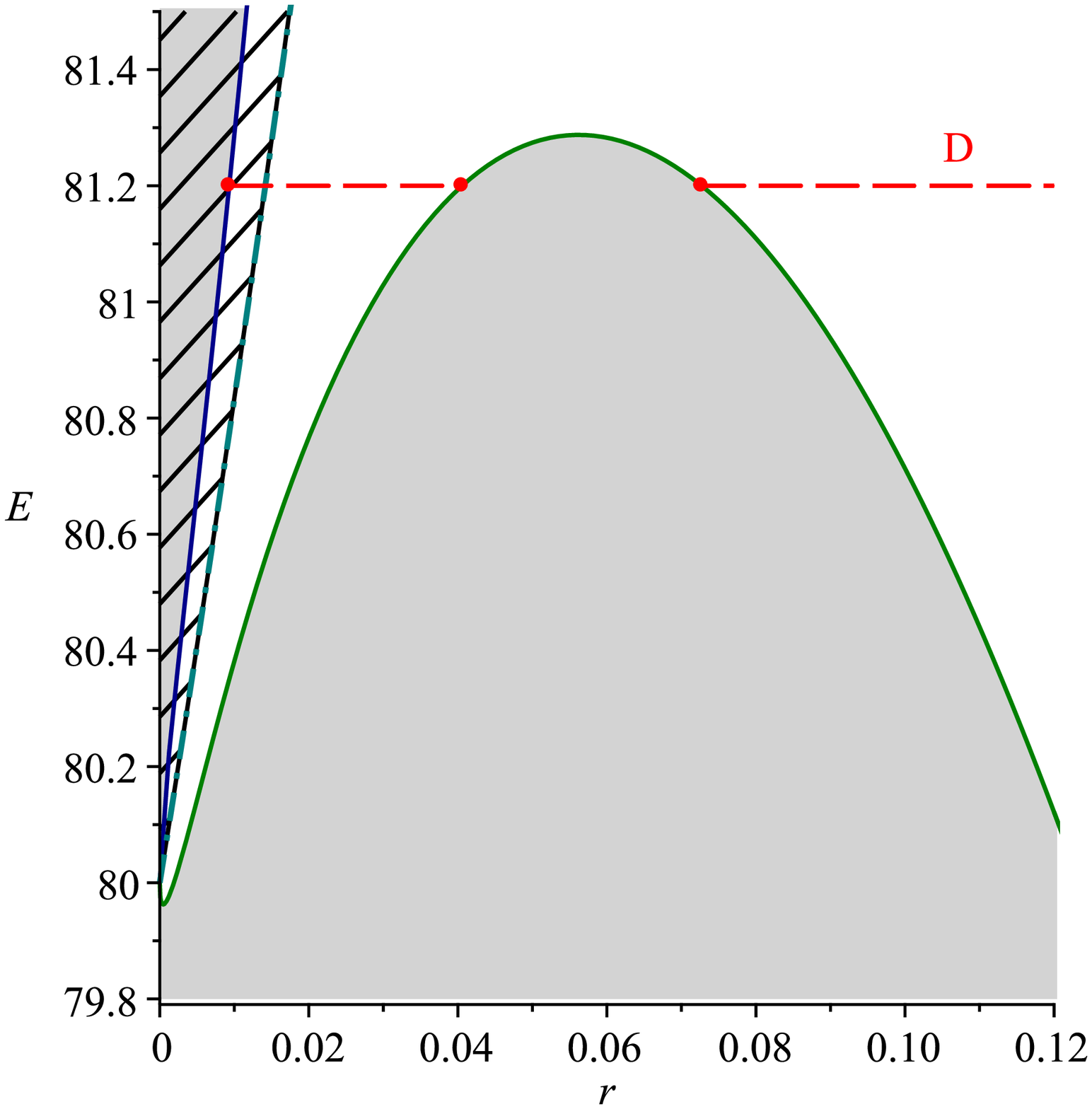}\label{fig:effpotr3d}%
	}
	\caption{\label{fig:effpotrd}%
		Effective potential $V_\pm$ of the $r$ motion for varying $\delta$ and $K =
		0,\, L = 80,\, l = 1$, which allows bound orbits of type D and F (see
		Table~\ref{tab:orbits}) behind the horizon. For a more detailed description
		see Fig.~\ref{fig:effpotrab}. Additionally, the turnaround energy
		$E_\text{turn}$ of the $\phi$ motion is shown as a dash-dotted line, which
		mostly overlaps with the border of the hatched area corresponding to
		$\dd{t}/\dd{\gamma}<0$.}
\end{figure*}

\begin{figure*} 
	\subfloat[$K = 0,\, L = 8,\, \delta=-1,\, l = 1$.]{%
		\includegraphics[width=.333\linewidth]{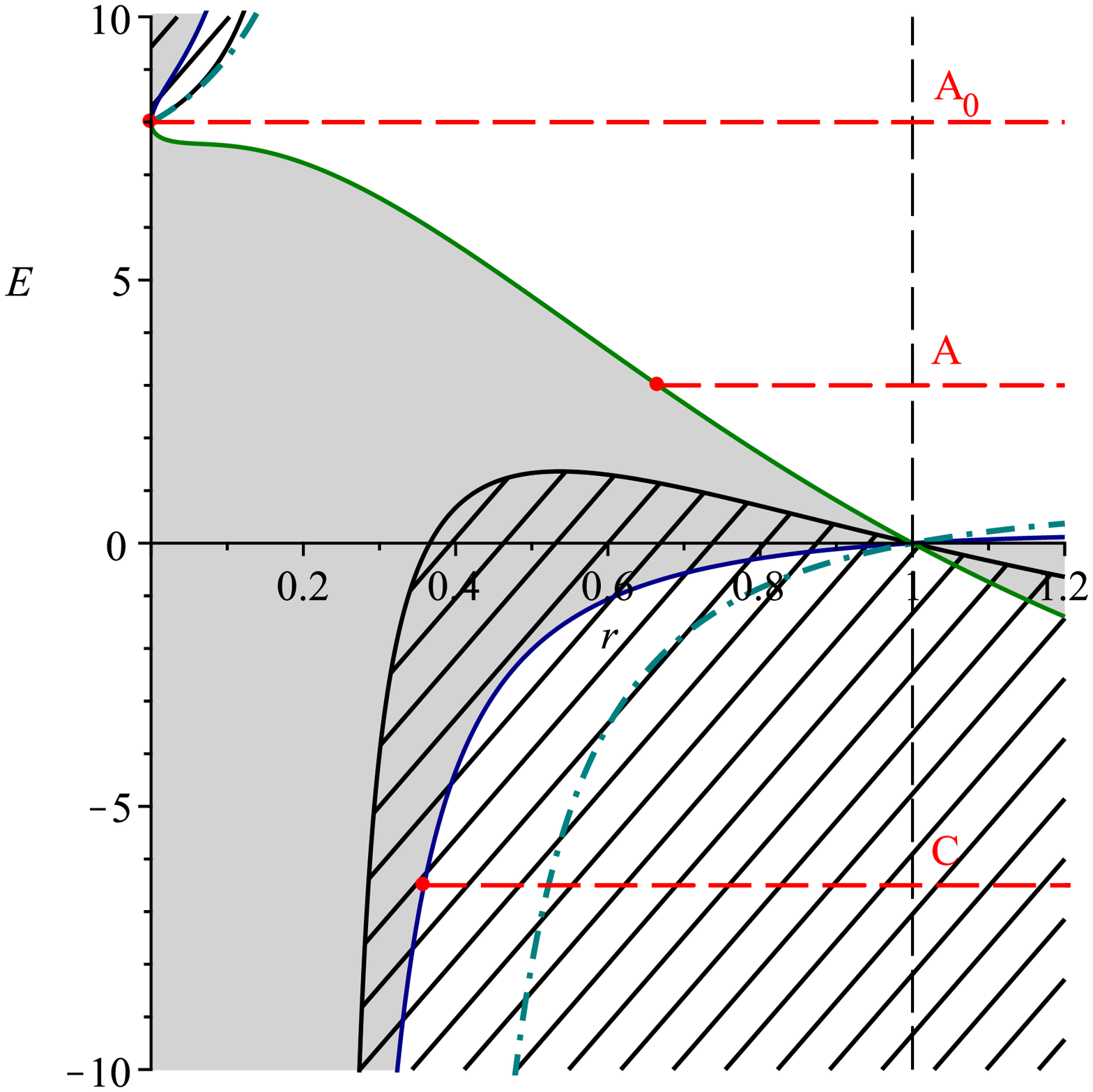}\label{fig:effpotr3f}%
	}
	\subfloat[$K = 0,\, L = 40,\, \delta=-1,\, l = 0.3$.]{%
		\includegraphics[width=.333\linewidth]{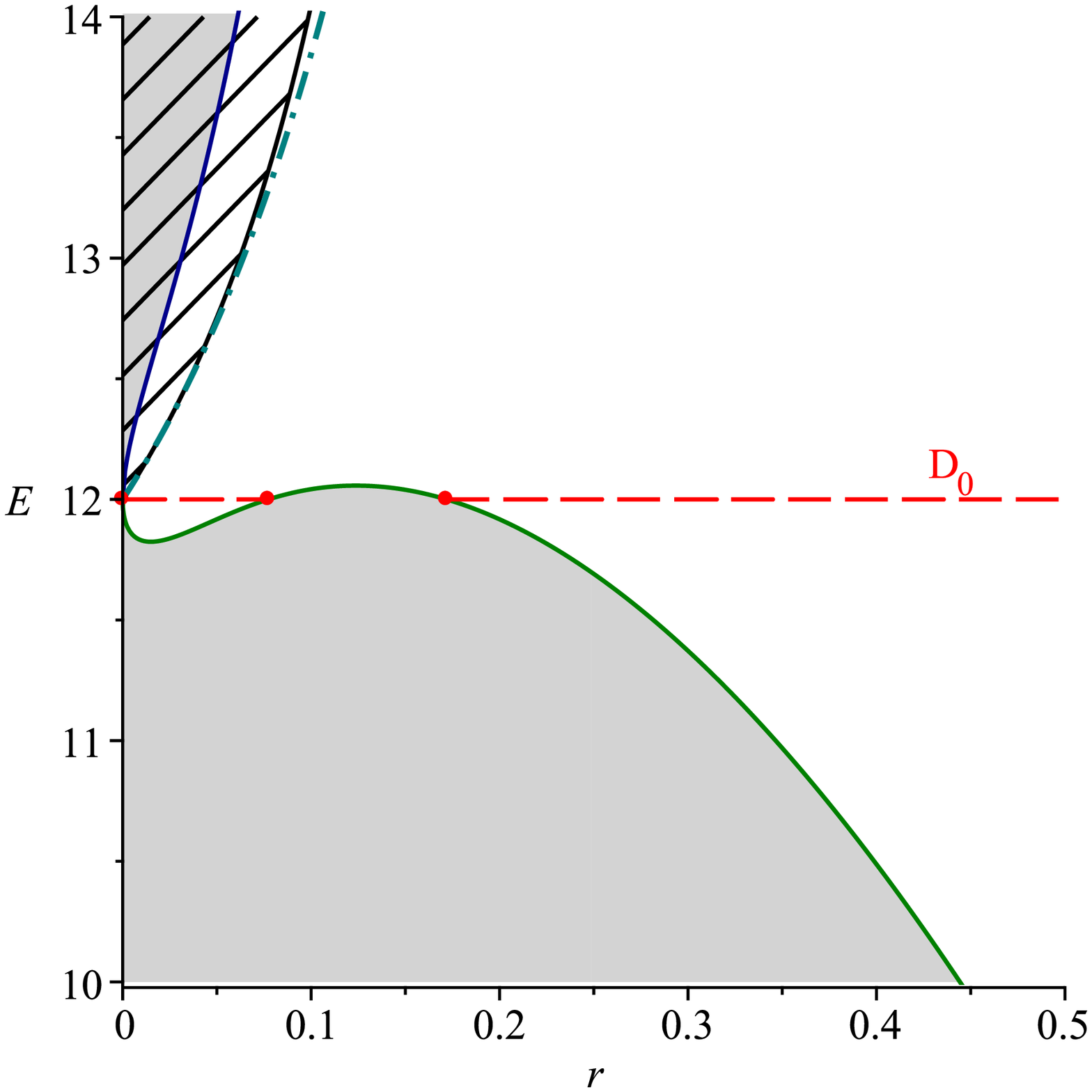}\label{fig:effpotr3h}%
	}
	\subfloat[$K = 0,\, L = 17.1,\, \delta = 1,\, l = 1$.]{%
		\includegraphics[width=.333\linewidth]{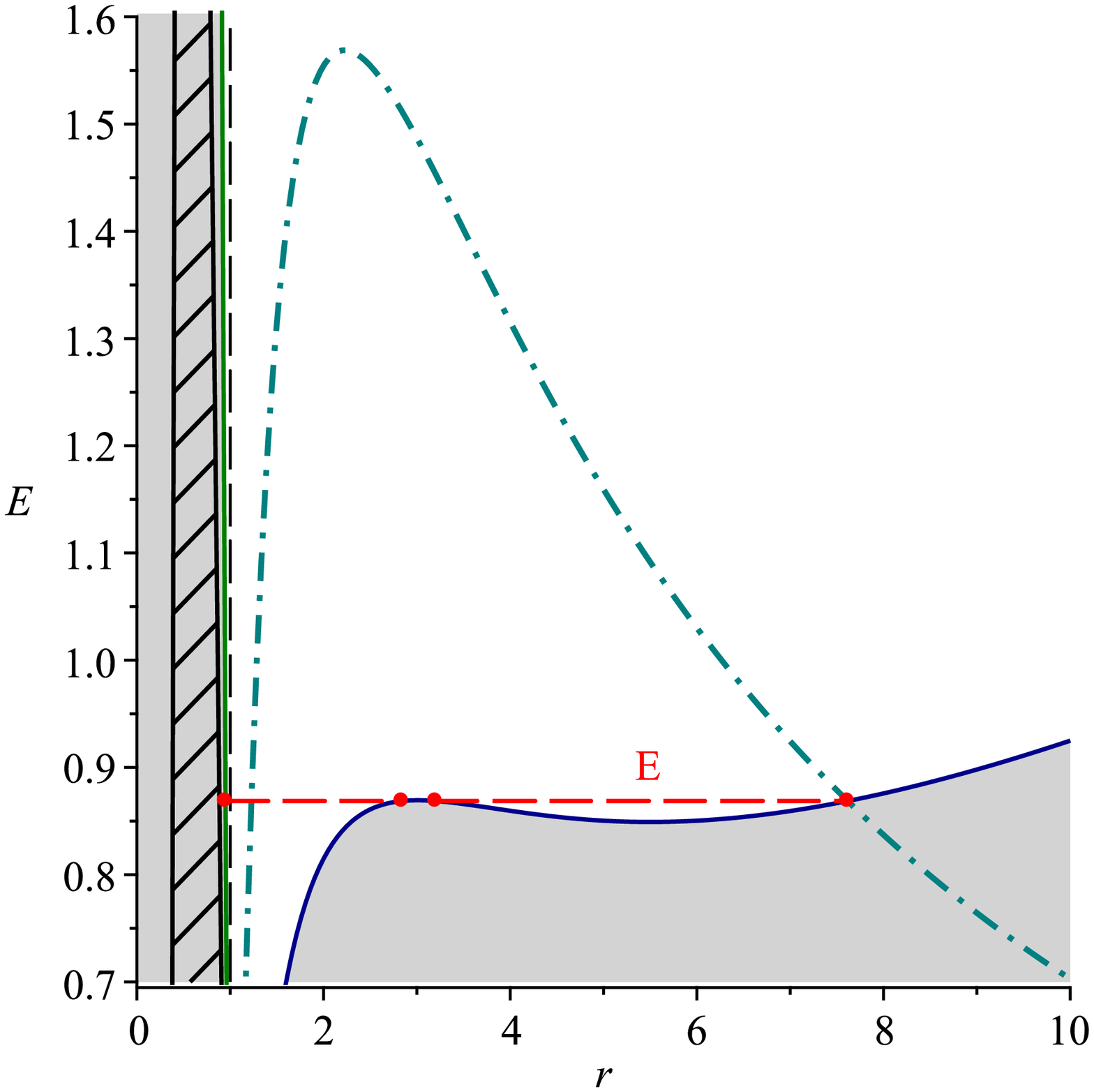}\label{fig:effpotr1g}%
	}
	\caption{\label{fig:effpotrfhg}%
		\csubref{fig:effpotr3f} and \csubref{fig:effpotr3h}: Effective potential
		$V_\pm$ of the $r$ motion for $\delta=-1$ with parameters chosen to allow for
		terminating orbits of type $\text{A}_0$ and $\text{D}_0$ (see
		Table~\ref{tab:orbits}). \csubref{fig:effpotr1g}: Effective potential $V_\pm$
		of the $r$ motion with parameters chosen to allow for a \textit{pointy petal}
		BO of type E. For a more detailed description see Fig.~\ref{fig:effpotrab}.
		Additionally, the turnaround energy $E_\text{turn}$ of the $\phi$ motion is
		shown as a dash-dotted line.}
\end{figure*}

Some effective potentials are shown in
Figs.~\ref{fig:effpotrab}--\ref{fig:effpotrfhg}. Since one can factor out
$(r-1)$ in $V_\pm$, $V_+$ and $V_-$ intersect on the horizon at $E=0$, given
that $V_\pm$ is real. Additionally, for the requirement that time should always
run forward, i.e., $\dd{t}/\dd{\gamma}\geq0$, Eq.~(\ref{eq:motiont}) is treated
in a similar way as Eq.~(\ref{eq:motionr}) for the effective potential and
corresponding regions are shown as well.

Since $\Rho<0$ for $\delta=1$ and $\Rho>0$ for $\delta=-1$ hold in the limit
$r\rightarrow\infty$, orbits of particles of positive mass are always bounded,
particles of imaginary mass can exist in unbound orbits for arbitrary energy.
More precisely, one finds for particles of non zero mass, i.e., $\delta=\pm1$,
in the limiting case $r\rightarrow\infty$ from $\Rho=0$
\begin{equation}
	V_\pm^\infty=\pm\frac{\sqrt{\delta r}}{|l|}\,.
\end{equation}

Similarly, for massless particles, i.e., $\delta=0$, it follows in the limit
$r\rightarrow\infty$
\begin{equation}
	V_\pm^\infty=-\frac{L}{2l}\pm\frac{\sqrt{L^2+K}}{2|l|}\,.
\end{equation}
Since here $V_+^\infty>0$ and $V_-^\infty<0$ hold (for $K>0$) and $\Rho<0$ for
$E=0$ in the limit $r\rightarrow\infty$, for every set of parameters a
physically forbidden region in the vicinity of $E=0$ can be found. This behavior
corresponds to the border of region (2) in Fig.~\ref{fig:paraplot1b} as varying
the energy $E$ to region (3)$_+$ allows for an additional unbound orbit.

In special cases, terminating orbits can be found. Due to the smoothness of the
polynomial $\Rho$, the condition $\Rho(0)\geq0$ has to be fulfilled, which
implies real $V_\pm(0)$, since $\Rho(0)$ opens downward with respect to $E$. It
follows
\begin{equation}
	V_\pm(0)=Ll\pm\sqrt{-Kl^2-2K}\,,
\end{equation}
giving $K=0$ and $E=Ll$ as first conditions. To allow for non trivial
terminating orbits with non zero range, additionally $\Rho(r)>0$ has to hold on
$r\in(0,\varepsilon)$ with $\varepsilon>0$. This implies that the lowest order
non vanishing coefficient of $\Rho$ must be positive, which is a condition that
is always met in case of $\delta=-1$ but never possible in other cases. The
influence of the choice of $\delta$ on the effective potential at the
singularity around $E=Ll$ is shown in Fig.~\ref{fig:effpotrd}. Here parameters
are chosen to allow for bound orbits behind the horizon for all particle masses
$\delta$. As regions with $\dd{t}/\dd{\gamma}<0$ are shown as well, it can be
seen that all depicted bound orbits cross regions where time is running
backwards. For a particle of non zero mass this can be avoided by lowering its
energy to the bottom of the potential well. A short-range terminating orbit can
be found in Fig.~\ref{fig:effpotr3d} at $E=Ll$. Additionally, the two possible
types of terminating orbits can be seen in Figs.~\ref{fig:effpotr3f} and
\ref{fig:effpotr3h}.

With the results of the parametric diagrams and the effective potentials we
summarize all combinations of zeros of $\Rho$ and show their corresponding types
of orbits in Table~\ref{tab:orbits}. Here the following connection of regions in
the parametric diagrams and types of orbits was found:
\begin{enumerate}
	\item Region (1): $\Rho$ has one non negative zero $r_1<1$, which corresponds
	to a TEO of type A. For particles of imaginary mass a TO of type
	$\text{A}_{0}$ is possible for an energy of $E=Ll$ [see, e.g.,
	Fig.~\ref{fig:effpotr3f}].
	\item Region (2): $\Rho$ has two positive zeros $r_1<1<r_2$, which corresponds
	to a MBO of type B.
	\item Region (3): $\Rho$ has three non negative zeros $r_i$.
	\begin{enumerate}
		\item Region $\text{(3)}_{+}$: It holds $r_1<1<r_2,r_3$, resulting in a MBO
		and an EO of type C.
		\item Region $\text{(3)}_{-}$: For all zeros it holds $r_i<1$, resulting in
		a BO and a TEO of type D. Again, for particles of imaginary mass a TO of
		type $\text{D}_{0}$ is possible for an energy of $E=Ll$ [see, e.g.,
		Fig.~\ref{fig:effpotr3h}].
	\end{enumerate}
	\item Region (4): $\Rho$ has four positive zeros $r_i$.
	\begin{enumerate}
		\item Region $\text{(4)}_{+}$: It holds $r_1<1<r_2,r_3,r_3$, resulting in a
		MBO and a BO of type E.
		\item Region $\text{(4)}_{-}$: It holds $r_1,r_2,r_3<1<r_4$, resulting in a
		BO and a MBO of type F.
	\end{enumerate}
\end{enumerate}

\def\mirror {\draw[draw=none] (0,0) circle (3pt); \draw[->](0,0)--(4,0);
\draw(0,-0.2)--(0,0.2) (1.98,-0.2)--(1.98,0.2) (2.02,-0.2)--(2.02,0.2);}
\begin{table*}
	\caption{\label{tab:orbits}%
		Types of orbits of light and particles in the spacetime of a supersymmetric
		AdS$_5$ black hole. Thick lines represent the range of the orbits and thick
		dots indicate their turning points. The vertical double line represents the
		horizon. The vertical single line corresponds to the singularity, which can
		only be reached by particles of imaginary mass and $K=0$, resulting in
		orbits of type $\text{A}_{0}$ and $\text{D}_{0}$. The column $\delta$
		distinguishes between particles with imaginary/positive mass ($-1$/1) and
		light (0).}
	\begin{tabular*}{\textwidth}{@{\extracolsep{\fill}}llllcl@{}}
		\toprule
		Type & Region & Zeros & $\delta$ & Range of $r$ & Orbit \\
		\midrule
		A	& (1) & 1 & $-1$, 0 &	\begin{tikzpicture}
		\mirror;
		\filldraw (1.5,0) circle (2pt);
		\draw[line width=1pt](1.5,0)--(4,0);
		\end{tikzpicture} & TEO \\
		$\text{A}_{0}$	& & & $-1$	&	\begin{tikzpicture}
		\mirror;
		\filldraw (0,0) circle (2pt);
		\draw[line width=1pt](0,0)--(4,0);
		\end{tikzpicture} & TO \\
		B	& (2) & 2 & 0, 1	&	\begin{tikzpicture}
		\mirror;
		\filldraw (1.5,0) circle (2pt);
		\draw[line width=1pt](1.5,0)--(2.5,0);
		\filldraw (2.5,0) circle (2pt);
		\end{tikzpicture} & MBO \\
		C	& $\text{(3)}_{+}$ & 3 & $-1$, 0	&	\begin{tikzpicture}
		\mirror;
		\filldraw (1.5,0) circle (2pt);
		\draw[line width=1pt](1.5,0)--(2.5,0);
		\filldraw (2.5,0) circle (2pt);
		\filldraw (3,0) circle (2pt);
		\draw[line width=1pt](3,0)--(4,0);
		\end{tikzpicture} & MBO, EO \\
		D	& $\text{(3)}_{-}$ & 3 & $-1$, 0	&	\begin{tikzpicture}
		\mirror;
		\filldraw (0.5,0) circle (2pt);
		\draw[line width=1pt](0.5,0)--(1,0);
		\filldraw (1,0) circle (2pt);
		\filldraw (1.5,0) circle (2pt);
		\draw[line width=1pt](1.5,0)--(4,0);
		\end{tikzpicture} & BO, TEO \\
		$\text{D}_{0}$	& & & $-1$	&	\begin{tikzpicture}
		\mirror;
		\filldraw (0,0) circle (2pt);
		\draw[line width=1pt](0,0)--(1,0);
		\filldraw (1,0) circle (2pt);
		\filldraw (1.5,0) circle (2pt);
		\draw[line width=1pt](1.5,0)--(4,0);
		\end{tikzpicture} & TO, TEO \\
		E	& $\text{(4)}_{+}$ & 4 & 1	&	\begin{tikzpicture}
		\mirror;
		\filldraw (1.5,0) circle (2pt);
		\draw[line width=1pt](1.5,0)--(2.5,0);
		\filldraw (2.5,0) circle (2pt);
		\filldraw (3,0) circle (2pt);
		\draw[line width=1pt](3,0)--(3.5,0);
		\filldraw (3.5,0) circle (2pt);
		\end{tikzpicture} & MBO, BO \\
		F	& $\text{(4)}_{-}$ & 4 & 1	&	\begin{tikzpicture}
		\mirror;
		\filldraw (0.5,0) circle (2pt);
		\draw[line width=1pt](0.5,0)--(1,0);
		\filldraw (1,0) circle (2pt);
		\filldraw (1.5,0) circle (2pt);
		\draw[line width=1pt](1.5,0)--(2.5,0);
		\filldraw (2.5,0) circle (2pt);
		\end{tikzpicture} & BO, MBO \\
		\bottomrule
	\end{tabular*}
\end{table*}

Note that for every radially allowed orbit an angular momentum $J$ according to
Section~\ref{sec:classificationtheta} can be chosen to allow for $\theta$ motion
as well.

\subsection{\label{sec:static}Static orbits}
It has been shown in \cite{Collodel:2017end} that some axisymmetric rotating
spacetimes possess a ring in the equatorial plane, on which stationary particles
remain stationary with respect to an asymptotic static observer. As this is
possible in higher dimensions as well (see \cite{Collodel:2017end} and
references therein), we check whether this is the case for the spacetime at
hand. Hence, parameters have to be chosen to allow for an extremum of $V_\pm$ at
$r_\text{st}$ and a double zero in Eq.~(\ref{eq:zerostheta}) at
$\cos\theta_\text{st}$. Additionally, the right hand sides of
Eqs.~(\ref{eq:motionphi}) and (\ref{eq:motionpsi}) have to vanish. We use
$\cos\theta_\text{st}=\frac{LJ}{L^2+K}$ from Eq.~(\ref{eq:zerostheta}), under
conditions discussed in Section~\ref{sec:classificationtheta}, together with
Eq.~(\ref{eq:motionpsi}), from which the requirement $K=0$ and $|J|\leq|L|$ can
be derived. With this, a turnaround energy $E_\text{turn}$ can be defined by
solving Eq.~(\ref{eq:motionphi}) for $E$ yielding
\begin{equation}
	E_\text{turn}(r)=\frac{(r-1)Ll}{2r^2+2r-1}\,.
\end{equation}
This is also shown in Figs.~\ref{fig:effpotrd} and \ref{fig:effpotrfhg}, since
here $K=0$ holds. To obtain a stationary point, $E_\text{turn}$ has to intersect
$V_\pm$ in an extremum of $V_\pm$. In this spacetime, we could not find such
intersections. However, \textit{pointy petal} BOs (in front of the horizon),
\textit{semi} BOs (behind the horizon) as well as MBOs, where the particle is
periodically at rest, are found at arbitrary intersections of $E_\text{turn}$
and $V_\pm$ for particles of positive mass. In case of a \textit{pointy petal}
BO, an effective $r$ potential and $\phi$ turnaround energy is shown in
Fig.~\ref{fig:effpotr1g}.

\subsection{\label{sec:spacelike}Spacelike geodesics and AdS/CFT}
Spacelike geodesics are usually not considered in the analysis of geodesic
motion, since they represent test particles with imaginary rest mass ($\delta
=-1$). In the context of AdS/CFT, however, there are applications for spacelike
geodesics. CFT correlators or Feynman propagators describe observables on the
asymptotic boundary of an AdS spacetime. Correlation functions of fields in the
bulk are related to correlation functions of CFT operators on the boundary.
Using the Green function, the correlator of two operators can be written as
\begin{equation}
	\langle\mathcal{O}(t,\boldsymbol{x})\,\mathcal{O}(t',\boldsymbol{x}')\rangle
		=\int\!\exp[i\Delta L(\mathcal{P})]\mathcal{DP}\,.
\end{equation}
Here $L(\mathcal{P})$ describes the proper length of the path $\mathcal{P}$
between the boundary points $(t,\boldsymbol{x})$ and $(t,\boldsymbol{x}')$. In
the case of spacelike trajectories $L(\mathcal{P})$ is imaginary, so that the
whole expression is real. $m$ is the mass of the bulk field, which is related to
the conformal dimension $\Delta=1+\sqrt{1+m^2}$ of the Operator $\mathcal{O}$.
For large masses, i.e., $\Delta\approx m$, the WKB approximation can be used to
calulate the operator, which is then described by the sum over all spacelike
geodesics between the boundary points
\begin{equation}
	\langle\mathcal{O}(t,\vec{x})\,\mathcal{O}(t',\vec{x}')\rangle
		=\sum_{g}\exp(-\Delta L_{{g}})\,.
\end{equation}
The real proper length of a geodesic $L_g$ diverges due to contributions near
the AdS boundary and has to be renormalized by removing the divergent part in
pure AdS. The sum is then dominated by the shortest spacelike geodesic between
the boundary points (see, e.g., \cite{Balasubramanian:1999zv},
\cite{Balasubramanian:2011ur}, \cite{Louko:2000tp}).

In this formalism we need geodesics that have endpoints on the boundary at
$r\rightarrow\infty$. This is the case for escape orbits (EOs) and two-world
escape orbits (TEOs) that have a single turning point and reach infinity. For
EOs both endpoints of the geodesic are located on a single boundary, since the
turning point is outside the horizons. The corresponding two-point correlators
can be used to calculate for example the thermalization time
\cite{Balasubramanian:2011ur} or the entanglement entropy \cite{Hubeny:2007xt},
\cite{AbajoArrastia:2010yt}.

Particles on TEOs cross the horizon and therefore the endpoints are on two
disconnected boundaries. TEOs can also be considered as propagators
\cite{Balasubramanian:1999zv,Louko:2000tp}. The boundary correlators can probe
the physics behind the horizon. This could be used to study the formation of
black holes \cite{Balasubramanian:1999zv}, the black hole singularity
\cite{Fidkowski:2003nf}, or the information paradox \cite{Papadodimas:2012aq}.

In the spacetime of a supersymmetric AdS$_5$ black hole, geodesics relevant for
AdS/CFT exist in the regions (1), (3)$_+$ and (3)$_-$, see
Fig.~\ref{fig:paraplot} and Table~\ref{tab:orbits}. Depending on the parameters
of the black hole, we find EOs or TEOs. In region (3)$_+$ EOs exist that return
to the same boundary where they started. In region (1) and (3)$_-$, there are
TEOs crossing the horizon with endpoints on two disconnected boundaries.

\section{\label{sec:solution}Solution of the geodesic equations}
In this section we solve the equations of motion
(\ref{eq:motionr})--(\ref{eq:motiont}) analytically.

\subsection{Solution of the \texorpdfstring{$\theta$}{theta} equation}
Eq.~(\ref{eq:derivationtheta}) can be solved by an elementary function. In case
of $a<0$ and $D>0$, this leads to
\begin{equation}
	\theta(\gamma)=\arccos\left(\frac{1}{2a}\left(\sqrt{D}\sin\left(\pm\sqrt{-a}
		(\gamma-\gamma_0)+\gamma_0^\theta\right)-b\right)\right)
\end{equation}
as a solution of Eq.~(\ref{eq:motiontheta}). Here $\gamma_0$ and $\theta_0$ are
the initial values of $\gamma$ and $\theta$, respectively, we set
$\gamma_0^\theta=\arcsin{\frac{2a\cos\theta_0+b}{\sqrt{D}}}$ and ``$\pm$''
denotes the sign of $\dv{\theta}{\gamma}{(\gamma_0)}$.

\subsection{\label{sec:solutionr}Solution of the \texorpdfstring{$r$}{r}
equation}
By substituting $r=\frac{1}{x}+r_\Rho$, with $r_\Rho$ chosen to be a zero of
$\Rho$, one can simplify Eq.~(\ref{eq:motionr}) to a differential equation of
the type $\big(\dv{x}{\gamma}\big)^{\!2}=\Rho_3^x$ with a polynomial
$\Rho_3^x\coloneqq\sum_{i=0}^{3}b_ix^i$ of third order on its right hand side
(this step is not necessary for $\delta=0$). The following substitution
$x=\frac{1}{b_3}\left(4y-\frac{b_2}{3}\right)$ transforms this to
\begin{equation}
	\left(\dv{y}{\gamma}\right)^{\!2}=4y^3-g_2y-g_3\eqqcolon\Rho_3^y\,,
	\label{eq:derivationr1}
\end{equation}
with coefficients
\begin{equation}
	g_2=\frac{b_2^2}{12}-\frac{b_1b_3}{4}\,, \quad
	g_3=\frac{b_1b_2b_3}{48}-\frac{b_0b_3^2}{16}-\frac{b_2^3}{216}\,.
\end{equation}
This elliptic differential equation is solved by the Weierstrass $\wp$ function
\cite{markushevich1967theory}, which leads to
\begin{equation}
	y(\gamma)=\wp\left(\gamma-\gamma'_0;g_2,g_3\right) \label{eq:derivationr2}
\end{equation}
as a solution of Eq.~(\ref{eq:derivationr1}). Here we set
$\gamma'_0\coloneqq\gamma_0+\int_{y_0}^\infty\frac{\dd{y}}{\sqrt{\Rho_3^y}}$
with $y_0=y(x(r_0))$ and the initial value $r_0$ of $r$. The solution of
Eq.~(\ref{eq:motionr}) is now obtained by
\begin{equation}
	r(\gamma)=\frac{b_3}{4\wp(\gamma-\gamma'_0;g_2,g_3)-\frac{b_2}{3}}+r_R\,.
\end{equation}

\subsection{\label{sec:solutionphi}Solution of the \texorpdfstring{$\phi$}{phi}
equation}
To integrate Eq.~(\ref{eq:motionphi}), we handle its two parts separately and
set
\begin{equation}
	\dv{\phi}{\gamma}\eqqcolon A(r)+B(\theta)\,. \label{eq:derivationphi1}
\end{equation}
Again, the substitution $\xi=\cos\theta$ is used, this time upon $B$. Together
with Eq.~(\ref{eq:derivationtheta}), one gets for the $\theta$ part
\begin{equation}
	\pm\dd{\phi_\theta}=\frac{\xi^2}{\xi^2-1}\frac{L\dd{\xi}}{\sqrt{\Theta_\xi}}
	-\frac{\xi }{\xi^2-1}\frac{J\dd{\xi}}{\sqrt{\Theta_\xi}}\,.
\end{equation}
Here the ``$\pm$'' indicates the sign of $\dv{\theta}{\gamma}{(\gamma_0)}$. Upon
the two fractions we apply partial fraction decompositions
\begin{equation}
	\pm\dd{\phi_\theta}=
		\frac{1}{2}\frac{L-J}{\xi-1}\frac{\dd{\xi}}{\sqrt{\Theta_\xi}}
		-\frac{1}{2}\frac{L+J}{\xi+1}\frac{\dd{\xi}}{\sqrt{\Theta_\xi}}
		+L\frac{\dd{\xi}}{\sqrt{\Theta_\xi}}\,.
\end{equation}
The first part can be easily integrated with the substitution $y=\xi-1$, the
second part with $y=\xi+1$. The third part is equal to $\mp L\dd{\gamma}$, as
can be seen from Eq.~(\ref{eq:derivationtheta}). This yields
\begin{align}
	\begin{split}
		\phi_\theta(\gamma)=&\pm\frac{1}{2}\frac{L-J}{\sqrt{-c_1}}\arctan
			\frac{2c_1+b_1y}{2\sqrt{-c_1\Theta_{y,1}}}\biggr|_{\xi_0-1}^{\xi(\gamma)-1}\\
		&\mp\frac{1}{2}\frac{L+J}{\sqrt{-c_2}}\arctan
			\frac{2c_2+b_2y}{2\sqrt{-c_2\Theta_{y,2}}}\biggr|_{\xi_0+1}^{\xi(\gamma)+1}\\
		&-L(\gamma-\gamma_0) \label{eq:derivationphi2}
	\end{split}\\
	\makebox[0pt][l]{$\eqqcolon I_1(\gamma)+I_2(\gamma)+I_3(\gamma)\,,$}
	\phantom{=}& \label{eq:phiintegrals}
\end{align}
where $\Theta_{y,i}\coloneqq ay^2+b_iy+c_i$, $i=1,2$ with $b_1=b+2a$,
$b_2=b-2a$, $c_1=a+b+c$, and $c_2=a-b+c$. Furthermore, $c_i<0$ was assumed.

Now we integrate the $r$ dependent part $A$ of Eq.~(\ref{eq:motionphi}). We use
the two substitutions applied in Section~\ref{sec:solutionr} and recall
Eq.~(\ref{eq:derivationr2}) to identify $A=A(r(y(\gamma)))$. A partial fraction
decomposition on $A(y)$ is performed to get
\begin{equation}
	\dd{\phi_r}=\left[K_0+\frac{K_1}{y(\gamma)-p_1}
		+\frac{K_2}{y(\gamma)-p_2}\right]\dd{\gamma}\,, \label{eq:derivationphi3}
\end{equation}
where $K_0,\,K_1$, and $K_2$ are constants of the partial fraction decomposition
and $p_1=\frac{(l^2+r_\Rho+2)b_2-3b_3}{12(l^2+r_\Rho+2)}$ and
$p_2=\frac{(r_\Rho-1)b_2-3b_3}{12(r_\Rho-1)}$ are poles of first order.
According to \cite{Willenborg:2018zsv}\footnote{Here an incorrect index has been
changed.}, the elliptic integrals of third kind
$\int_{\nu_0}^{\nu}\frac{\dd{\nu}}{\wp(\nu)-p}$ can be solved in terms of the
Weierstrass $\sigma$ and $\zeta$ functions by using addition theorems. One
obtains with $\nu\coloneqq\gamma-\gamma_0$
\begin{equation}
	\begin{split}
		\phi_r(\gamma)={}&K_0(\nu-\nu_0)\\
		&+\sum\limits_{i=1}^{2}\frac{K_i}{\wp'(\nu_{i})}
			\bigg[2\zeta(\nu_i)(\nu-\nu_0)\\
		&\phantom{+\sum\limits_{i=1}^{2}\frac{K_i}{\wp'(\nu_{i})}\bigg[}
			+\ln\frac{\sigma(\nu-\nu_i)}{\sigma(\nu_0-\nu_i)}
			-\ln\frac{\sigma(\nu+\nu_i)}{\sigma(\nu_0+\nu_i)}\bigg]\,.
		\label{eq:derivationphi4}
	\end{split}
\end{equation}
Here $\nu_i$ is a Weierstrass transformed pole $p_i$, which solves
$\wp(\nu_i)=p_i$ for $i=1,2$ in the fundamental parallelogram of $\wp(\nu)$.

The solution of Eq.~(\ref{eq:derivationphi1}) is given by the sum of this
expression, Eq.~(\ref{eq:derivationphi2}), and the initial value $\phi_0$.

\subsection{Solution of the \texorpdfstring{$\psi$}{psi} equation}
To obtain the $\psi$ motion from Eq.~(\ref{eq:motionpsi}), we proceed similarly
to the solution of the $\theta$ part of the $\phi$ motion in
Section~\ref{sec:solutionphi}. Again, with the substitution $\xi=\cos\theta$ and
partial fraction decompositions, one gets
\begin{equation}
	\pm\dd{\psi}=-\frac{1}{2}\frac{L-J}{\xi-1}\frac{\dd{\xi}}{\sqrt{\Theta_\xi}}
		-\frac{1}{2}\frac{L+J }{\xi+1}\frac{\dd{\xi}}{\sqrt{\Theta_\xi}}\,.
\end{equation}
Using the definitions of Eq.~(\ref{eq:phiintegrals}), we find with the initial
value $\psi_0$ of $\psi$
\begin{equation}
	\psi(\gamma)=-I_1(\gamma)+I_2(\gamma)+\psi_0\,.
\end{equation}

\subsection{Solution of the \texorpdfstring{$t$}{t} equation}
The integral of the right hand side of Eq.~(\ref{eq:motiont}) can be derived
similarly to that of $A(r)$ in Section~\ref{sec:solutionphi}. Here the partial
fraction decomposition leads to
\begin{equation}
	\dd{t}=\left[H_0+\frac{H_1}{y(\gamma)-p_1}+\frac{H_2}{y(\gamma)-p_2}
		+\frac{H_3}{(y(\gamma)-p_2)^2}\right]\dd{\gamma}
\end{equation}
with $H_0,\, H_1,\, H_2$, and $H_3$ given by the partial fraction decomposition
and $p_1$ and $p_2$ are poles of first and second order, respectively, that are
identical to those in Eq.~(\ref{eq:derivationphi3}). The integral of the first
three terms yields an expression of the type obtained in
Eq.~(\ref{eq:derivationphi4}). Again, according to
\cite{Willenborg:2018zsv}\footnote{Here two incorrect signs have been changed.},
the elliptic integral of type
$\int_{\nu_0}^{\nu}\frac{\dd{\nu}}{(\wp(\nu)-p)^2}$ can be solved in terms of
the Weierstrass $\sigma$ and $\zeta$ functions. One obtains with
$\nu\coloneqq\gamma-\gamma_0$ and the initial value $t_0$ of $t$
\begin{equation}
	\begin{split}
		t(\gamma)={}&t_0+H_0(\nu-\nu_0)\\
		&+\sum\limits_{i=1}^{2}\frac{H_i}{\wp'(\nu_{i})}
			\bigg[2\zeta(\nu_i)(\nu-\nu_0)\\
		&\phantom{+\sum\limits_{i=1}^{2}\frac{H_i}{\wp'(\nu_{i})}\bigg[}
			+\ln\frac{\sigma(\nu-\nu_i)}{\sigma(\nu_0-\nu_i)}
			-\ln\frac{\sigma(\nu+\nu_i)}{\sigma(\nu_0+\nu_i)}\bigg]\\
		&-\frac{H_3\wp''(\nu_2)}{(\wp'(\nu_2))^3}\bigg[2\zeta(\nu_2)(\nu-\nu_0)\\
		&\phantom{-\frac{H_3\wp''(\nu_2)}{(\wp'(\nu_2))^3}\bigg[}
			+\ln\frac{\sigma(\nu-\nu_2)}{\sigma(\nu_0-\nu_2)}
			-\ln\frac{\sigma(\nu+\nu_2)}{\sigma(\nu_0+\nu_2)}\bigg]\\
		&-\frac{H_3}{(\wp'(\nu_2))^2}\bigg[2\wp(\nu_2)(\nu-\nu_0)
			+2[\zeta(\nu)-\zeta(\nu_0)]\\
		&\phantom{-\frac{H_3}{(\wp'(\nu_2))^2}\bigg[}
			+\frac{\wp'(\nu)}{\wp(\nu)-\wp(\nu_2)}
			-\frac{\wp'(\nu_0)}{\wp(\nu_0)-\wp(\nu_2)}\bigg]\,.
	\end{split}
\end{equation}

\section{\label{sec:orbits}The orbits}
To plot the spatial coordinates $(\phi,\psi,\theta,r)$ of the set of analytical
solutions for the particle's motion in a Cartesian coordinate system, a
coordinate transformation has to be chosen. The angular line element
$\dd{\Omega^2}=\frac{1}{R^2}\sum_{\mu,\nu=1}^3g_{\mu\nu}\dd{x^\mu}\dd{x^\nu}$
for $x^\mu=(t,\phi,\psi,\theta,R)$ of the metric in
Eq.~(\ref{eq:metricequation}) yields in case of $R\rightarrow\infty$
\begin{equation}
	\dd{\Omega^2}=\frac{1}{4}\left[\dd{\theta^2}+\dd{\psi^2}+\dd{\phi^2}
		+2\cos\theta\dd{\phi} \dd{\psi}\right]\,. \label{eq:angularlineelement1}
\end{equation}
On the other hand, the angular line element of flatspace in biazimuthal
coordinates $(\varphi_1,\varphi_2,\vartheta,R)$ is
\begin{equation}
	\dd{\Omega^2}=\dd{\vartheta^2}+\sin^2\vartheta\dd{\varphi_1^2}
		+\cos^2\vartheta\dd{\varphi_2^2}\,. \label{eq:angularlineelement2}
\end{equation}
A comparison of Eqs.~(\ref{eq:angularlineelement1}) and
(\ref{eq:angularlineelement2}) suggests the coordinate transformation
\begin{align}
	\phi&=\varphi_1+\varphi_2\,, & \psi&=\varphi_2-\varphi_1\,, &
		\theta&=2\vartheta\,,
\end{align}
which, together with the biazimuthal coordinates and $r=R^2$, transforms
$(\phi,\psi,\theta,r)$ to $(x,y,z,w)$ by
\begin{subequations}
	\begin{align}
		x&=\sqrt{r}\sin\frac{\theta}{2}\cos\frac{\phi-\psi}{2}\,,\\
		y&=\sqrt{r}\sin\frac{\theta}{2}\sin\frac{\phi-\psi}{2}\,,\\
		z&=\sqrt{r}\cos\frac{\theta}{2}\cos\frac{\phi+\psi}{2}\,,\\
		w&=\sqrt{r}\cos\frac{\theta}{2}\sin\frac{\phi+\psi}{2}\,.
	\end{align}
\end{subequations}

Furthermore, we choose a projection in the 3-dimensional Cartesian subspace of
the coordinates $(x,y,z)$ by setting $w=0$. This preserves the notion of a
horizon on the surface of a sphere with radius $R=1$, but as a consequence we
have to set $\psi=-\phi$, so only analytical solutions of $\phi,\,\theta,$ and
$r$ are used. Another consequence is that the $\theta$ motion on the interval
$[0,\pi]$ is now mapped to the upper hemisphere only, thereby, as the horizon is
shown in the following as a complete sphere of radius $1$, the lower half sphere
is of no physical relevance.

\begin{figure*}
	\subfloat[MBO and its projections. The $x$-$y$ projection (bottom right) shows
	the $\phi$ turning points at $\theta=\pi/2$ as a dotted circle.]{%
		\includegraphics[width=.5\linewidth]{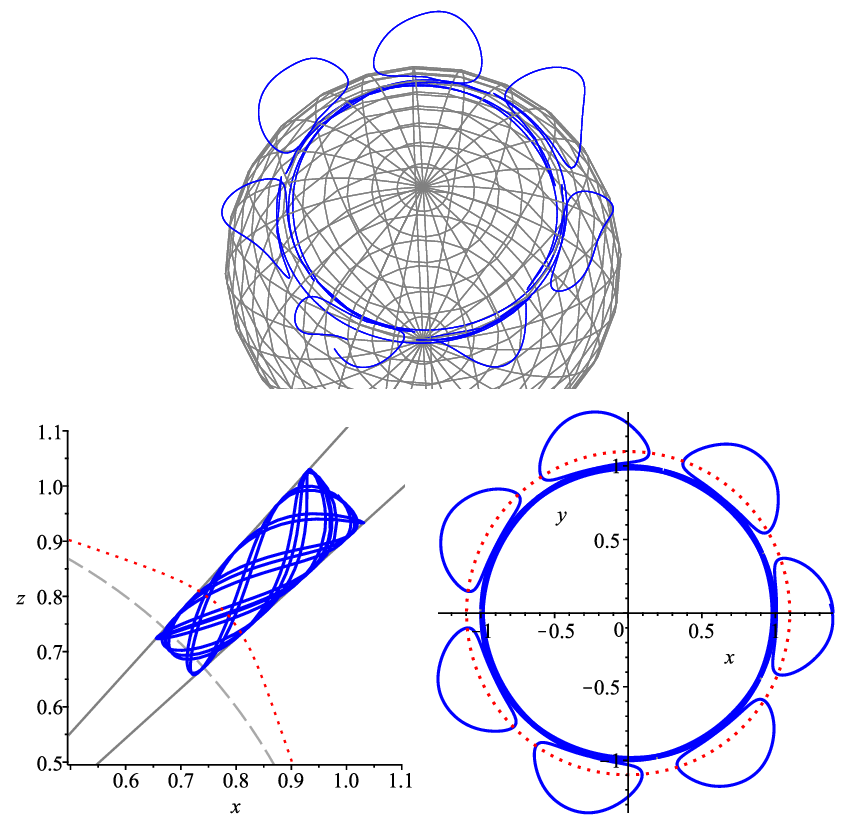}%
		\label{fig:orbit1_1}%
	}
	\subfloat[BO and its projections.]{%
		\includegraphics[width=.5\linewidth]{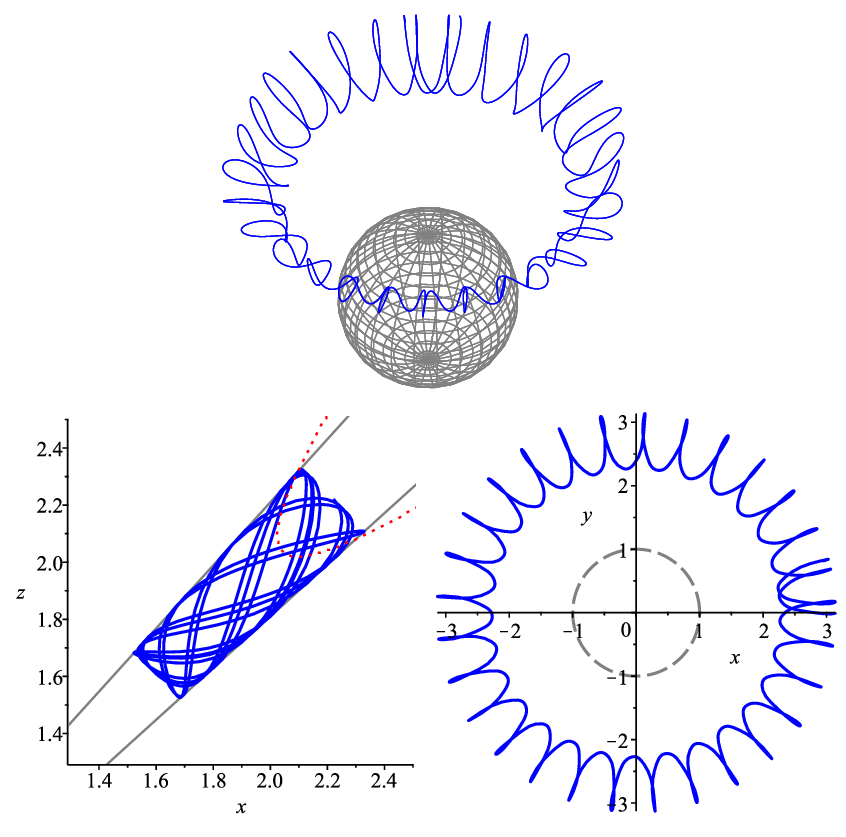}%
		\label{fig:orbit1_2}%
	}
	\caption{\label{fig:orbit1}%
		MBO and BO for parameters $K = 4,\, L = 20,\, \delta = 1,\, l = 1,\, J = 0$,
		and energy $E=0.95$. The projections onto the $x$-$z$ plane (bottom left)
		show the turning points of the $\phi$ motion as a dotted line. Projections
		onto the $x$-$y$ plane are shown at the bottom right. The spheres and dashed
		circles show the horizon at $R=1$. The corresponding effective potential of
		the $r$ motion can be found in Fig.~\ref{fig:effpotr1b}.}
\end{figure*}

\begin{figure*}
	\subfloat[MBO and its projections.]{%
		\includegraphics[width=.5\linewidth]{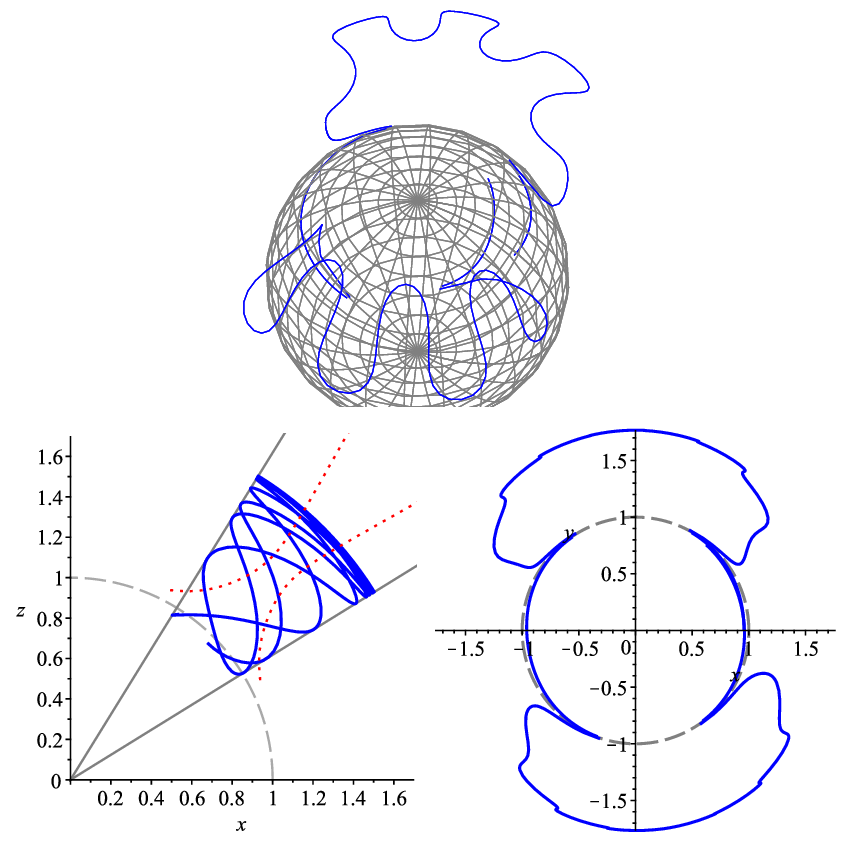}%
		\label{fig:orbit2_1}%
	}
	\subfloat[EO and its projections.]{%
		\includegraphics[width=.5\linewidth]{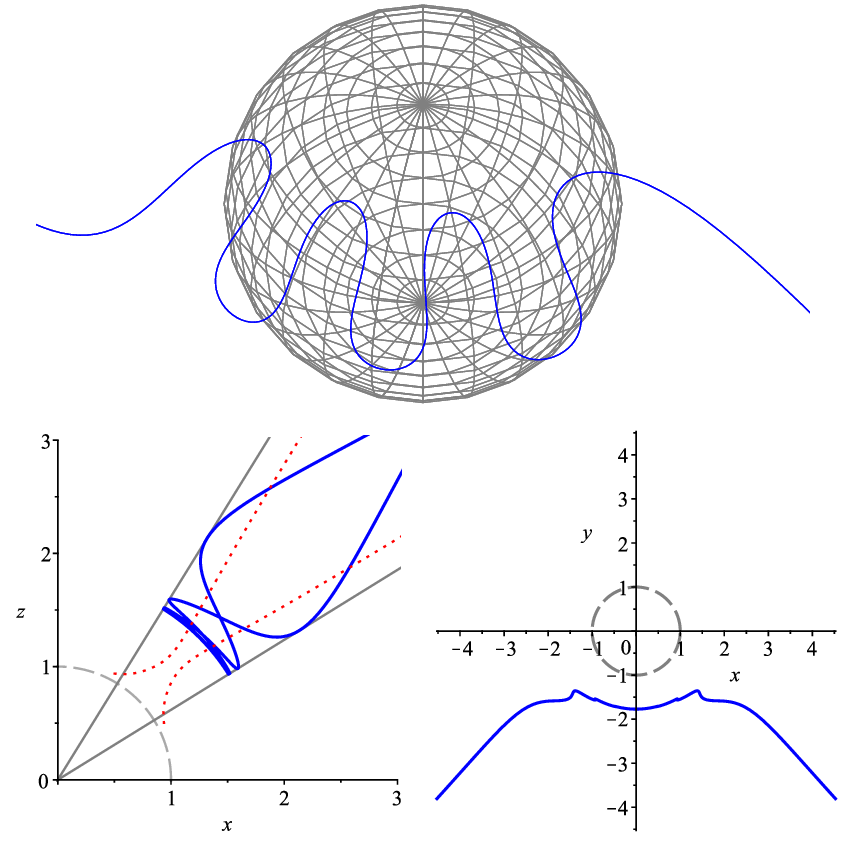}%
		\label{fig:orbit2_2}%
	}
	\caption{\label{fig:orbit2}%
		MBO and EO for parameters $K = 4,\, L=4,\, \delta = 0,\, l = 1,\, J = 0$,
		and energy $E = 0.386699$. The projections onto the $x$-$z$ plane (bottom
		left) show the turning points of the $\phi$ motion as a dotted line.
		Projections onto the $x$-$y$ plane are shown at the bottom right. The
		spheres and dashed circles show the horizon at $R=1$.}
\end{figure*}

\begin{figure*}
	\subfloat[MBO.]{%
		\includegraphics[width=.25\linewidth]{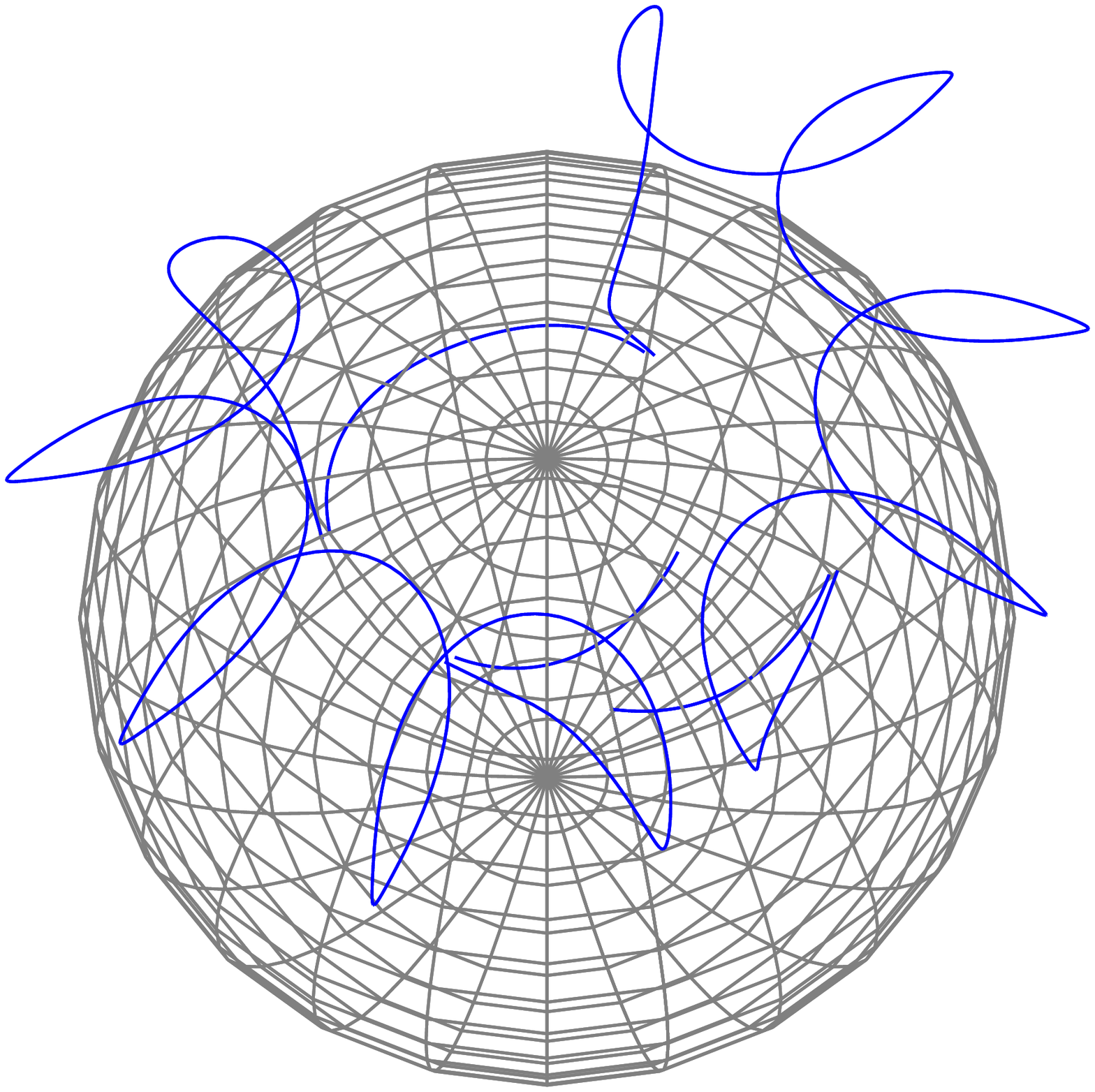}\label{fig:orbit3_1}%
	}
	\subfloat[Projection of \csubref{fig:orbit3_1} onto the $x$-$z$ plane.]{%
		\includegraphics[width=.25\linewidth]{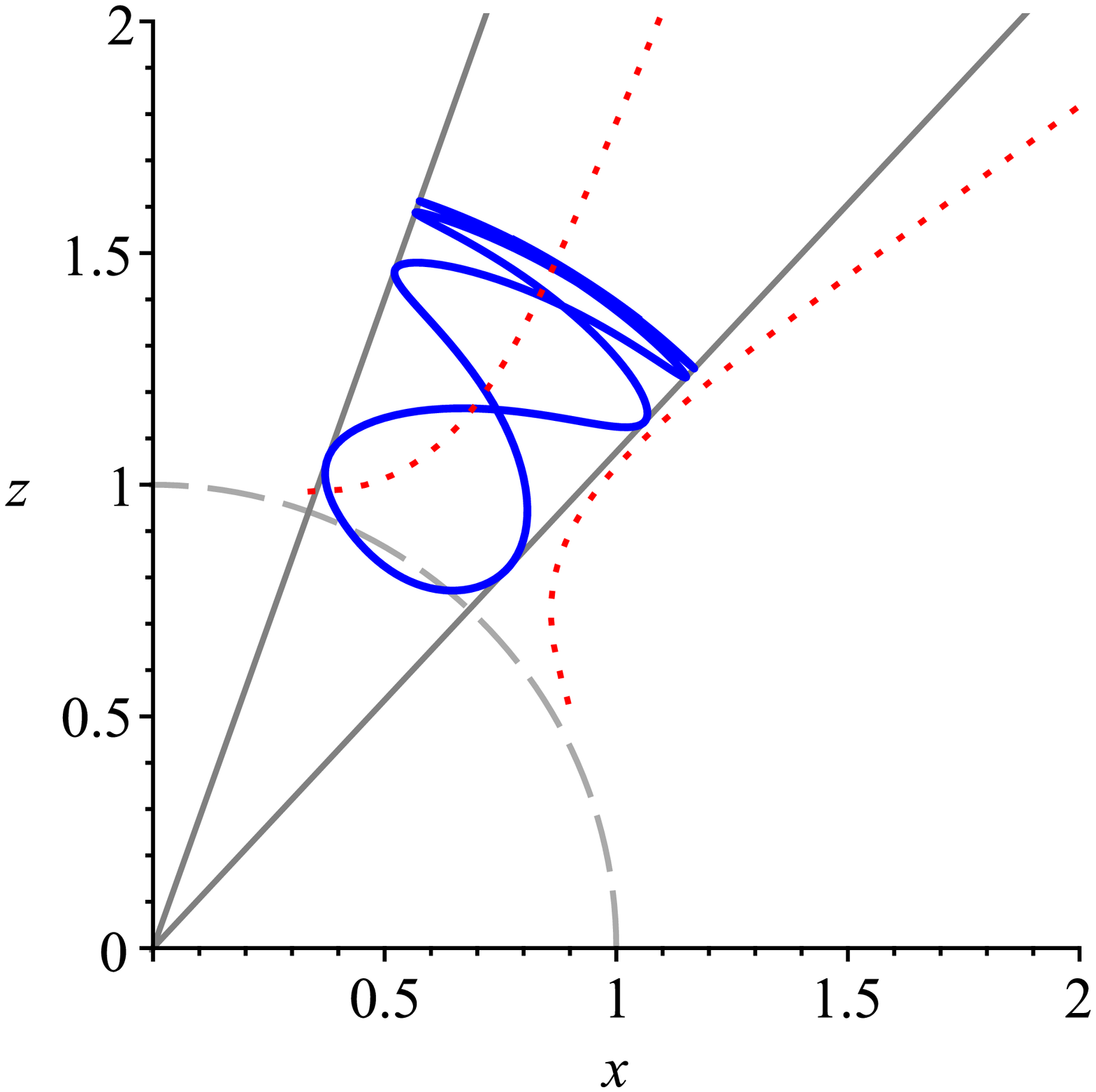}%
		\label{fig:orbit3_1xz}%
	}
	\subfloat[EO.]{%
		\includegraphics[width=.25\linewidth]{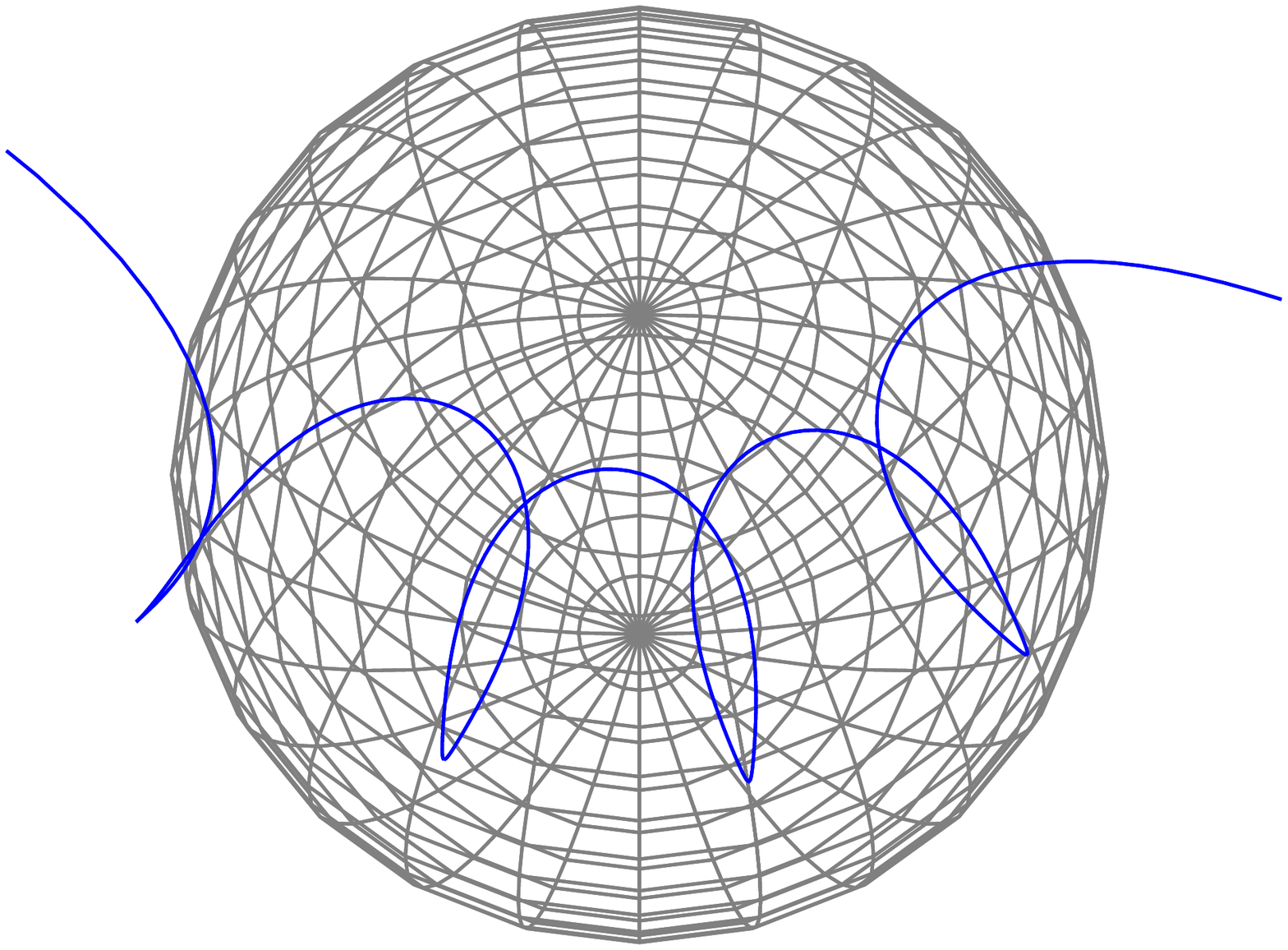}\label{fig:orbit3_2}%
	}
	\subfloat[Projection of \csubref{fig:orbit3_2} onto the $x$-$z$ plane.]{%
		\includegraphics[width=.25\linewidth]{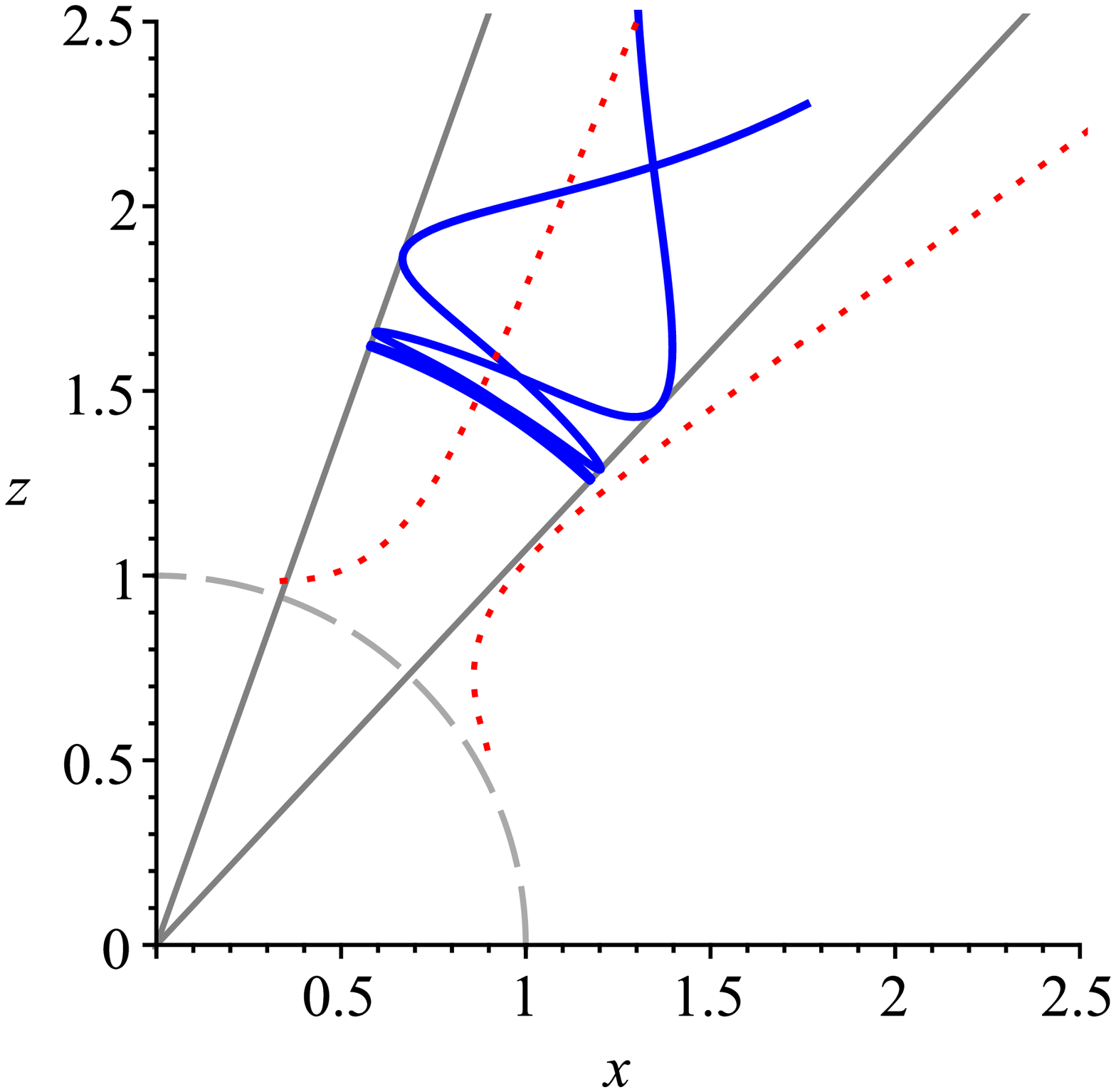}%
		\label{fig:orbit3_2xz}%
	}
	\caption{\label{fig:orbit3}%
		MBO and EO for parameters $K = 12,\, L = 8,\, \delta = 0,\, l = 1,\, J = 4$,
		and energy $E = 0.6672265$. The trajectory is confined to an area above
		$\theta=\pi/2$. In \csubref{fig:orbit3_1xz} and \csubref{fig:orbit3_2xz} the
		dotted curves denote the turning points of the $\phi$ motion. The spheres
		and dashed circles show the horizon at $R=1$.}
\end{figure*}

The figures below show the particle's motion for various sets of parameters. The
first orbit of a particle of positive mass is of type E and is shown in
Fig.~\ref{fig:orbit1}, its effective potential can be found in
Fig.~\ref{fig:effpotr1b}. For similar parameters ($L$ was decreased) a lightlike
orbit of type C is shown in Fig.~\ref{fig:orbit2}. Both of those orbits pass the
equatorial plane. An orbit that is confined to one hemisphere has to satisfy
$K<J^2$, as was discussed in Section~\ref{sec:classificationtheta}. A lightlike
orbit of this kind is shown in Fig.~\ref{fig:orbit3} for type C.

\begin{figure*}
	\subfloat[MBO.]{%
		\includegraphics[width=.25\linewidth]{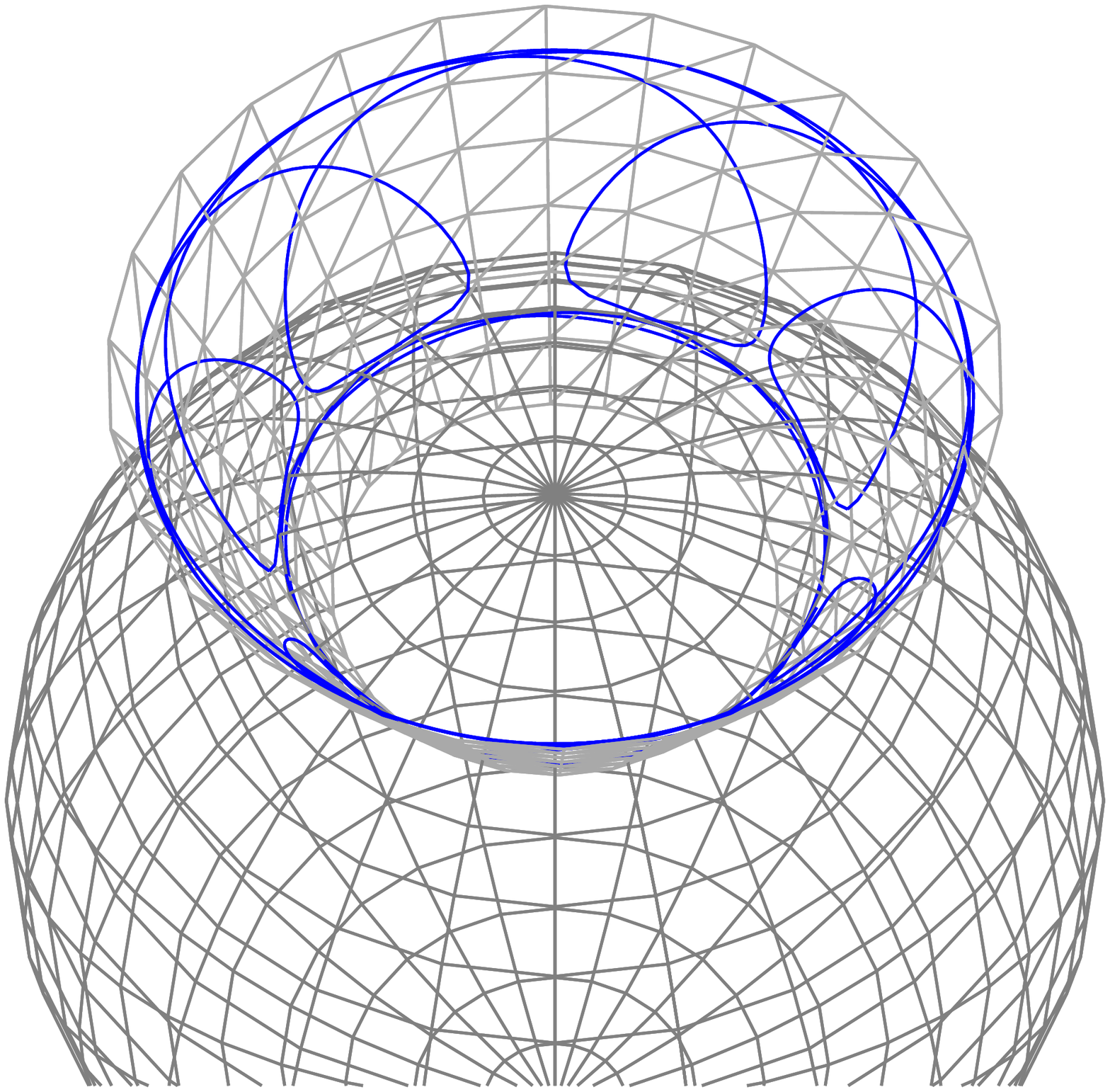}\label{fig:orbit4_1}%
	}
	\subfloat[Projection of \csubref{fig:orbit4_1} onto the $x$-$y$ plane.]{%
		\includegraphics[width=.25\linewidth]{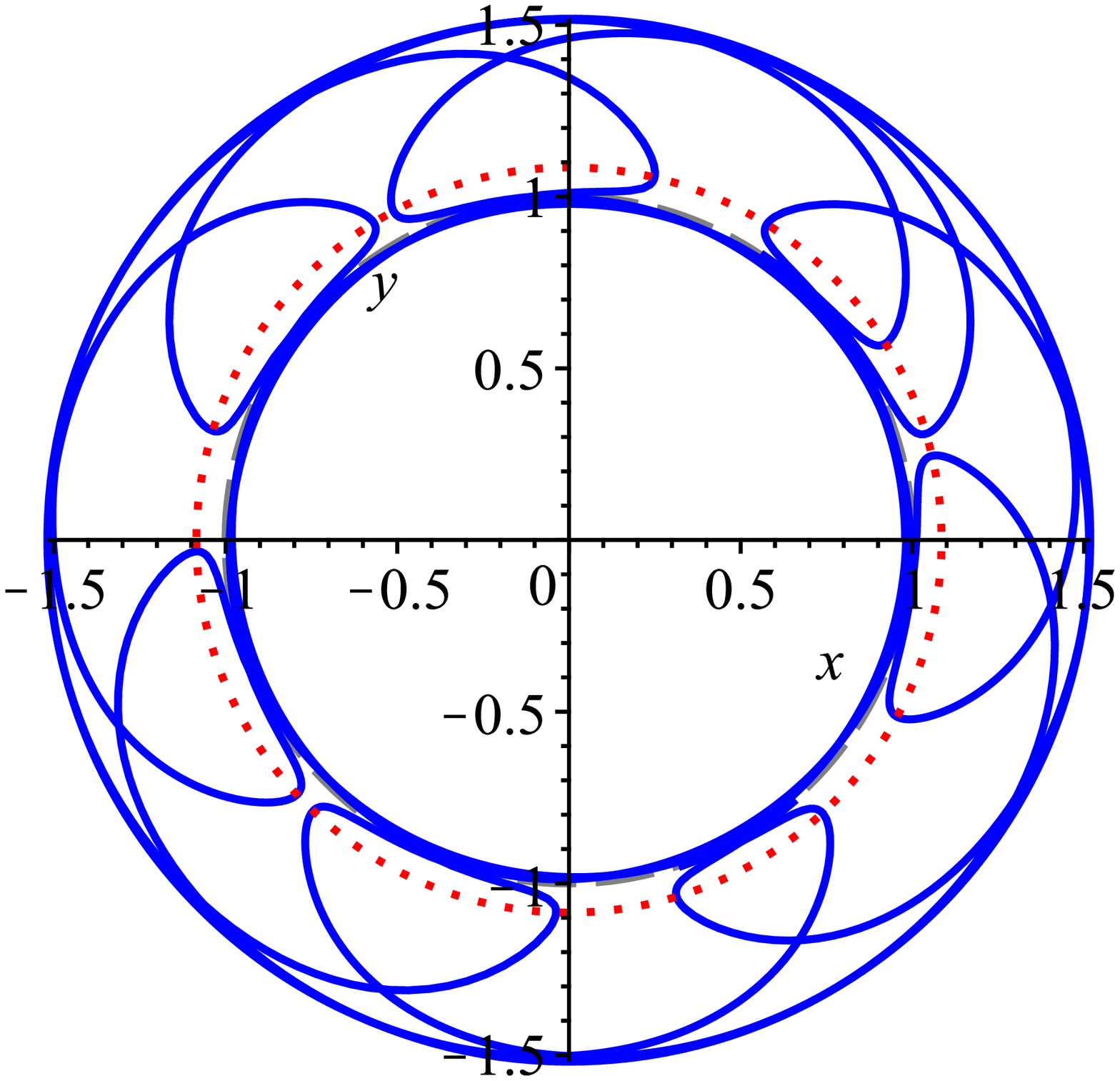}%
		\label{fig:orbit4_1xy}%
	}
	\subfloat[EO.]{%
		\includegraphics[width=.25\linewidth]{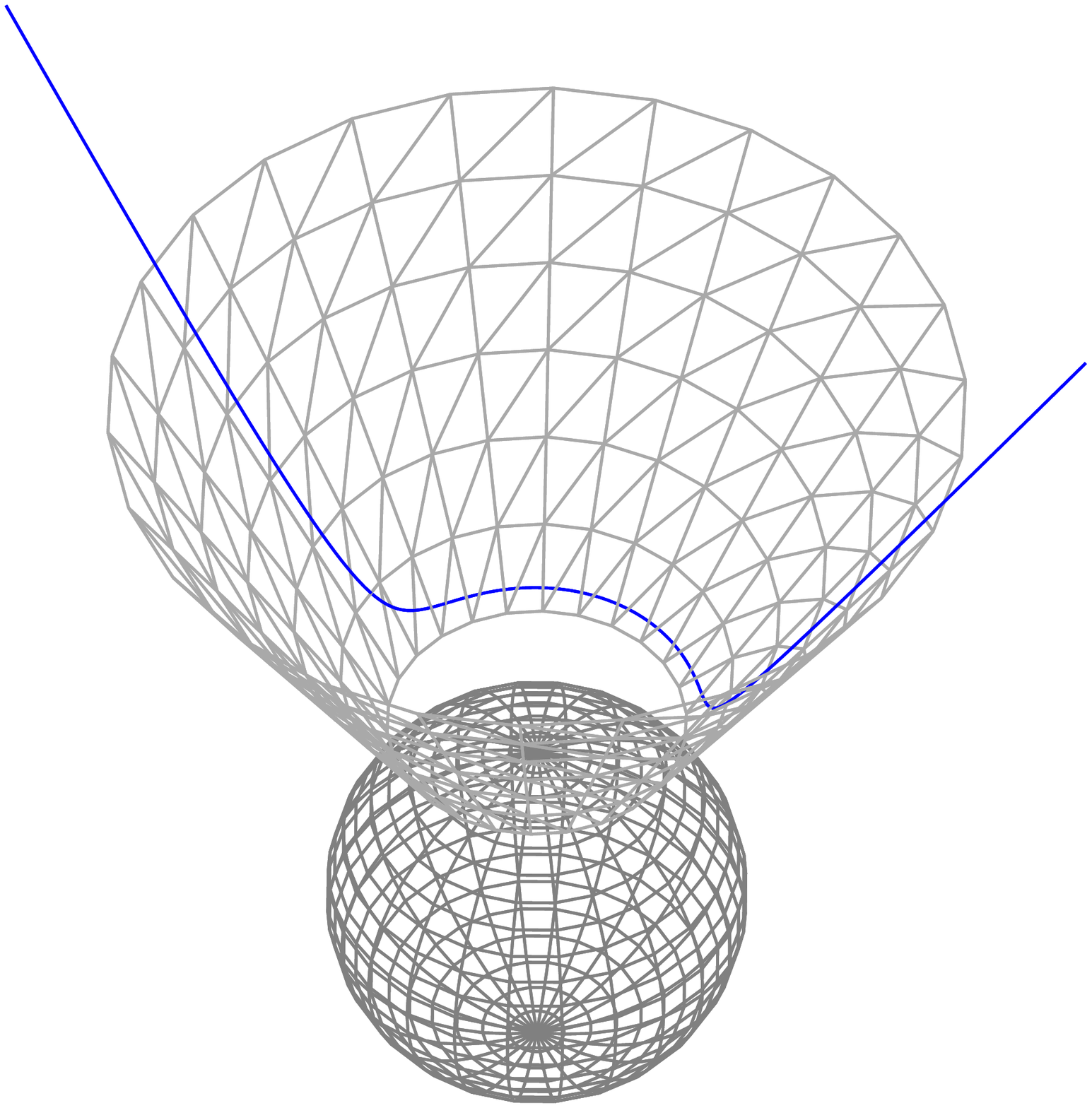}\label{fig:orbit4_2}%
	}
	\subfloat[Projection of \csubref{fig:orbit4_2} onto the $x$-$y$ plane.]{%
		\includegraphics[width=.25\linewidth]{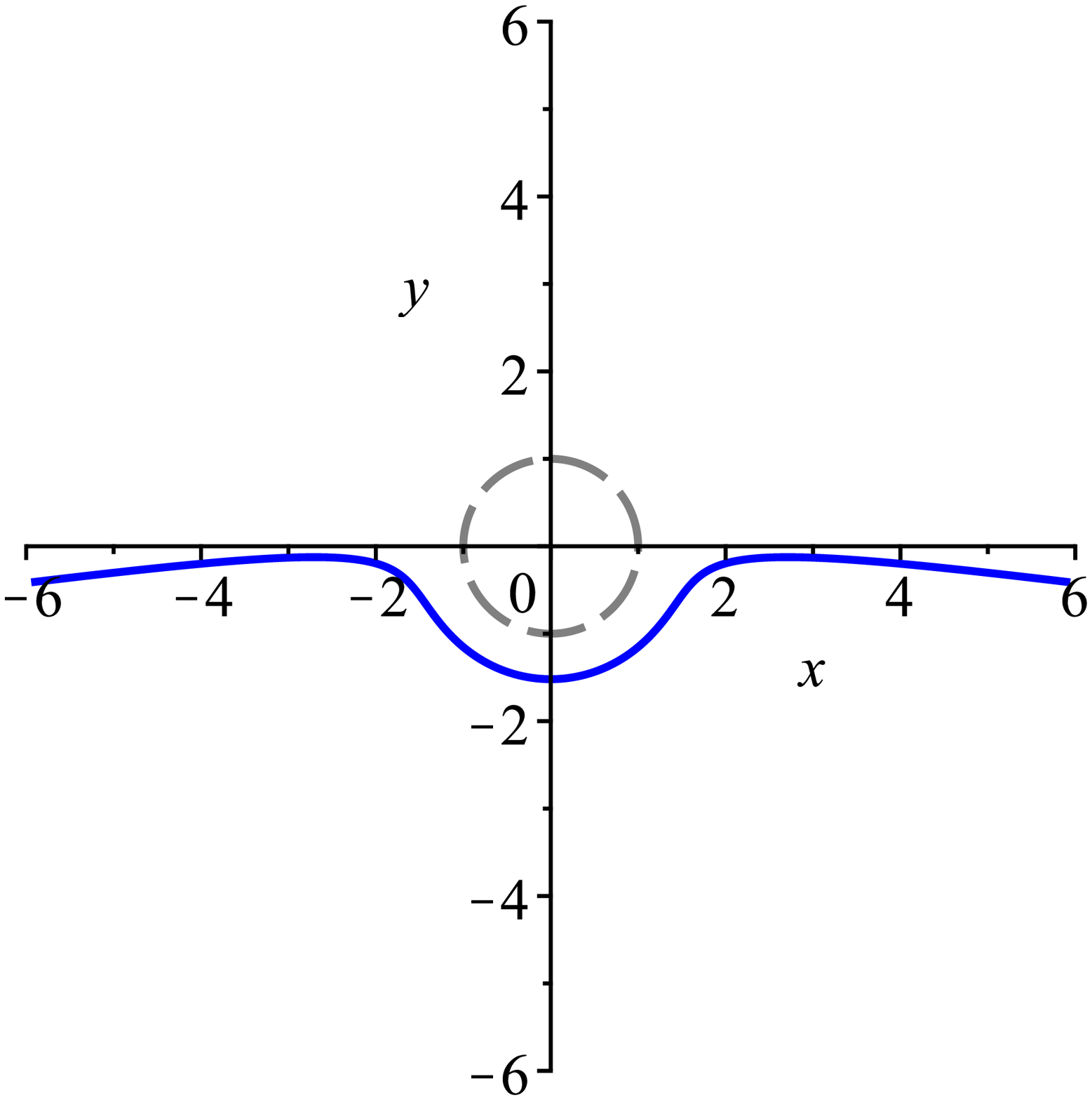}%
	}
	\caption{\label{fig:orbit4}%
		MBO and EO for parameters $K = 0,\, L = 8,\, \delta = 0,\, l = 1,\, J = 4$,
		and energy $E = 0.343913$. The trajectory lies on a cone with opening angle
		$\theta<\pi/2$. In \csubref{fig:orbit4_1xy} the dotted circle indicates the
		$\phi$ turning points. The spheres and dashed circles show the horizon at
		$R=1$.}
\end{figure*}

\begin{figure*}
	\subfloat[TEO.]{%
		\includegraphics[width=.25\linewidth]{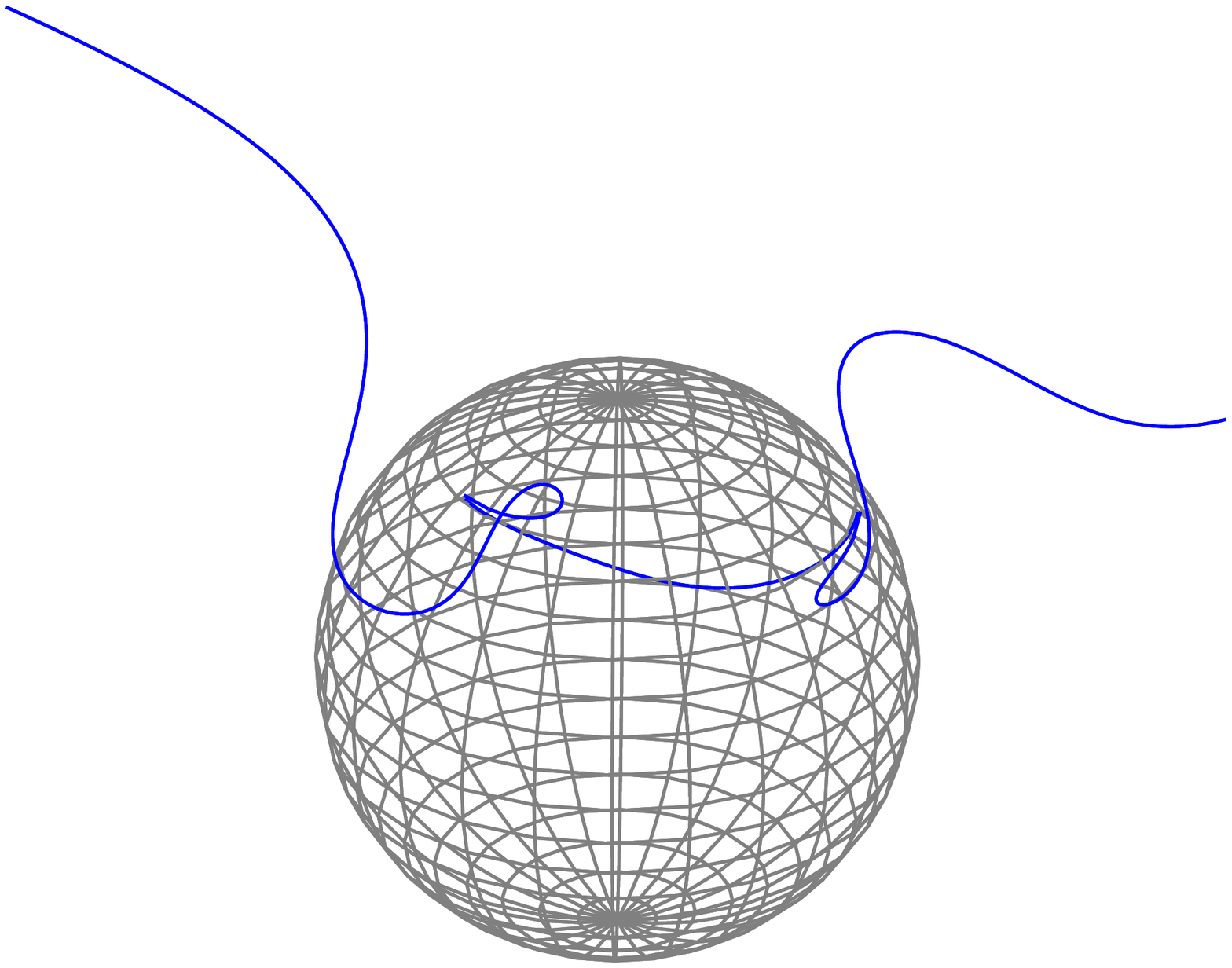}\label{fig:orbit5_1}%
	}
	\subfloat[Projection of \csubref{fig:orbit5_1} onto the $x$-$z$ plane.]{%
		\includegraphics[width=.25\linewidth]{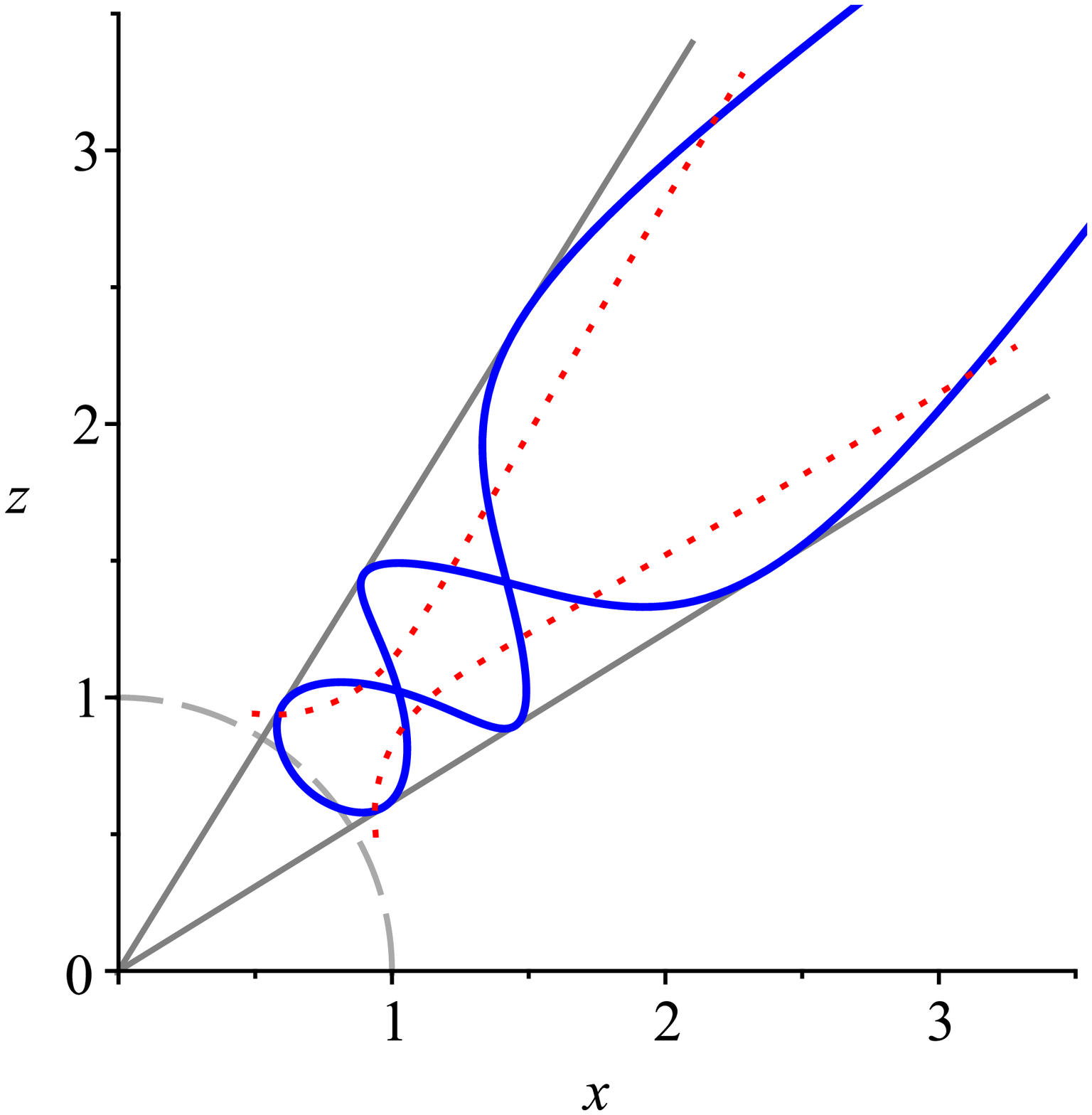}%
		\label{fig:orbit5_1xz}%
	}
	\subfloat[TO.]{%
		\includegraphics[width=.25\linewidth]{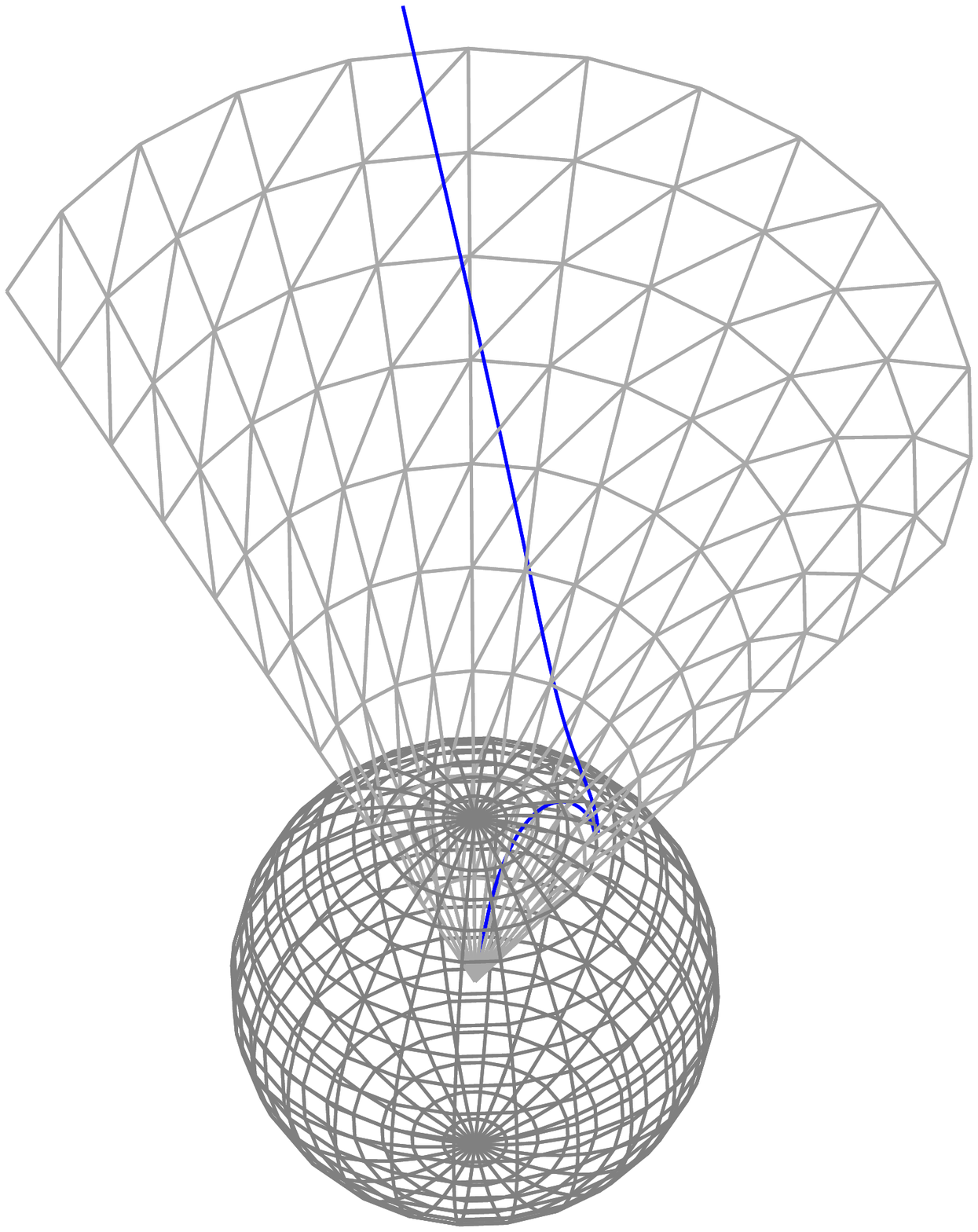}\label{fig:orbit6_1}%
	}
	\subfloat[Projection of \csubref{fig:orbit6_1} onto the $x$-$y$ plane.]{%
		\includegraphics[width=.25\linewidth]{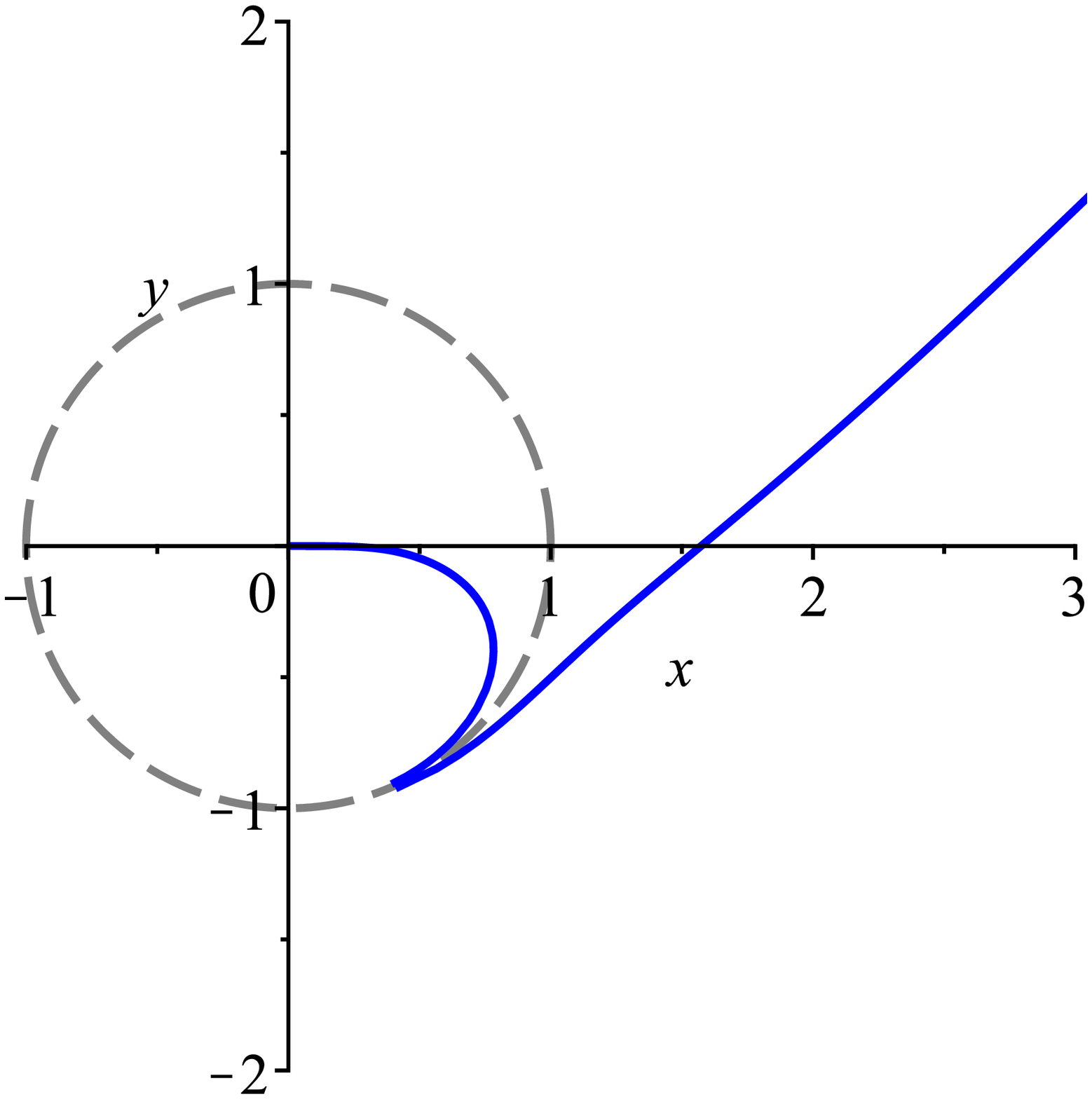}%
		\label{fig:orbit6_1xy}%
	}
	\caption{%
		\csubref{fig:orbit5_1} and \csubref{fig:orbit5_1xz}: TEO for parameters $K =
		4,\, L = 4,\, \delta = 0,\, l = 1,\, J = 0$, and energy $E = 0.4$. In
		\csubref{fig:orbit5_1xz} the dotted curve denotes the turning points of the
		$\phi$ motion. \csubref{fig:orbit6_1} and \csubref{fig:orbit6_1xy}: TO for
		parameters $K = 0,\, L = 8,\, \delta = -1,\, l = 1,\, J = 4$, and energy $E
		= Ll = 8$. The trajectory lies on a cone with opening angle $\theta<\pi/2$.
		The corresponding effective potential of the $r$ motion can be found in
		Fig.~\ref{fig:effpotr3f}. The spheres and dashed circles show the horizon at
		$R=1$.}
\end{figure*}

In case of $K=0$ and $\kappa>0$, one obtains an orbit confined to a cone of
fixed opening angle $\arccos\frac{J}{|L|}$. This is shown in
Fig.~\ref{fig:orbit4} for type C. A two-world escape orbit of type A can be
found in Fig.~\ref{fig:orbit5_1}. This is obtained by increasing the energy
slightly above the effective potential's maximum while the other parameters are
unchanged compared to Fig.~\ref{fig:orbit2}. By the increase in energy, the
many-world periodic bound orbit and the escape orbit merged into one two-world
escape orbit.

\begin{figure*}
	\subfloat[TO.]{%
		\includegraphics[width=.25\linewidth]{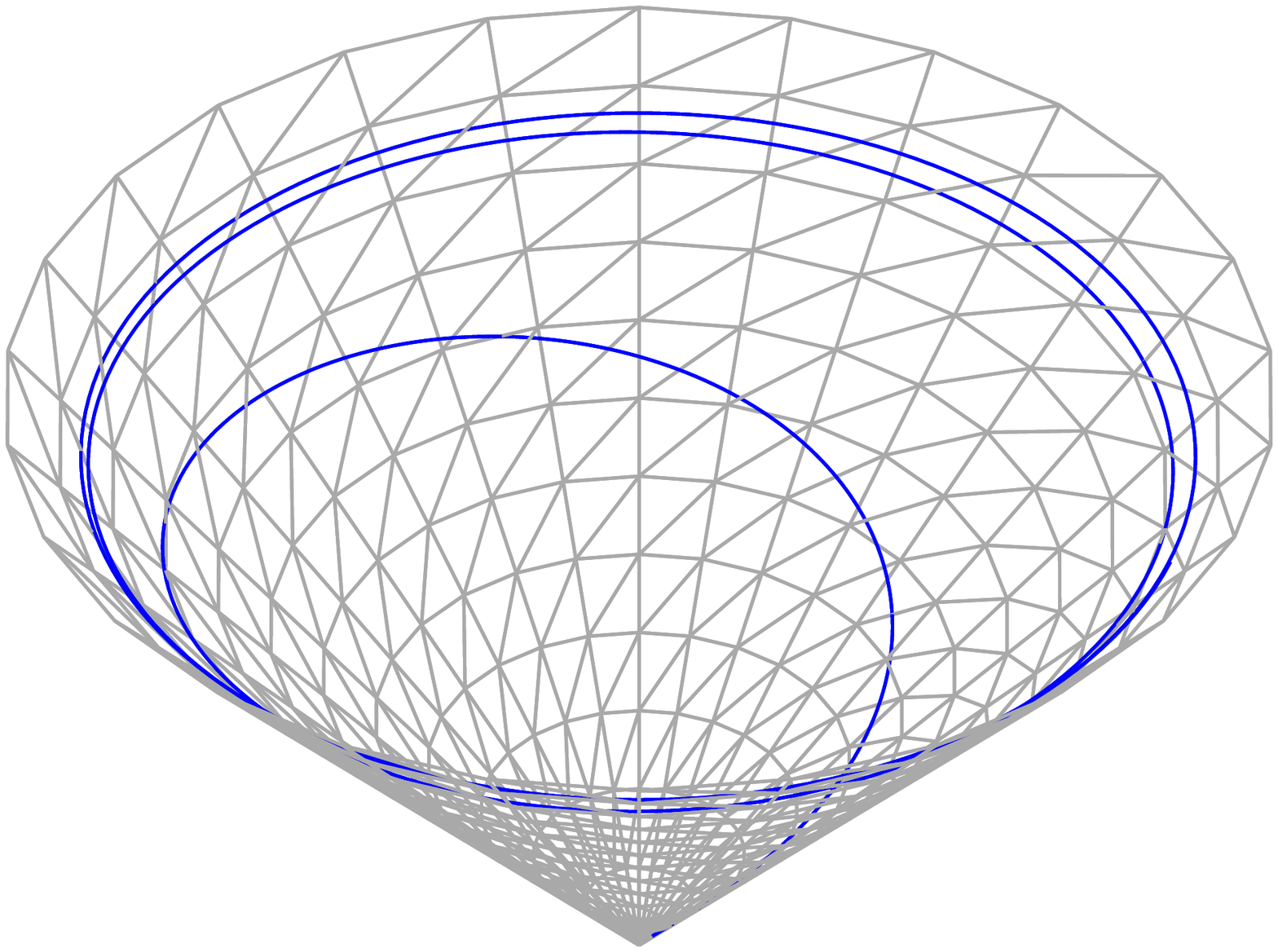}\label{fig:orbit7_1}%
	}
	\subfloat[Projection of \csubref{fig:orbit7_1} onto the $x$-$y$ plane.]{%
		\includegraphics[width=.25\linewidth]{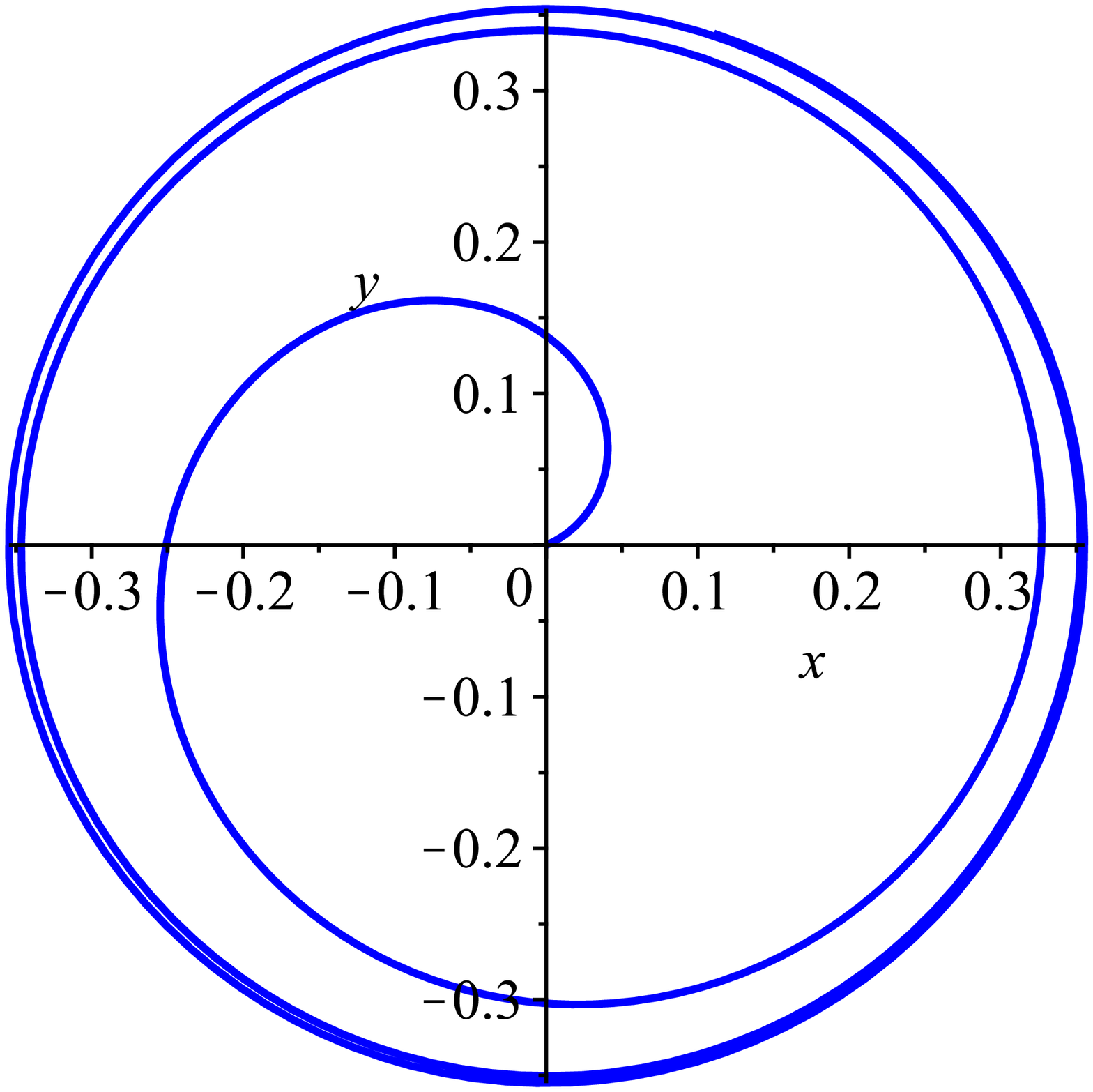}%
	}
	\subfloat[TEO.]{%
		\includegraphics[width=.25\linewidth]{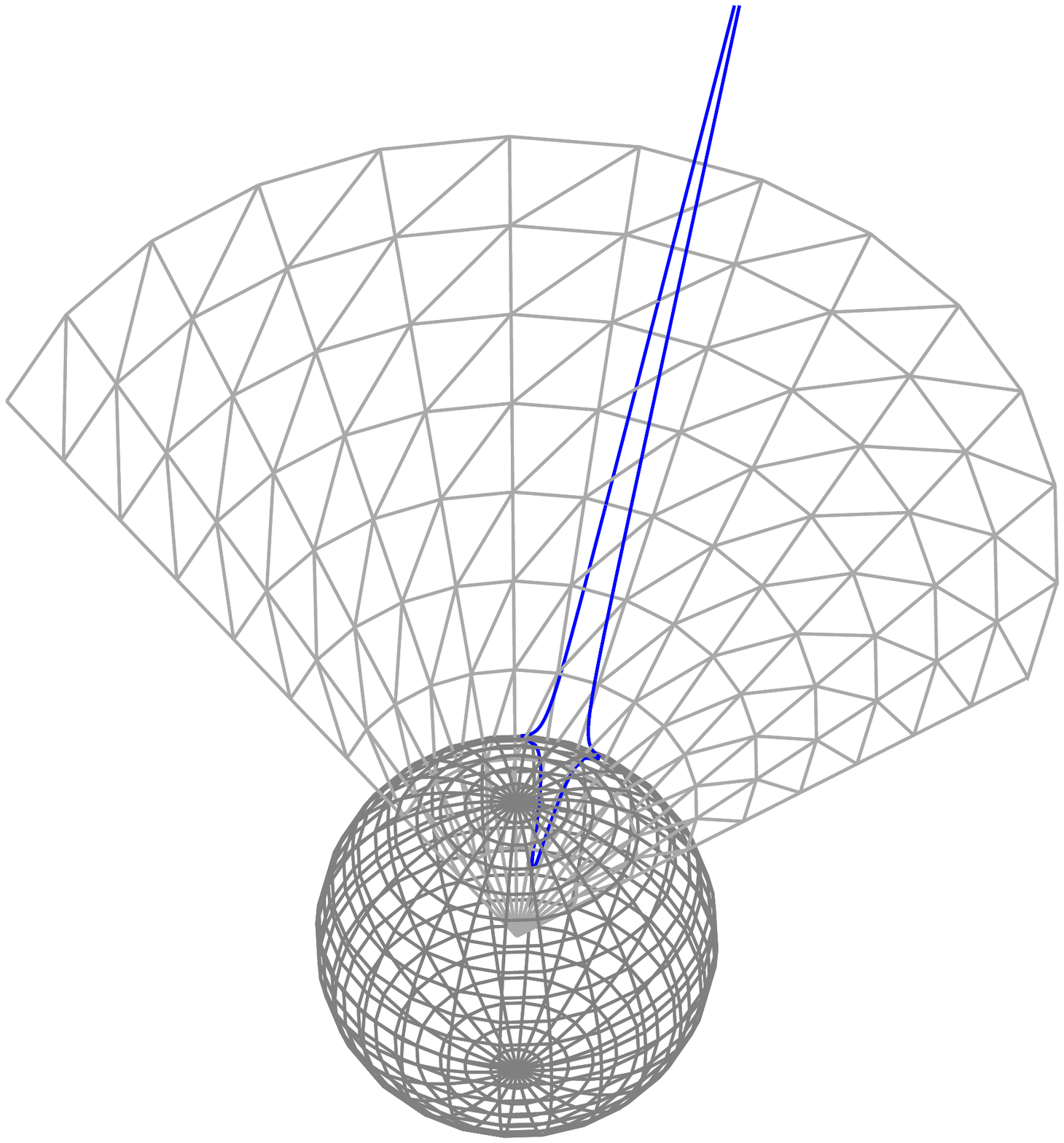}\label{fig:orbit7_2}%
	}
	\subfloat[Projection of \csubref{fig:orbit7_2} onto the $x$-$y$ plane.]{%
		\includegraphics[width=.25\linewidth]{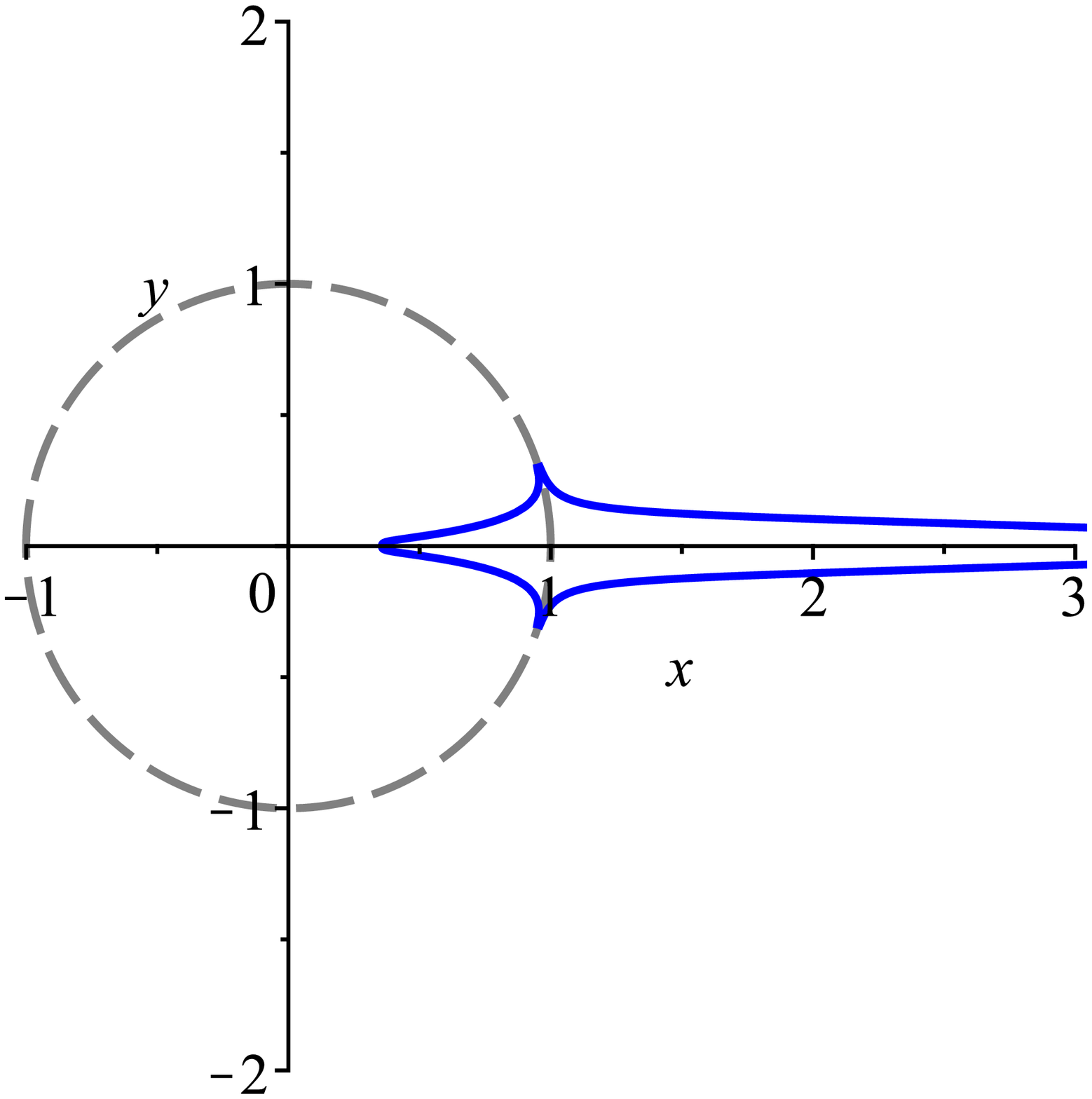}%
	}
	\caption{\label{fig:orbit7}%
		TO and TEO for parameters $K = 0,\, L = 36.39992,\, \delta = -1,\, l =
		0.3,\, J = 4$, and energy $E = Ll= 10.919976$. The trajectory lies on a cone
		with opening angle $\theta<\pi/2$. The sphere and dashed circle show the
		horizon at $R=1$. The corresponding effective potential of the $r$ motion is
		very similar to the one shown in Fig.~\ref{fig:effpotr3h}.}
\end{figure*}

Now we present orbits of particles with imaginary mass that reach the boundary
at infinity and therefore are of relevance for AdS/CFT. A terminating orbit of
type $\text{A}_0$, with its effective potential depicted in
Fig.~\ref{fig:effpotr3f}, can be found in Fig.~\ref{fig:orbit6_1}. In
Fig.~\ref{fig:orbit7} we show a terminating orbit and a two-world escape orbit
of type $\text{D}_0$ of a particle with imaginary mass in an effective potential
very similar to the one shown in Fig.~\ref{fig:effpotr3h}.

\begin{figure*}
	\centering
	\subfloat[\textit{Pointy petal} BO.]{%
		\includegraphics[width=.25\linewidth]{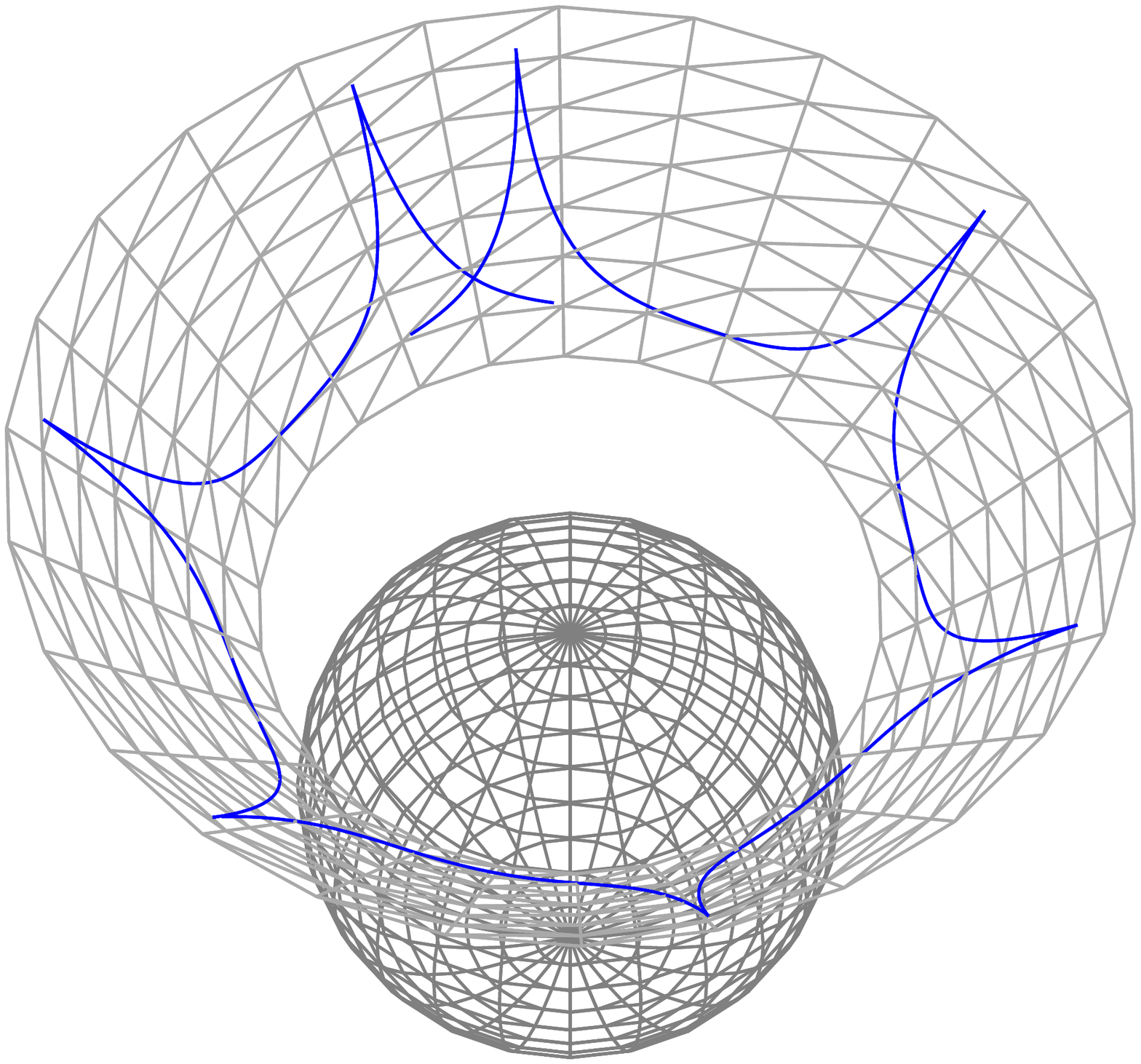}\label{fig:orbit9_1}%
	}\quad
	\subfloat[Projection of \csubref{fig:orbit9_1} onto the $x$-$y$ plane.]{%
		\includegraphics[width=.25\linewidth]{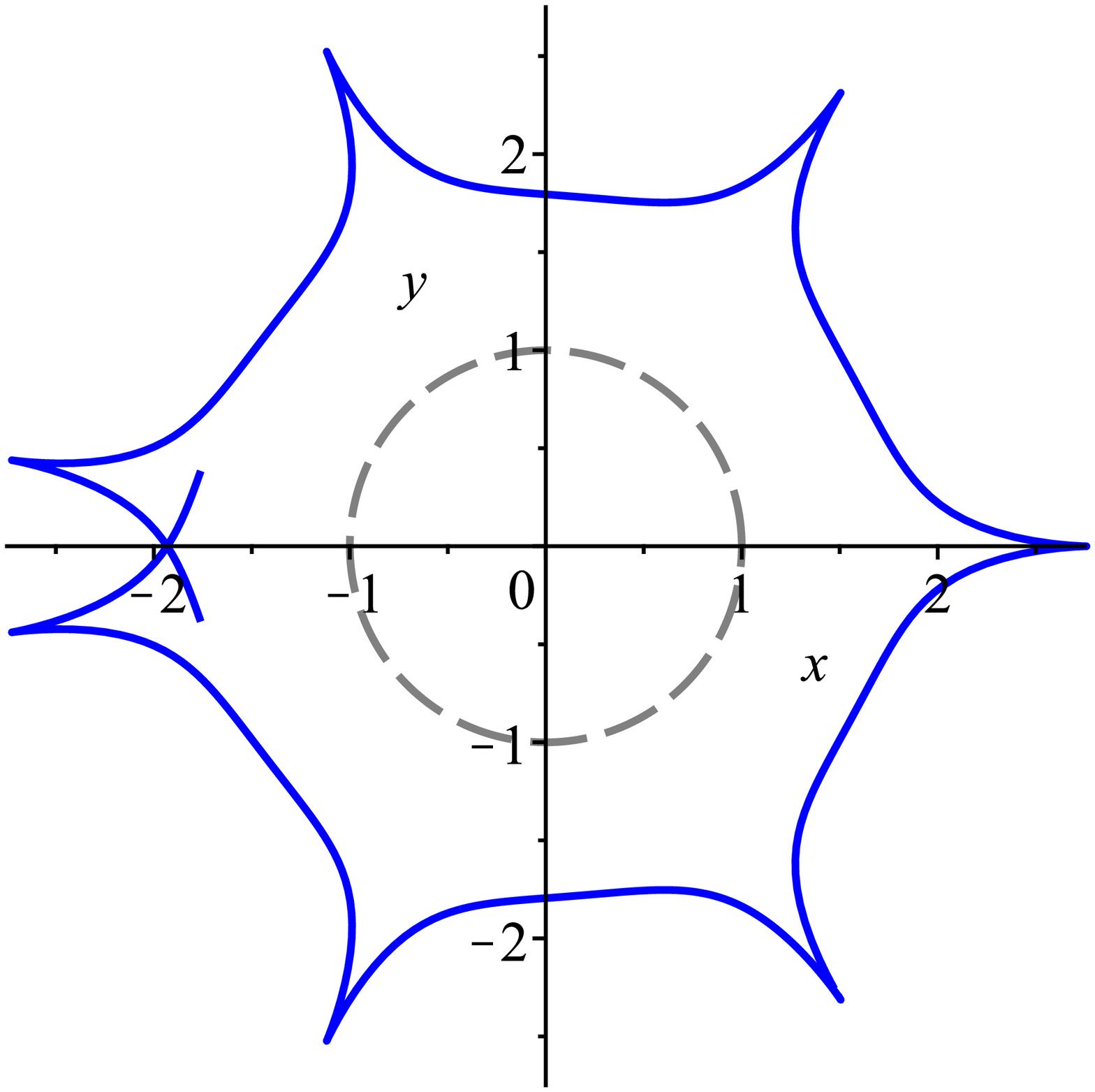}%
	}
	\caption{\label{fig:orbit9}%
		\textit{Pointy petal} BO for parameters $K = 0,\, L = 17.1,\, \delta = 1,\,
		l = 1,\, J = 0$, and energy $E = 0.8686915$. The trajectory lies on a cone
		with opening angle $\theta=\pi/2$. The sphere and dashed circle show the
		horizon at $R=1$. The corresponding effective potential and turnaround
		energy of the $r$ and $\phi$ motion, respectively, can be found in
		Fig.~\ref{fig:effpotr1g}. The particle periodically stops at the outer turning
		points.}
\end{figure*}

The \textit{pointy petal} bound orbit of type E of a particle that is
periodically at rest is depicted in Fig.~\ref{fig:orbit9}, its effective
potential in Fig.~\ref{fig:effpotr1g}. Here parameters are chosen by determining
the intersection of the effective potential $V_\pm$ and turnaround energy
$E_\text{turn}$ while $\theta$ is, for simplicity, chosen to be constant, as
discussed in Section~\ref{sec:static}.

\begin{figure*}
	\centering
	\subfloat[BO.]{%
		\includegraphics[width=.25\linewidth]{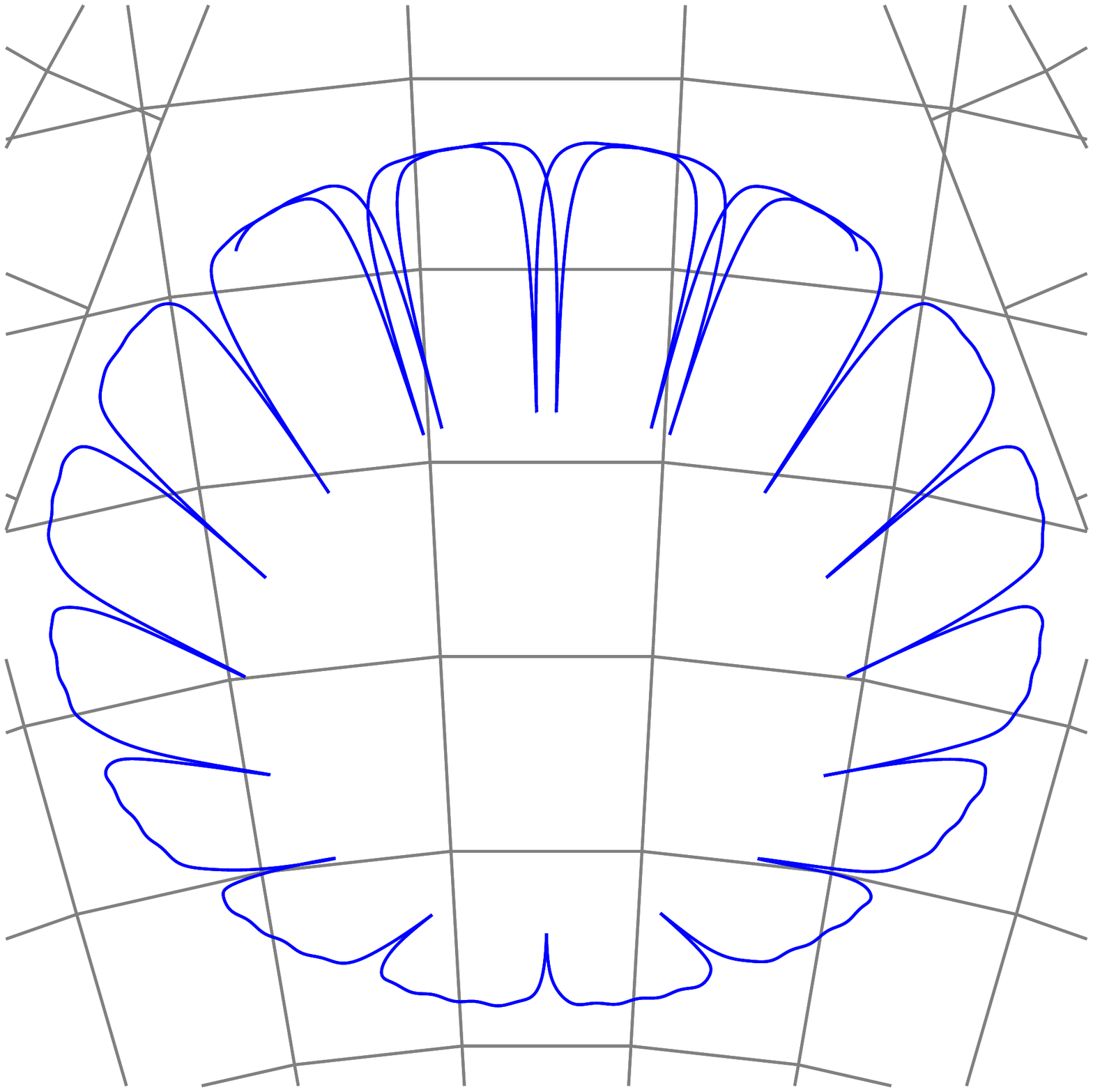}\label{fig:orbit8_1}%
	}\quad
	\subfloat[Projection of \csubref{fig:orbit8_1} onto the $x$-$y$ plane.]{%
		\includegraphics[width=.25\linewidth]{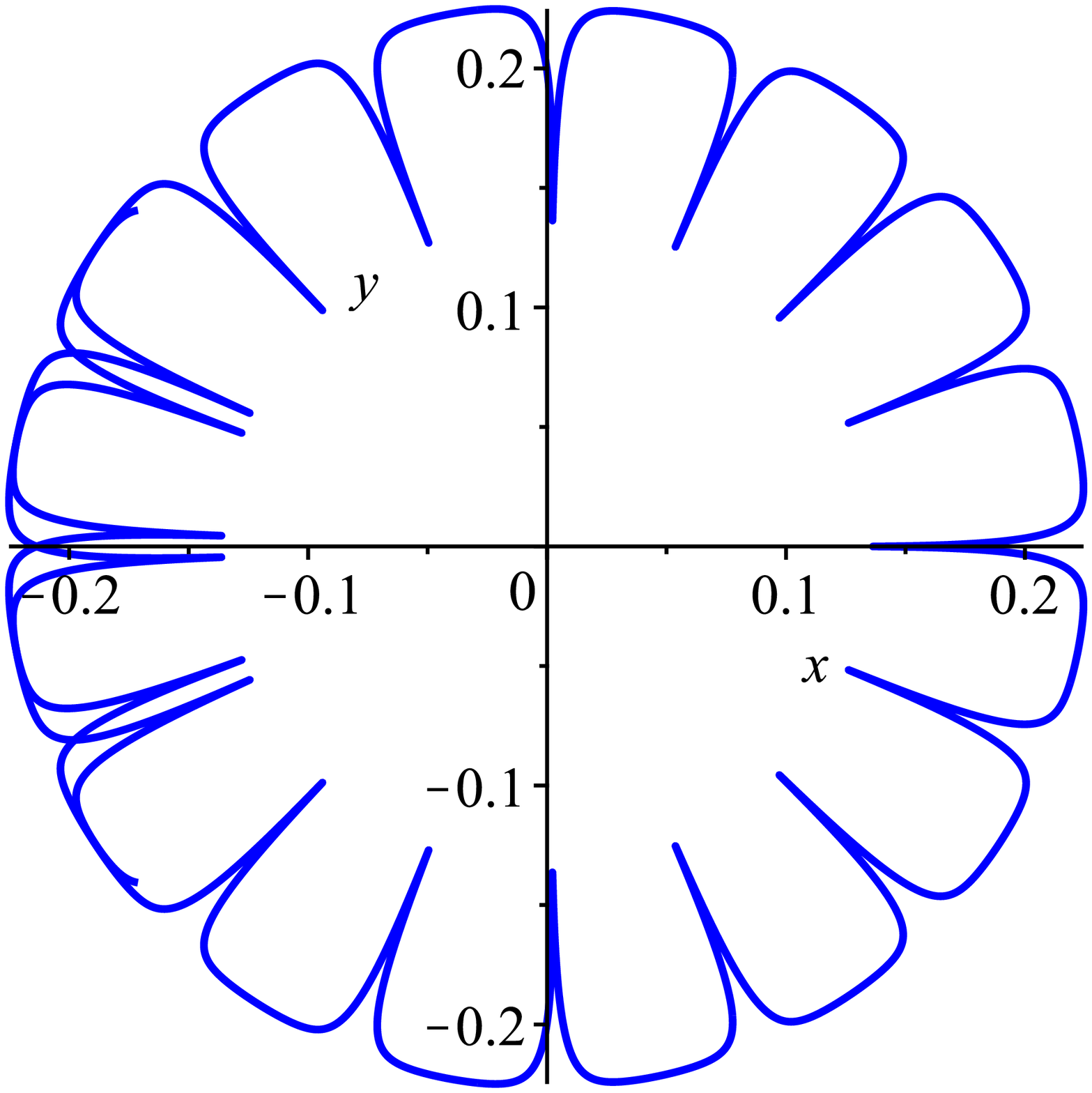}%
	}
	\caption{\label{fig:orbit8}%
		BO for parameters $K = 0.1,\, L = 80,\, \delta = 1,\, l = 1,\, J = 0$, and
		energy $E = 81.5082984$. The corresponding effective potential of the $r$
		motion is very similar to the one shown in Fig.~\ref{fig:effpotr1d}.}
\end{figure*}

Finally, a bound orbit behind the horizon for a particle of positive mass (type
F) is shown in Fig.~\ref{fig:orbit8}. This requires small $K$ and large $L$. Its
effective potential is very similar to the one shown in Fig.~\ref{fig:effpotr1d}
but its trajectory is not confined to a cone of fixed opening angle $\theta$.

\section{\label{sec:conclusion}Conclusion}
In this article the spacetime of a supersymmetric AdS$_5$ black hole was studied
by analyzing the geodesics (elliptic) equation of motion and deriving its
analytical solutions in terms of the Weierstrass $\wp$, $\sigma$, and $\zeta$
functions. Effective potentials and parametric diagrams were used to classify
possible types of orbits, which are characterized by the particle's energy,
angular momenta, Carter constant, and mass parameter as well as the metric's AdS
radius.

We showed that timelike orbits are always bounded and thus do not reach the AdS
boundary. For lightlike and spacelike geodesics multiple types of orbits with a
boundary at infinity and therefore with relevance for AdS/CFT were found. This,
for spacelike geodesics and for a specific energy, includes the possibility of a
terminating orbit. Bound orbits behind the horizon (for large angular momentum
$L$) and many-world periodic bound orbits are possible independently of the
particle's mass. However, stable bound orbits outside of the horizon are only
possible for particles of positive mass.

Future work might focus on extending the equations of motion to particles with
electric and magnetic charge. Additionally, the orbits' observables, e.g., in
case of a bound orbit, its periastron shift or, in case of a lightlike escape
orbit, its light deflection or the black hole's shadow, might be calculated
similarly as in \cite{Hackmann:2010zz} by making use of the analytical
solutions.

\begin{acknowledgements}
	We would like to thank Jutta Kunz and Lucas G. Collodel for fruitful
	discussions. We gratefully acknowledge support by the DFG (Deutsche
	Forschungsgemeinschaft/German Research Foundation) within the Research
	Training Group 1620 ``Models of Gravity''.
\end{acknowledgements}

\bibliographystyle{spphys}
\bibliography{references}
\end{document}